\documentclass[letterpaper, 10 pt, conference]{IEEEtran}  

\IEEEoverridecommandlockouts                              

\usepackage{cite}
\usepackage{amsmath,amssymb,amsfonts}
\usepackage{algorithmic}
\usepackage{mathtools, nccmath}
\usepackage{graphicx}
\usepackage{color,soul}
\usepackage{textcomp}
\usepackage{xcolor}
\usepackage{gensymb}
\usepackage{nidanfloat}
\usepackage{gensymb}
\usepackage{enumerate}
\usepackage{cite}
\usepackage{subcaption}

\sethlcolor{yellow}

\begin{document}


\title{SCARE: \underline{S}ide \underline{C}hannel \underline{A}ttack on In-Memory Computing for \underline{R}everse \underline{E}ngineering }

\author{\IEEEauthorblockN{Sina Sayyah Ensan, Karthikeyan Nagarajan, Mohammad Nasim Imtia Khan, and Swaroop Ghosh}
\IEEEauthorblockA{\textit{School of Electrical Engineering and Computer Science, Pennsylvania State University, University Park, PA 16802 USA} \\
(sxs2541, kxn287, muk392, szg212)@psu.edu}
}

\maketitle

\begin{abstract}
In-memory computing (IMC) architectures provide a much needed solution to energy-efficiency barriers posed by Von-Neumann computing due to the movement of data between the processor and the memory. The functions implemented in such in-memory architectures are often proprietary and constitute confidential Intellectual Property (IP). Our studies indicate that IMC architectures implemented using Resistive RAM (RRAM) are susceptible to Side Channel Attack (SCA). Unlike the conventional SCAs that are aimed to leak private keys from cryptographic implementations, SCARE (\underline{SCA} on IMC for \underline{r}everse \underline{e}ngineering) can reveal the sensitive IP implemented within the memory. Therefore, the adversary does not need to perform invasive Reverse Rngineering (RE) to unlock the functionality. We demonstrate SCARE by taking recent IMC architectures such as Dynamic Computing In Memory (DCIM) and Memristor Aided Logic (MAGIC) as test cases. Simulation results indicate that $AND$, $OR$, and $NOR$ gates (which are the building blocks of complex functions) yield distinct power and timing signatures based on the number of inputs making them vulnerable to SCA. Although process variations can obfuscate the signatures due to significant overlap, we show that adversary can use statistical modeling (using foundry-calibrated simulations or fabricating known functions in test chips) and analysis to identify the structure of the implemented function (e.g., $x_1 \overline{x_2} + x_3$). Furthermore, SCARE can find the implemented IP by testing limited number of patterns. For example, the proposed technique reduces the number of patterns by 64\% compared to a brute force attack for $a+bc$ function. Additionally, analysis shows improvement in SCARE’s detection model due to adversarial change in supply voltage for both DCIM and MAGIC. We also propose countermeasures such as redundant inputs and expansion of literals. Redundant inputs can mask the IP with 25\% area and 20\% power overhead. However, functions can be found by greater RE effort. Expansion of literals incurs 36\% power overhead. However, it imposes brute force search by the adversary for which the RE effort increases by 3.04X.



\end{abstract}

\section{Introduction}

In Von-Neuman architecture, transistor scaling has improved memory and processing units asymmetrically, leading to gaps in performance/energy consumption. Faster processing units need frequent and energy-intensive data-transfers from slower memory, which imposes leads to a degradation in overall system performance. Processing data within the memory will avoid Von-Neumann bottleneck and improves system performance and power consumption.

In-Memory Computing (IMC) is a promising compute model to minimize the data transfer between processor and memory by confining data within compute-capable memory. Several works have been proposed to move compute logic closer to the main memory or infuse the processing ability into memory cells \cite{machine6t, hamid-iccd, deepsram, bitstt, magicjo, pinatubo}. IMC can perform specific tasks such as dot-products that are used for recognition \cite{deepsram} and search \cite{ftcam}, or support a wide range of logic \cite{mpim} and arithmetic operations (e.g. matrix multiplication \cite{prime}). The compute capability of conventional memories such as Static RAM (SRAM) and Dynamic RAM (DRAM) have been heavily studied \cite{X_SRAM}, \cite{DRAM-OM1}, \cite{DRAM-OM2}. IMC is also achievable by using emerging Non-Volatile Memories (NVMs)  e.g., Resistive RAM (RRAM), Spin Transfer Torque (STT) Magnetic RAM, Phase Change Memory, etc. RRAM based IMC architectures in particular, has exhibited significant promise due to low power consumption, fast operation, and high integration density ($4F^2$ footprint in crossbar architecture \cite{imaging}). RRAM provides higher sense-margin and many analog states compared to other NVMs e.g., STTRAM. Although RRAM may not be preferred for storage applications due to poor write endurance, it is useful for IMC since write-operation is only needed once (to program the function). Furthermore, unlike read operation of RRAM, IMC does not require constant DC current, which leads to higher endurance.
In this paper, we propose SCARE (\underline{SCA} on IMC for \underline{r}everse \underline{e}ngineering) taking RRAM based IMC architectures as test cases which can reveal IMC Intellectual Property (IP).

\begin{figure} [t]
	\centering
    \includegraphics[trim=4 8 2 4,clip,width=\linewidth]{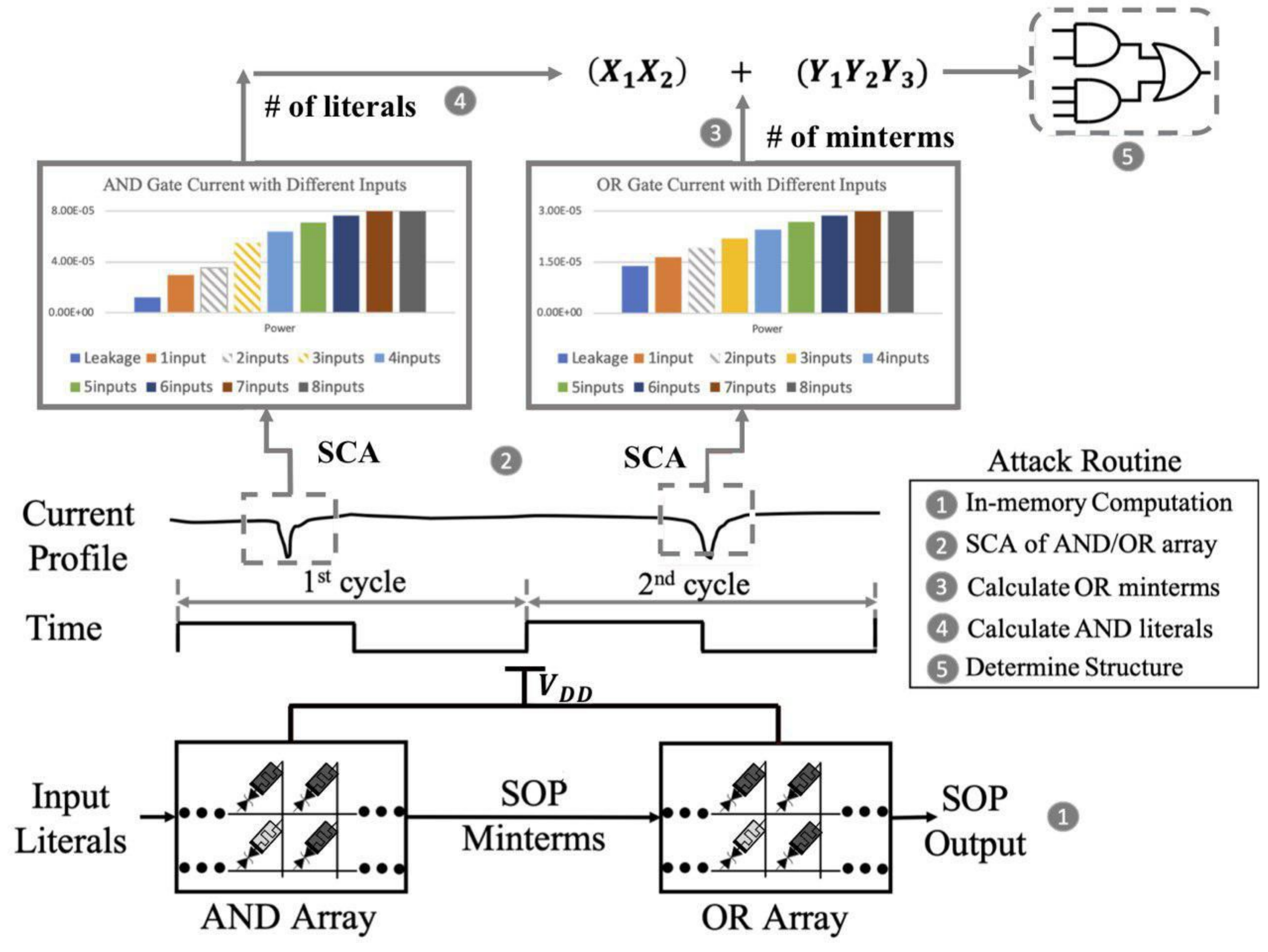}
	 \vspace{-6mm}
	\caption{Example of SCA to extract the structure of implemented function in IMC.}
	\label{Example}
	 \vspace{-6mm}
\end{figure}

Existing IMC architectures such as, Dynamic Computing In Memory (DCIM) \cite{hamid-iccd}, Memristor Aided Logic (MAGIC) \cite{magic}, material implication (IMPLY) \cite{imply}, etc. are only capable of implementing functions with limited number of inputs. Implementing large functions using DCIM reduces Sense Margin (SM) \cite{hamid-iccd}. For MAGIC and IMPLY, the delay/power consumption increases exponentially with function size.  An adversary can leverage power/current signature by Side Channel Attack (SCA) to Reverse Engineer (RE) the implemented function i.e., the number of minterms and the number of input literals per minterm. The final goal of the adversary is to find the function implemented in IMC (example is shown in Fig. \ref{Example}).

\textbf{Example of SCARE attack}: A simplified example of the $SCARE$ attack is shown in Fig. {\ref{Example}}. It considers that the IMC operation is carried out in two cycles to compute a function in the Sum-of-Product (SOP) form. The first cycle computes the $AND$ and the second cycle computes the $OR$ {\cite{hamid-iccd}}. In this example, $SCARE$ involves the following steps: 
(1) extraction of current profile  during the compute cycles; 
(2) matching the $2^{nd}$ cycle current profile with one of the pre-calculated current-profile models to determine the number of minterms implemented within the $OR$ array. Here, the minterms represent the fanin of the $OR$ gate; 
(3) matching the $1^{st}$ cycle current profile with one of the pre-calculated current-profile models to determine the number of literals in the $AND$ array i.e., the fanin of the $AND$ gates. Knowing the number of minterms from the $2^{nd}$ cycle allows the adversary to determine the number of input literals per SOP minterm; 
(4) after finding the function structure, extracting the implemented function by applying a limited number of patterns to the chip and validating using a golden chip. 

\textbf{Baseline attack model:} We have assumed that adversary can obtain the pre-computed current profile models developed through foundry-calibrated simulations or by fabricating small known functions in a test chip (more details in Section \ref{modeling}) to aid in the RE process. For the sake of brevity, this paper focuses on two RRAM based IMC architectures, namely DCIM \cite{hamid-iccd} and MAGIC \cite{magic} to demonstrate $SCARE$. However, the attack can also be implemented on other emerging IMC with minor changes. Note that process variation can lead to a large variation in power/timing profile of implemented gates. This, in turn, can make RE challenging due to the overlap of signatures from two different gates (e.g., 2-input AND gate power can overlap with 3-input AND due to variations). However, adversary can use statistical analysis of power/delay to filter the correct function (details in Section \ref{DCIM-attack} and \ref{MAGIC-attack}). IMC can improve power-efficiency by cutting down sensor data movement in IoT/mobile devices whereas in servers, it can improve performance and reduce cooling costs by lowering power dissipation. Our attack model is suited for IoT/mobile where the users/adversaries will have access to the devices/power port.

\textbf{Distinctions from memory SCA}: SCAs on traditional memories have been studied in the past. Side channel leakage of ASIC components is investigated in \cite{ASIC}. It shows that SRAMs have geometrically uniform structures and their leakage closely follows a generic hamming distance. The observation is based on SRAM write operation that flips the data stored in the memory cell. The data dependent write current can be leveraged to launch simple power analysis attacks. Emerging memories provide better scope of SCA due to higher write and read currents compared to SRAM. Differential Power Analysis (DPA) on STTRAM have been investigated in \cite{nasim-SCA} to decode the memory contents. Note that the write current based SCA will not work for all IMC architectures due to the absence of write operation (e.g., DCIM) during computing. Furthermore, IMC computing is different than the read operation due to the absence of static current during computing. Therefore, read current based SCA is not directly applicable to IMC. Furthermore, the objective of SCA based RE is to extract the implemented function in contrast to SCA based key extraction which only identifies the hamming weight of the data.

\begin{figure*} [!t] 
        \centering \hspace{-2mm}
        \begin{subfigure}[b]{0.54\textwidth}
                \centering
                \includegraphics[trim=4 10 4 4,clip, width=0.99\linewidth]{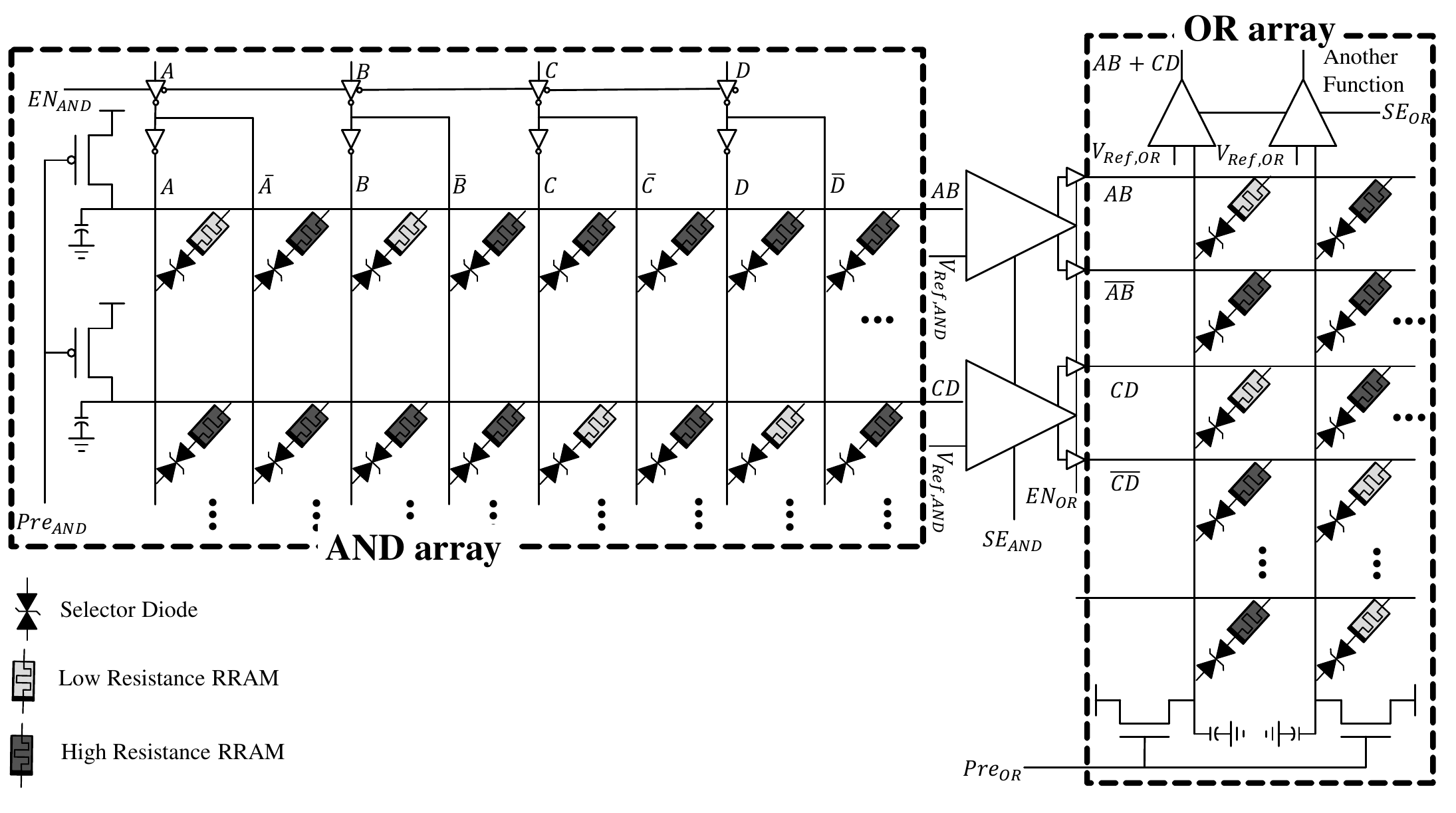}
                \caption{}
                \label{DCIM-a}
        \end{subfigure}%
        \hspace{-2mm} 
        \begin{subfigure}[b]{0.40\textwidth}
                \centering
                \includegraphics[trim=2 2 2 2,clip, width=0.99\linewidth]{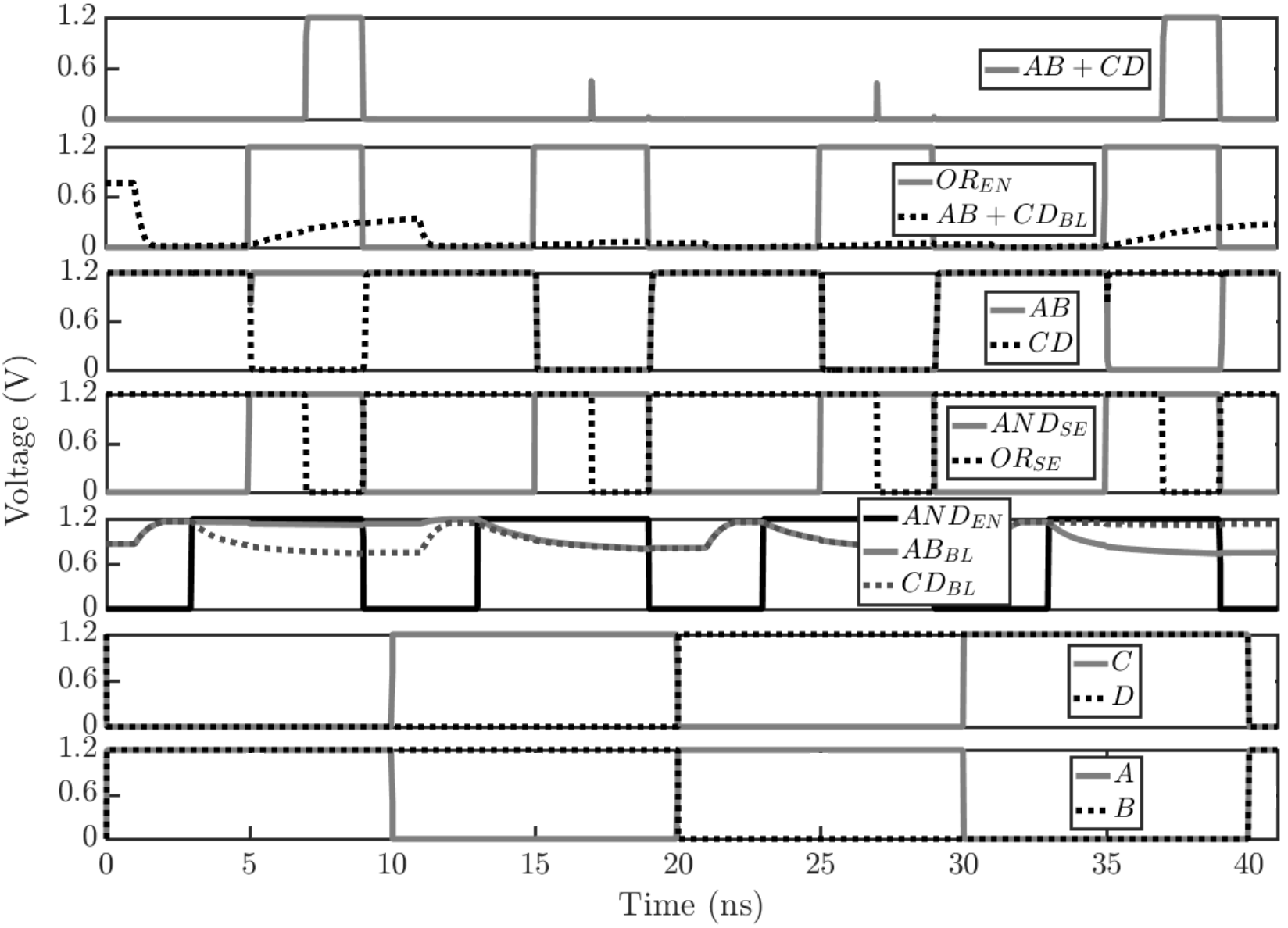}
                \caption{}
                \label{DCIM-b}
        \end{subfigure}%
        \hspace{-2mm} \vspace{-2mm}
        \caption{(a) Implementation of $AB + CD$ using DCIM \cite{hamid-iccd} in RRAM crossbar; and, (b) timing diagram of function implementation.}
        \label{DCIM}
        \vspace{-6mm}
\end{figure*}

\textbf{Distinctions from conventional RE}: RE is generally an invasive and destructive form of analyzing integrated circuits (IC) where an adversary grinds away each layer of an IC and captures optical images. The base layer provides gate types and upper layers provide their connectivity. By combining the information, the IP could be unlocked. In contrast, $SCARE$ is a non-invasive RE approach that exploits SCA to extract IP implemented in emerging IMC. This eliminates the need for highly expensive and invasive forms of RE for emerging IMC based computing. With SCARE, the extracted structure reveals the number of input literals and the number of SOP minterms. In essence, the structure reveals the number and input-type of each of the $AND$ and $OR$ gates present in the function. This extracted structure can be further used to design a limited number of input patterns to determine the overall SOP function (not covered in this paper). \textit{Based on our literature survey, $SCARE$ is the first work on RE of IMC based IP}.

In particular, we make the following contributions in this paper. We,
\begin{enumerate} [(a)]
\item investigate SCA on IMC architectures for non-invasive RE of IP;
\item exploit side channel current profiles to identify the gate structures of the implemented
functions; 
\item propose two attack models for DCIM and MAGIC, respectively. One works for true inputs only and other one works for both, true and complementary inputs;
\item conduct PV analysis of the IMC architectures to develop an SCA comparison model;
\item propose countermeasures such as, redundant inputs and expansion of literals to protect from SCARE.
 \end{enumerate}

The rest of the paper is organized as follows: Section II introduces the basics of RRAM, DCIM, and MAGIC IMC architectures, background on SCA and simulation setup; Sections III and IV describe the proposed attack models on DCIM and MAGIC, respectively and the results; Section V presents countermeasures; Section VI presents discussion; Finally, Section VII draws the conclusions.

\section{Background and Related Work}
\subsection {Basics of RRAM}
RRAM is a two terminal device, which resistive switching layer is sandwiched between two electrodes. Switching from Low Resistance State (LRS) to High Resistance State (HRS) is called `reset' process and switching from HRS to LRS is called `set' process. The resistance of the insulator layer changes from LRS (HRS) to HRS (LRS) depending on the voltage polarity between the two terminals \cite{MIM}.

\subsection {Basics of IMC}

\subsubsection{DCIM}
 We have implemented the DCIM architecture proposed in \cite{hamid-iccd}. Fig. \ref{DCIM-a} shows the implementation of $AB + CD$ as an example and  Fig. \ref{DCIM-b} shows the corresponding timing waveforms. Each memory cell consists of an RRAM connected in series with a selector diode. Functions are implemented in SOP form using pre-programmed memory arrays. Separate arrays for $AND$ and $OR$ operations are needed to implement the logical functions. Inputs are given to the arrays through Wordlines (WL). Final Bitline (BL) voltages are considered as outputs ($AND$ array BLs implement minterms e.g., $AB$ and $CD$, and, $OR$ array BLs implement functions). Any LRS RRAM in $AND$ array's BLs is considered as a literal to the respective BL and any LRS RRAM in $OR$ array's BLs serves as a minterm for the implemented function. 

Initially, $AND$ array's BLs are pre-charged to $V_{DD}$ by asserting $Pre_{AND}$. Then, inputs are applied to the RRAMs by activating $EN_{AND}$ signal. If one of the literals of $AND$ array's BLs is `0', the respective BL gets discharged and its voltage drops below the Sense Amplifier (SA) reference voltage ($V_{Ref-AND}$) and the minterm is considered as `0'. If all the inputs are logically `1', the BL holds its pre-charged value which is higher than the reference voltage and is considered as `1'.

The BLs of $OR$ array are initially pre-discharged to 0v. After activating $OR_{EN}$ signal if one of the input literals in a BL is logically `1' it can charge up the BL to a value higher than $OR$ array's reference voltage ($V_{Ref-OR}$). Finally, the voltage of $OR$ array's BLs are compared against $V_{Ref-OR}$ at the edge of $SE_{OR}$ and output is generated.

\subsubsection{MAGIC} 
\begin{figure} [t]
	\centering
	\includegraphics[width=0.85\linewidth]{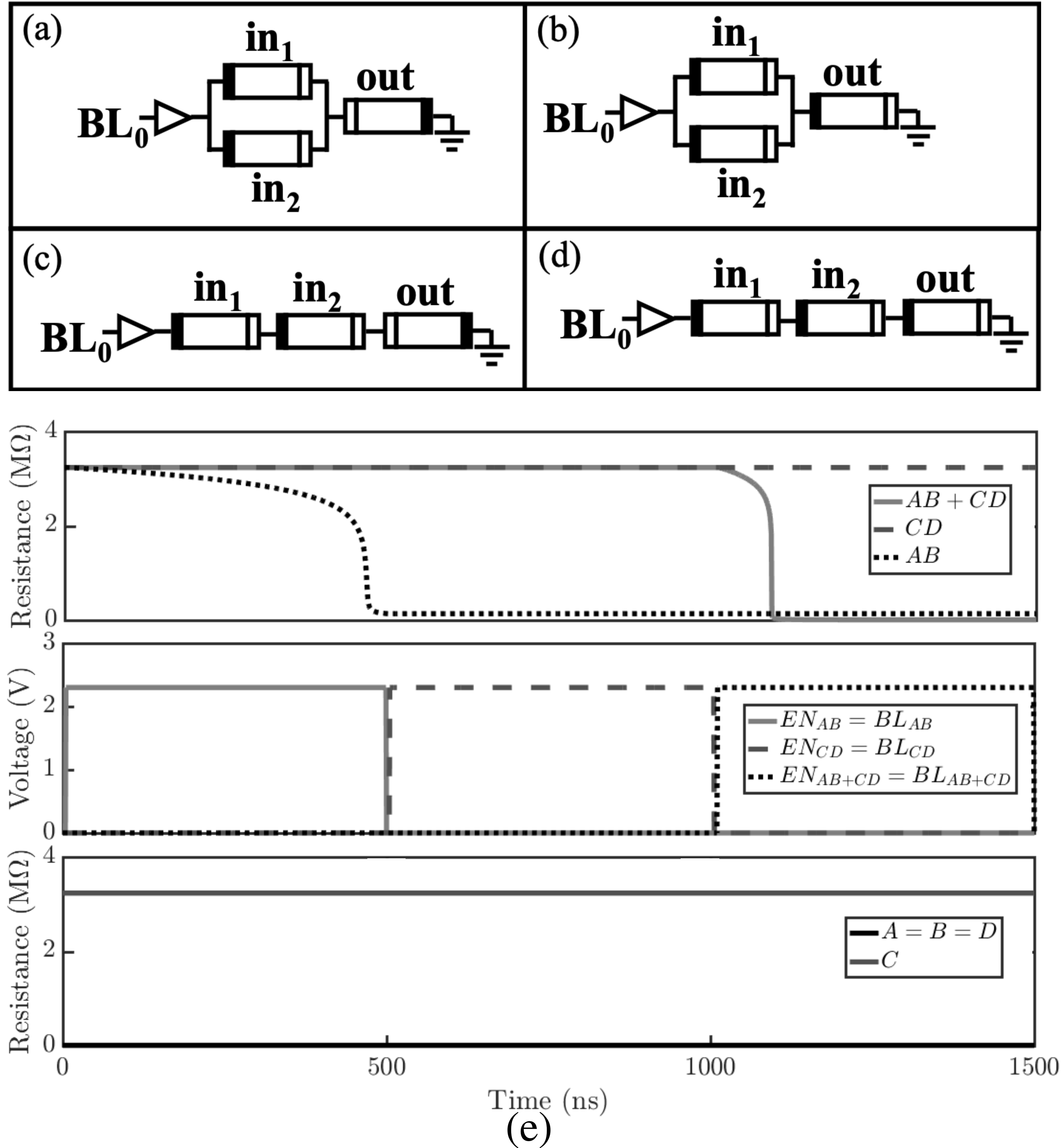} \vspace{-2mm}
	\caption{Implementation of MAGIC gates, (a) NOR; (b) OR; (c) NAND; and, (d) AND.}
	\label{MAGIC_Gates}
		\vspace{-2mm}
\end{figure}

{\label{MAGIC}}
We have also implemented MAGIC architecture proposed in \cite{magic} that employs memristors (RRAM in this paper) to implement logic gates. A number of memristors serve as inputs with previously stored data while an additional memristor acts as the output. Gates including MAGIC- $NOR$, $AND$, $OR$, and $NAND$ are shown in Fig. \ref{MAGIC_Gates} (a)-(d), respectively. MAGIC's logical state is represented as a resistance, where the HRS and LRS represent logical `0' and `1' respectively. Fig. \ref{MAGIC_Gates} (e) shows the implementation of $AB + CD$ as an example. It consists of two 2-fanin AND gates and one 2-fanin OR gate. Here, the input RRAMs, $A, B,$ and $D$, are initialized to logical `1' (LRS) and RRAM $C$ is initialized to logical `0' (HRS). All output RRAMs are initialized to `0' (HRS). In the first cycle, $AB$ is computed by asserting its bitline driver ($BL_{AB}$) using the enable ($EN_{AB}$) signal. Since $AB = 1$, $AB$'s output RRAM switches from `0' (HRS) $\rightarrow$ `1' (LRS). Similarly, during the second cycle, when  $BL_{CD}$ is asserted using the $EN_{CD}$ signal, $CD$'s output RRAM remains at `0' (HRS). During the final cycle, the bitline driver for the $OR_2$ operation ($BL_{AB+CD}$) is asserted using the $EN_{AB+CD}$ signal. Fig. \ref{MAGIC_Gates} (e) shows that the final output RRAM switches from `0'$\rightarrow$`1' and reflects the correct output of $AB+CD = 1.1+0.1 = 1$.

\subsection{Background on Side Channel Attack}
SCA \cite{ASIC} is a powerful threat which targets weak implementation of systems on the chips. 
SCA exploits the unintentional signature observed in physical channels like timing \cite{CACHE}, power consumption \cite{ASIC} and electromagnetic emanation \cite{Electromagnetic} etc. with an objective to recover the sensitive data being processed e.g., cryptographic keys. Since different data bits exhibit different physical signatures (power consumption, delay), SCA can unveil the data. 
SCA on memory components targets hamming distance which is equal to the number of bit transitions. Then, a statistical dependency is tested between the hypothetical leakage computed (using simulations, test chips, and leakage models) and the measured leakage to guess the stored data. 

\subsection{Adversarial Modeling of IMC Power/Timing} {\label{modeling}}
In order to correlate the IMC power/timing (extracted using SCA) with the appropriate gate type and fanin value, the adversary requires a pre-calculated power and timing model. This model may be easily developed if the adversary has access to the foundry calibrated device models. If not, the adversary can order a limited number of test chips that implement multiple small known functions using IMC. Such opportunity is available through shuttle programs of vendors like CMP and MOSIS. The adversary can then proceed to develop a model based on the power and timing distributions calculated for different gates and input sizes.

\subsection{Simulation Setup}
Simulations are performed in HSPICE with 65nm PTM technology \cite{ptm}, ASU RRAM model \cite{asu} and bi-directional selector diode model \cite{selector}. Detailed parameters of the devices employed for simulations are shown in Table \ref{parameters}. The test chips obtained by the adversary with the known functions/gates are subject to process variations. Therefore, the adversary will only have a distribution of power profiles and operation times. In order to represent this situation, we introduce process variations to the $SCARE$ power profile/operation-time modeling by performing Monte Carlo simulations with the parameters listed in Table \ref{PV}.

\begin{table}
    \centering
    \caption{Simulation Parameters}
    \vspace{-2mm}
    \begin{tabular}{|c|c|}\hline
        Parameter & Value \\ \hline
        MOSFET Gate Length & 65 nm \\ \hline
        NMOS/PMOS Threshold Voltage & 423/-365 mV  \\ \hline
        BL Capacitance & 100 fF \\ \hline
        RRAM Gap Min/Max/Oxide Thickness & 0.1/1.7/5 nm \\ \hline
        Atomic Energy: Vacancy Generation/Recombination & 1.501/1.5 eV \\ \hline
        RRAM Write Latency & 25 ns\\ \hline
        RRAM HRS/LRS at 1.2V & 6.7M/58.9K $\Omega$ \\
        \hline\end{tabular}
        \label{parameters}
\end{table}

\begin{table}
    \centering
   
    \caption{Monte Carlo simulation parameters}
     \vspace{-2mm}
    \begin{tabular}{|c|c|c|c|}\hline
        Parameter & Real Value & Variation & STD. Deviation \\ \hline
        RRAM LRS Gap & 0.1 nm & 7\% & $3\sigma$\\ \hline
        RRAM HRS Gap & 1.7 nm & 7\% & $3\sigma$ \\ \hline
        MOS Oxide Thickness& 1.2 nm & 10\% & $3\sigma$ \\ \hline
        MOS Gate Length & 65nm & 10\% & $3\sigma$ \\ 
        \hline\end{tabular}
        \label{PV}
\end{table}

\section{Attack on DCIM Architecture}{\label{DCIM-attack}}
Adversary can distinguish power drawn by OR and AND arrays of DCIM by looking at their power signature. The AND array is pre-charged by a PMOS transistor during pre-charge phase, therefore, its power signature is negative (power is drawn from voltage supply). The OR array is pre-discharged by an NMOS transistor during pre-discharge phase, therefore, the power is signature is positive (power is dissipated by the ground node).

\subsection{Attack Model 1} {\label{DCIM-true}}
\subsubsection{Leveraging the power drawn by OR and AND array}

In DCIM, computations are performed in two cycles between $EN$ signal and $SE$ signal ($EN_{AND}$ and $SE_{AND}$, and, $EN_{OR}$ and $SE_{OR}$) activation. $SCARE$ especially considers the peaks in the power profile for three reasons: (a) $EN$ signal activates large buffers (that are upsized to charge/discharge the BLs); (b) $SE$ signal activates SAs (significant capacitive component); (c) short circuit current from $V_{DD}$ to ground through buffers/SAs (when both pull up/down networks are on).

Initially, the adversary chooses two consecutive time periods between $EN$ and $SE$ in each cycle to launch the attack. This is followed by asserting all the input signals to logical `$1$'s and recording the power profile. 
The recorded power during the second cycle (i.e., the $OR$ function) is matched with the power-profile reading from the models developed using multiple test chips/simulations. This step allows the adversary to determine the number of SOP minterms processed in the $2^{nd}$ cycle ($OR$ function input literals). By setting all the input literals to logical `$1$', the adversary ensures that each SOP minterm in the function also equals a logical `$1$'.  The BLs in the DCIM $OR$ array are pre-discharged to `$0$'. Thus, by applying logical `$1$' to all SOP minterms, $SCARE$ ensures that the BLs are charged as fast as possible (leading to the highest possible power consumption).

Once the number of SOP minterms is determined, the adversary analyzes the power profile of the $1^{st}$ cycle to determine the number of input literals in each minterm. Each of the $AND$ array input literals are set to logical `$0$' to ensure that each SOP minterm equals to `$0$'. By applying `$0$' to each minterm, the adversary ensures that the BL, which is pre-charged to $V_{DD}$, discharges at the highest possible rate. Finally, by analyzing the power profile of the operation, the adversary determines the number of input literals for each $AND$ gate (Fig. \ref{Example}). Note that the leakage power must be subtracted to analyze the function-dependent power. 

The above attack model works only with true inputs. If the function consists of complementary inputs, the attack will fail since adversary cannot confirm if all the minterms are `1' by forcing all inputs to `1'.

\begin{figure}[t]
        \centering 
        \begin{subfigure}[b]{0.5\linewidth}
                \centering
                \includegraphics[width=0.99\linewidth]{{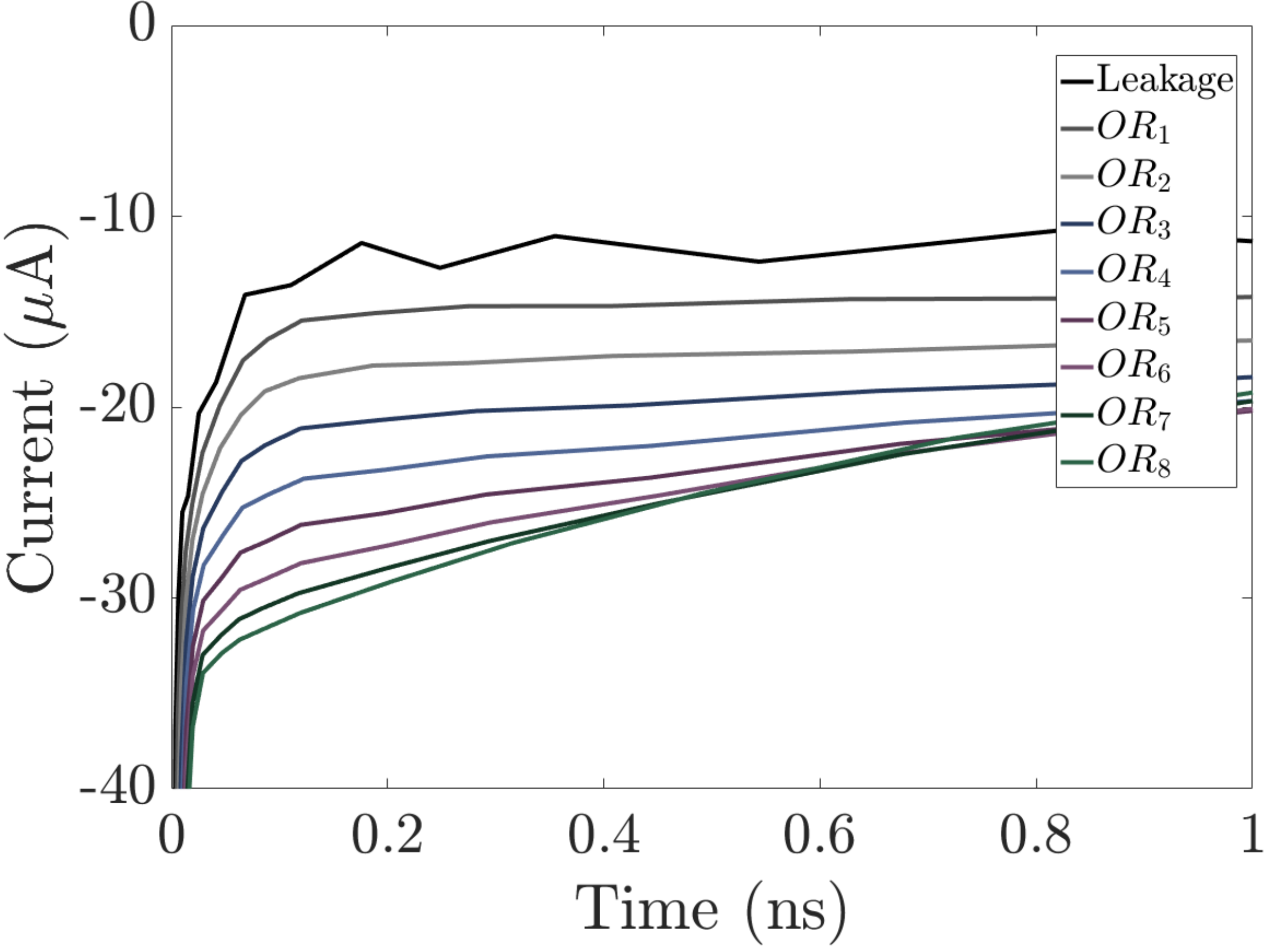}}
                \caption{}
                \label{OR-True-a}
        \end{subfigure}%
        \begin{subfigure}[b]{0.5\linewidth}
                \centering
                \includegraphics[width=0.99\linewidth]{{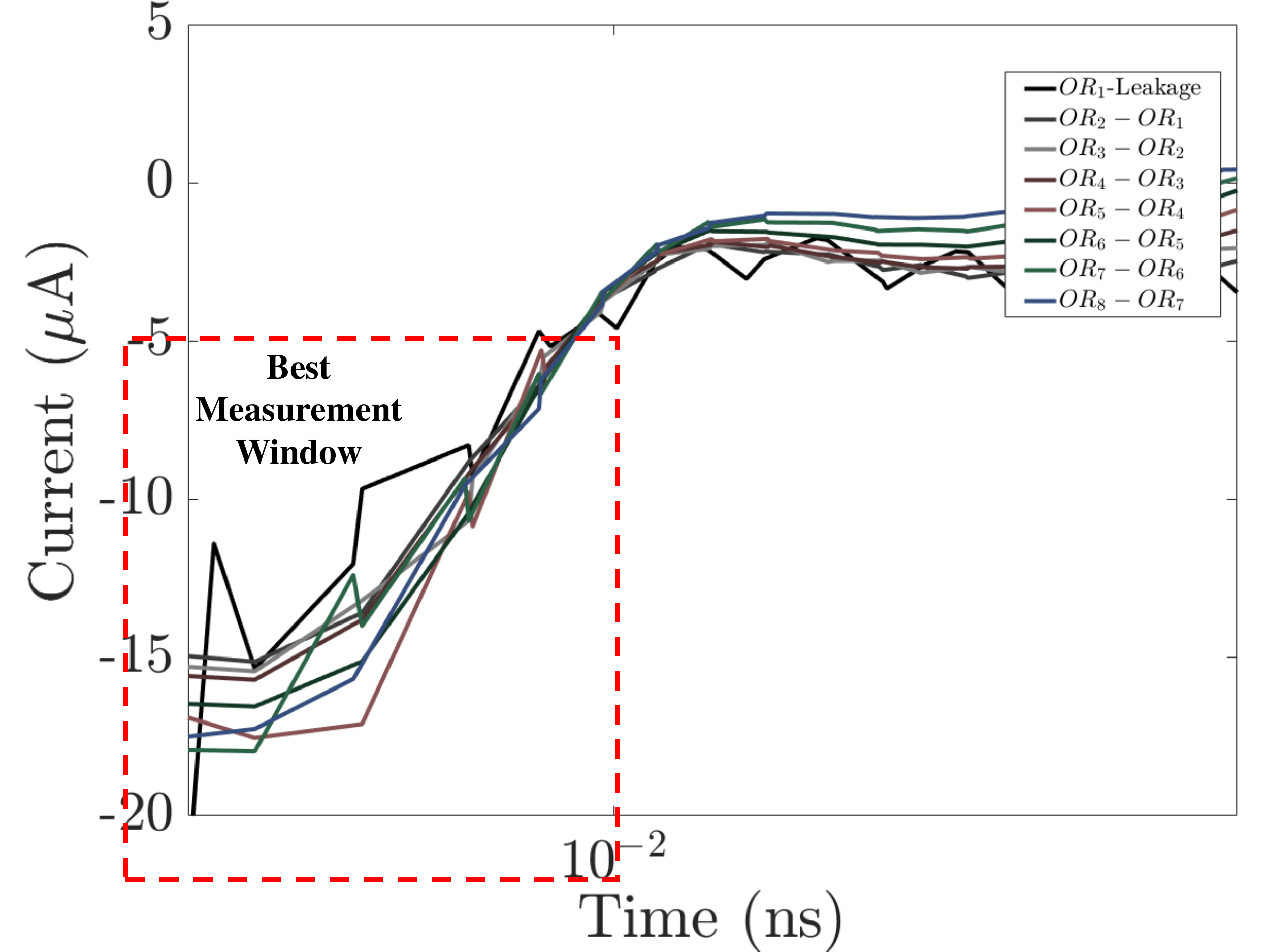}}
                \caption{}
                \label{OR-True-b}
        \end{subfigure}\\ 
              \begin{subfigure}[b]{0.5\linewidth}
                \centering
                \includegraphics[width=0.99\linewidth]{{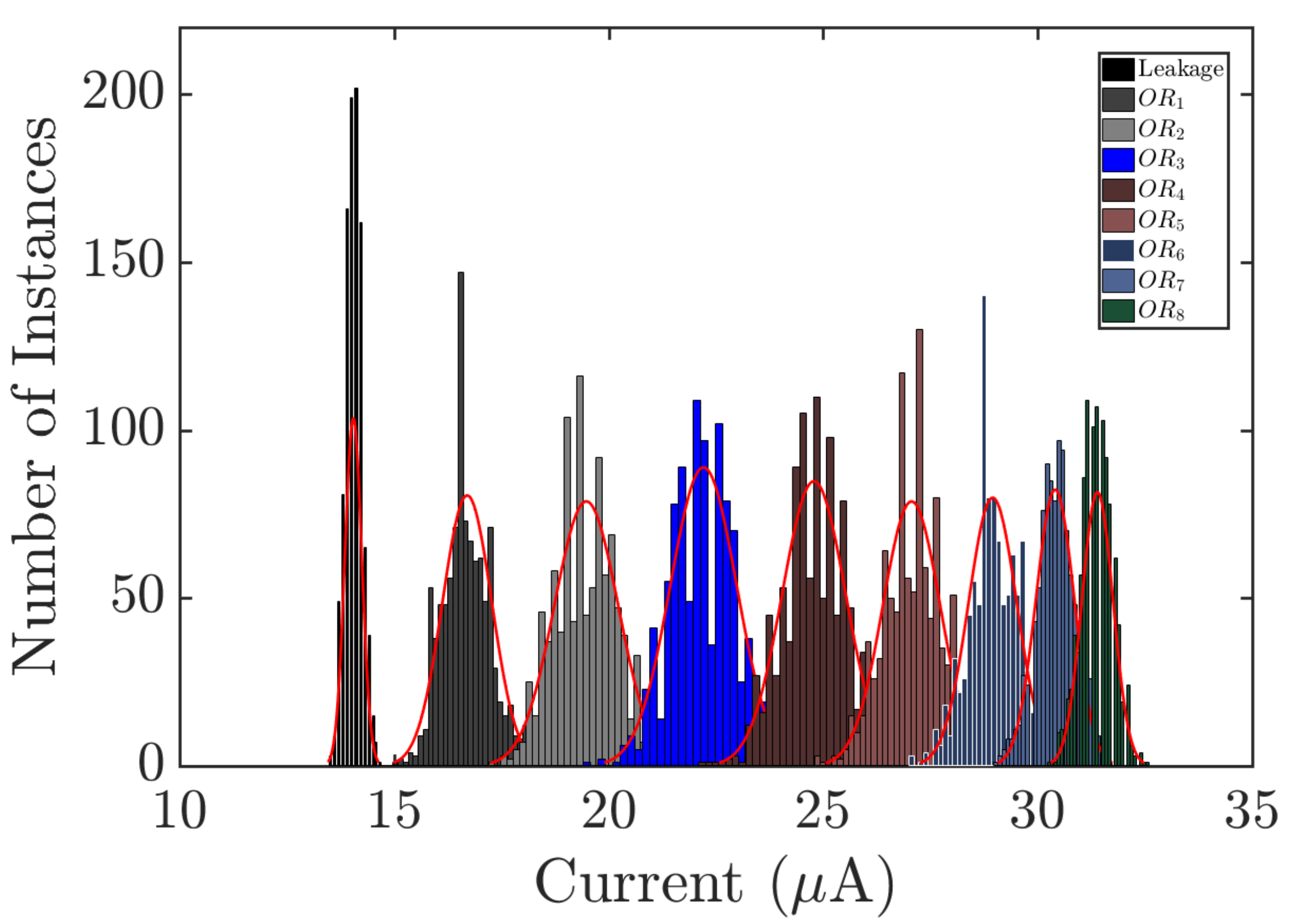}}
                \caption{}
                \label{OR-True-c}
        \end{subfigure}%
        \begin{subfigure}[b]{0.5\linewidth}
                \centering
                \includegraphics[width=0.99\linewidth]{{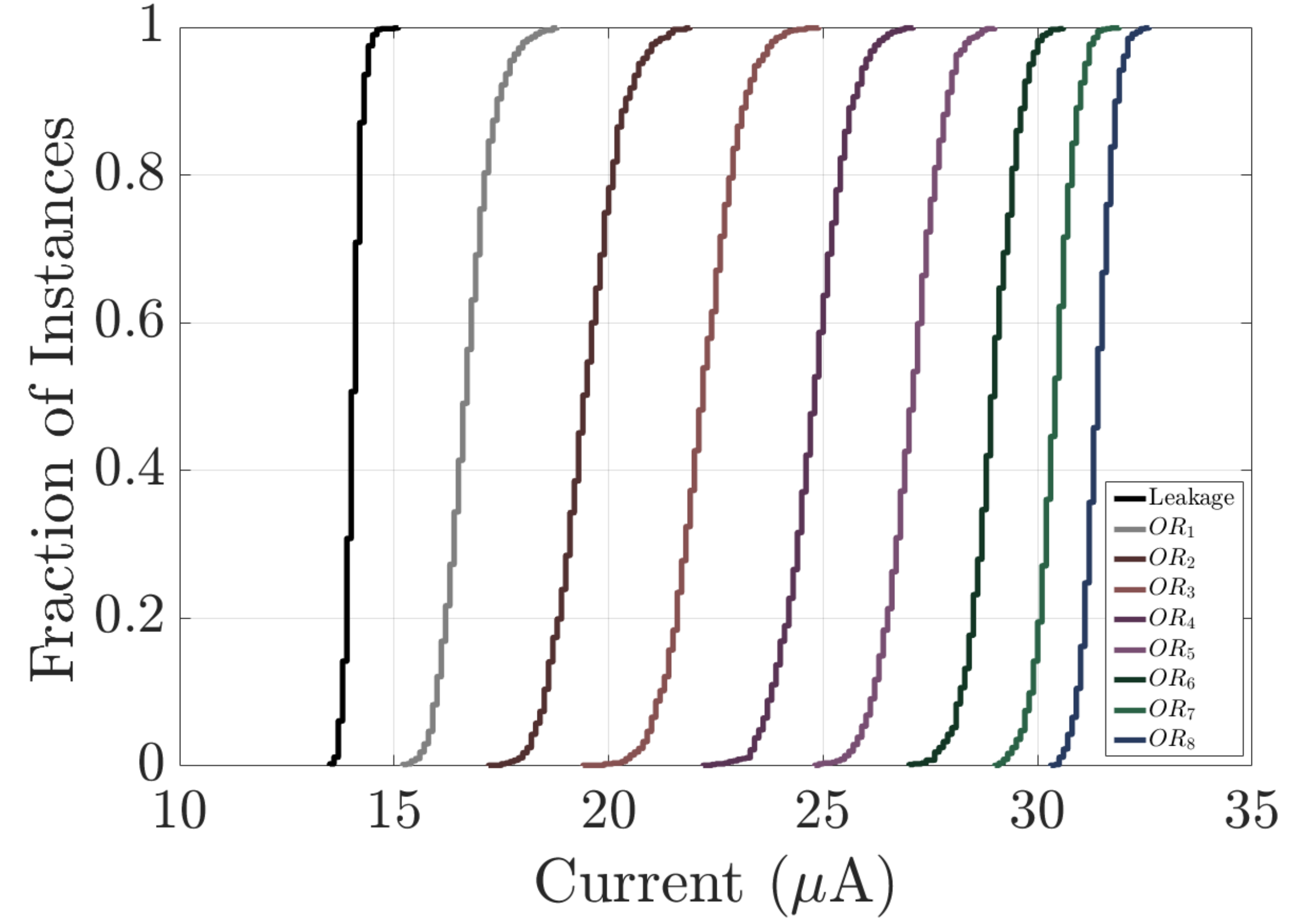}}
                \caption{}
                \label{OR-True-d}
        \end{subfigure}\\
              \begin{subfigure}[b]{0.5\linewidth}
                \centering
                \includegraphics[width=0.99\linewidth]{{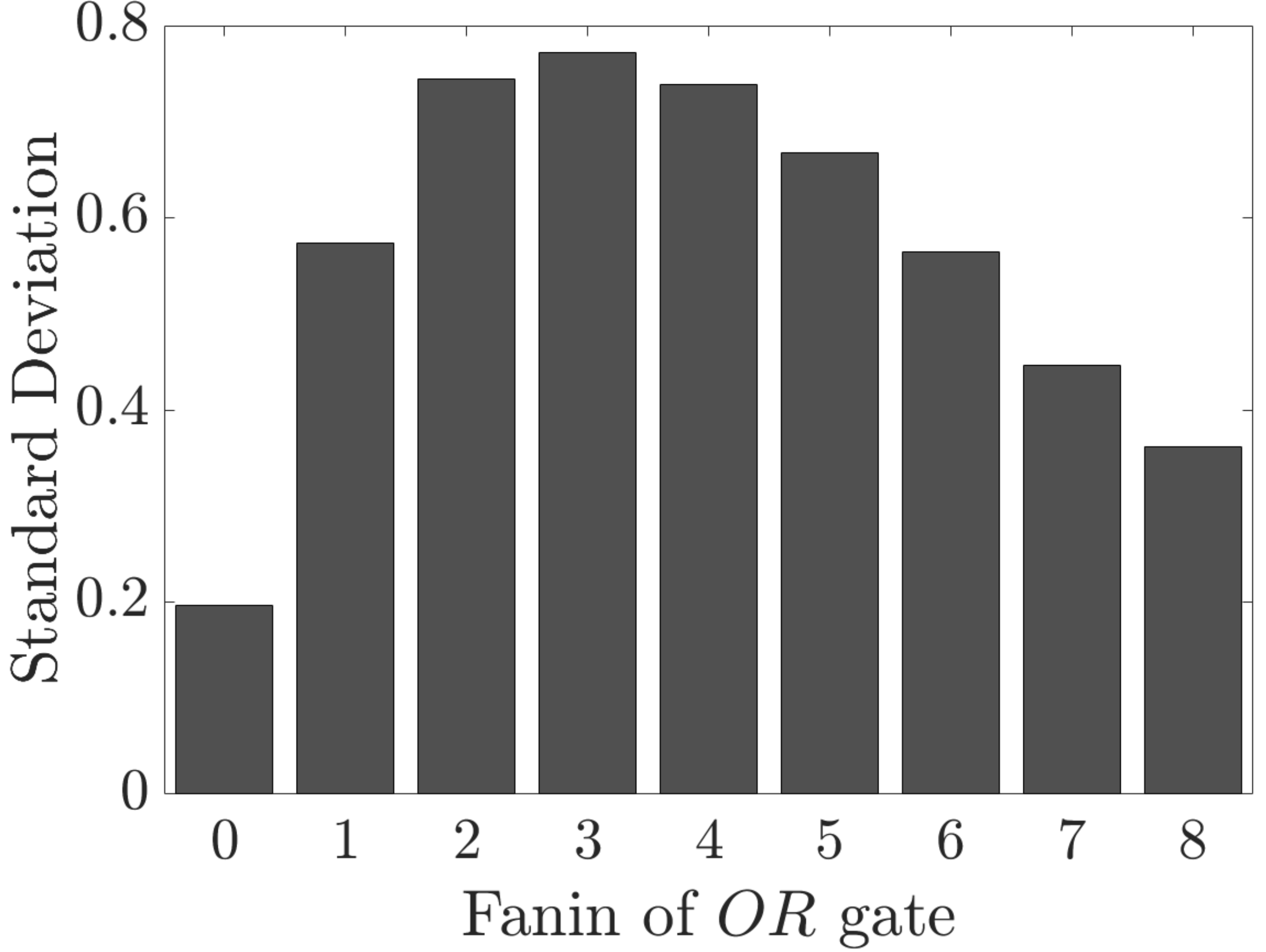}}
                \caption{}
                \label{OR-True-e}
        \end{subfigure}%
        \begin{subfigure}[b]{0.5\linewidth}
                \centering
                \includegraphics[width=0.99\linewidth]{{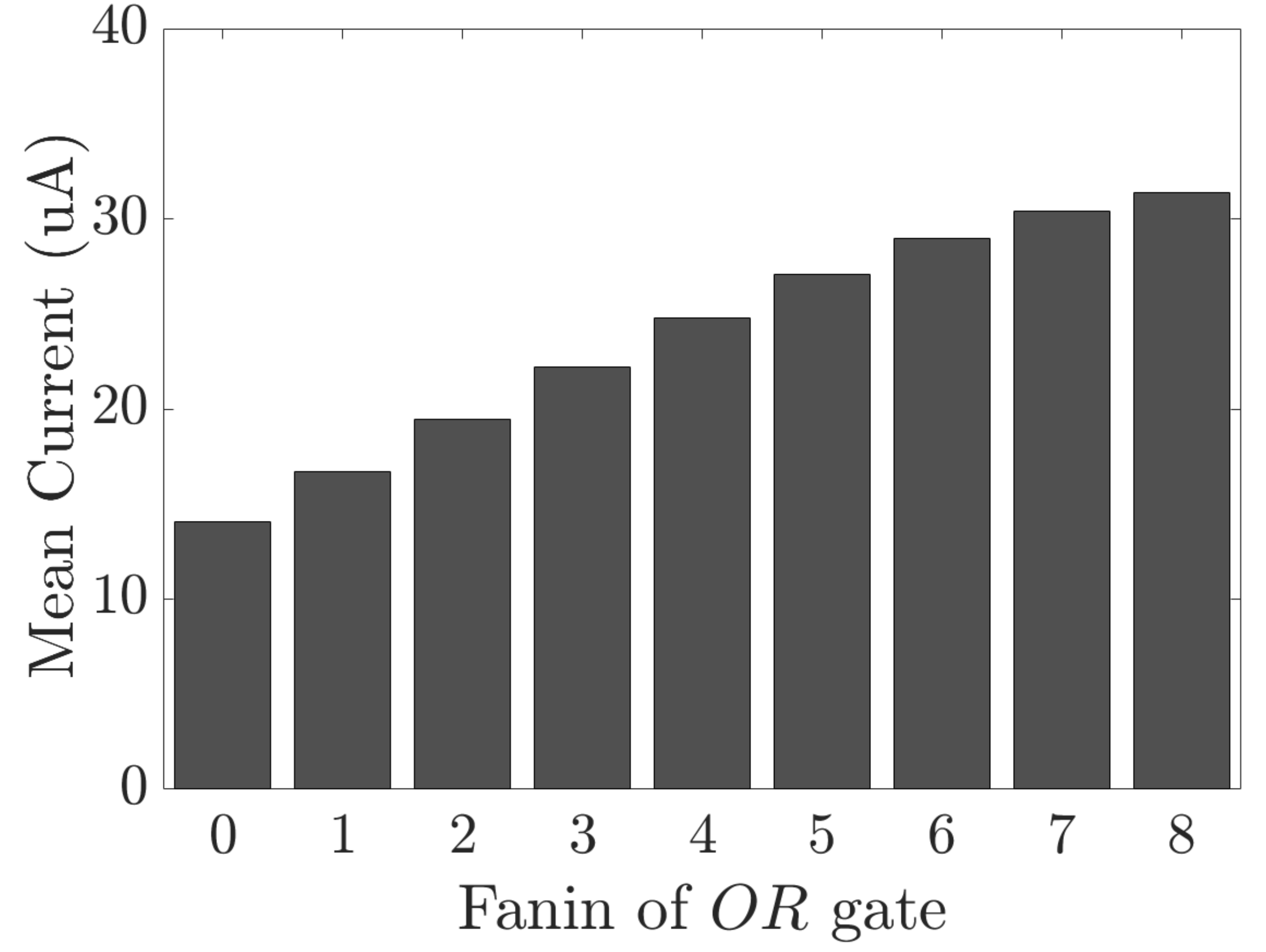}}
                \caption{}
                \label{OR-True-f}
        \end{subfigure} \vspace{-3mm}
        \caption{(a) Analysis of $OR$ gate current profile with various fanins; (b) current difference with fanin of n and n+1 to find the measurement window; (c) PDF, (d) CDF, (e) STD and, (f) mean current of $OR$ gates.}
	\vspace{-6mm}
        \label{DCIM-OR-Current}
\end{figure}

\begin{figure}[t]
        \centering 
        \begin{subfigure}[b]{0.5\linewidth}
                \centering
                \includegraphics[width=0.99\linewidth]{{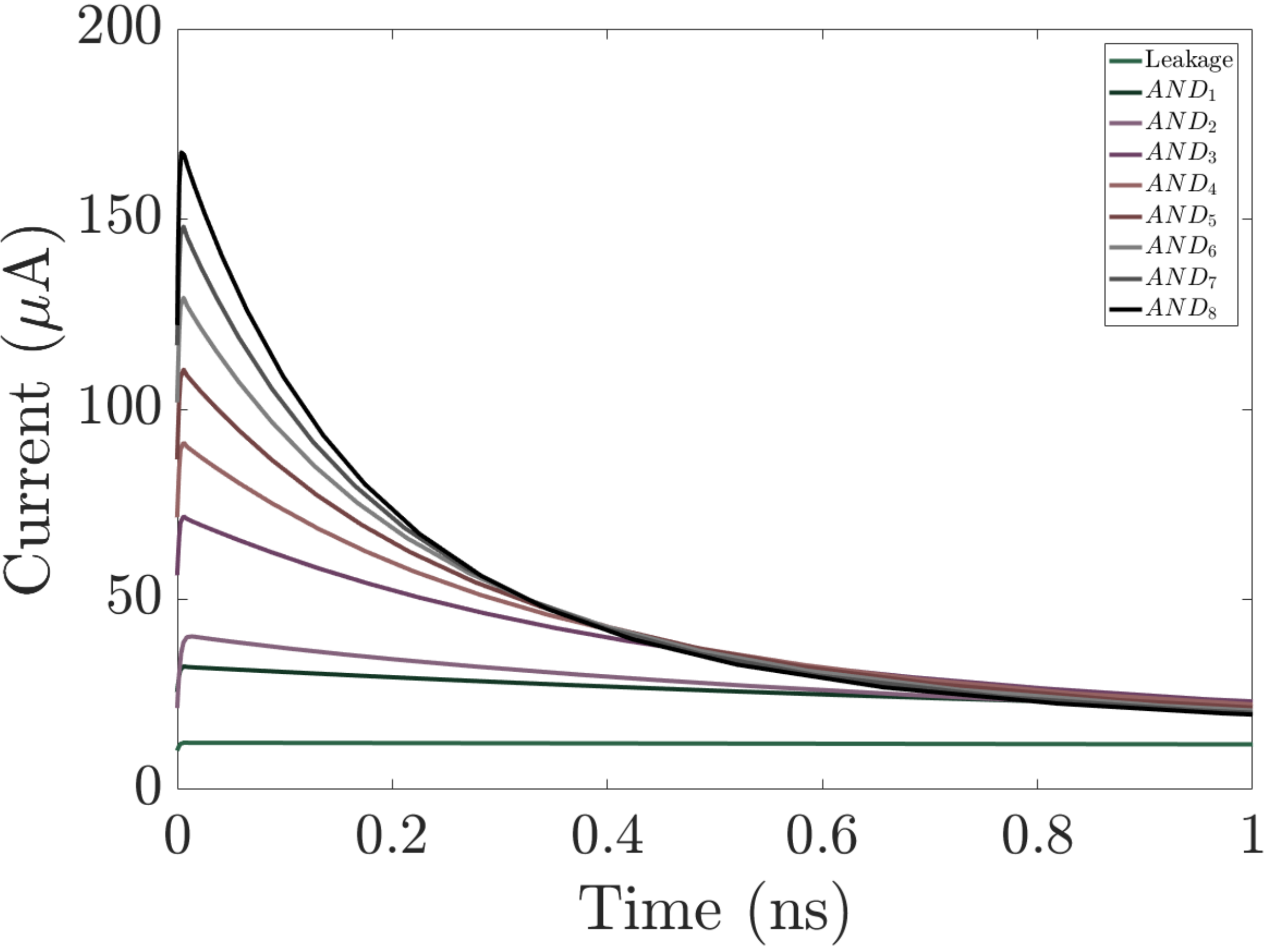}}
                \caption{}
                \label{AND-True-a}
        \end{subfigure}%
        \begin{subfigure}[b]{0.5\linewidth}
                \centering
                \includegraphics[width=0.99\linewidth]{{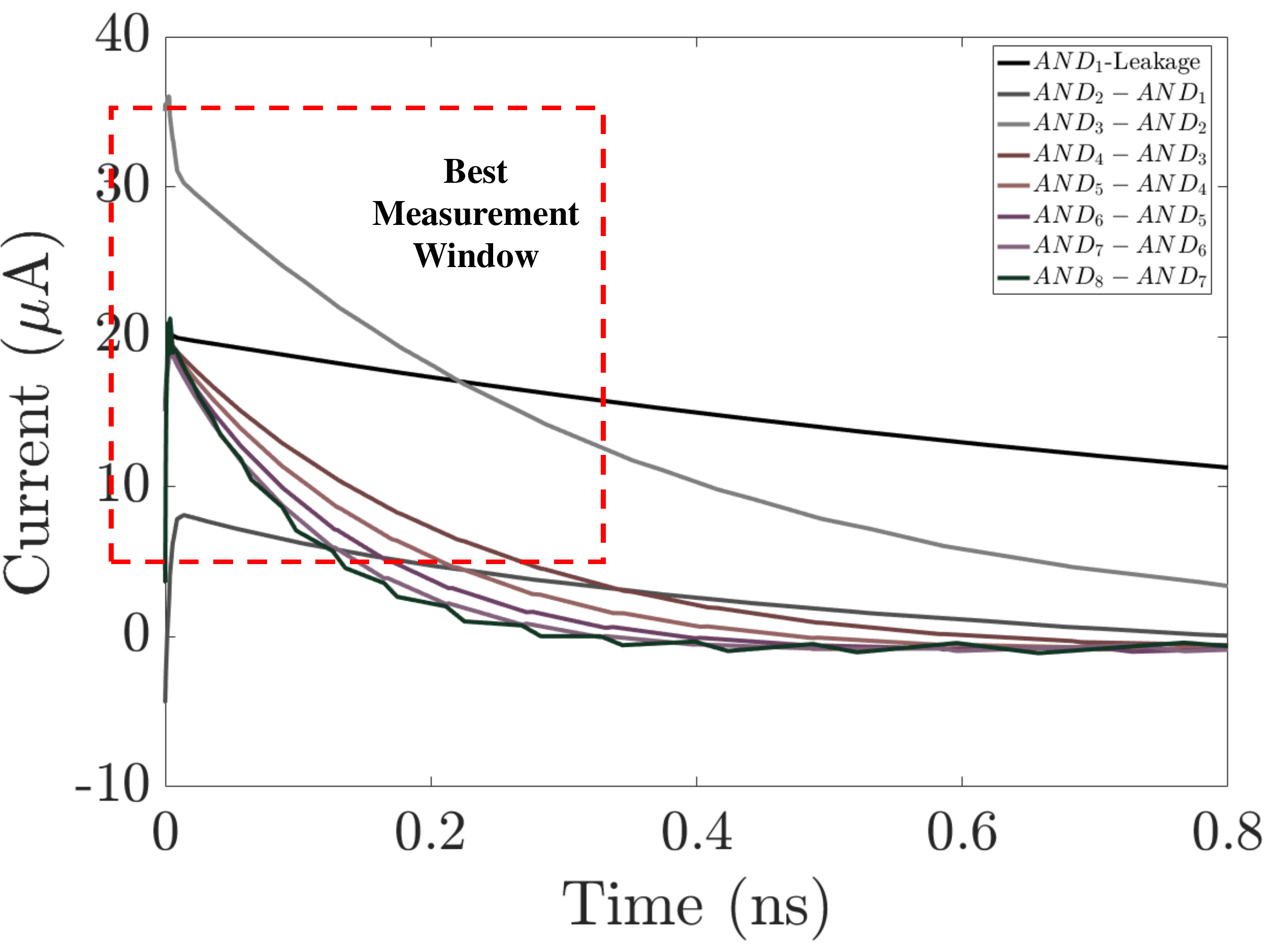}}
                \caption{}
                \label{AND-True-b}
        \end{subfigure}\\ 
              \begin{subfigure}[b]{0.5\linewidth}
                \centering
                \includegraphics[width=0.99\linewidth]{{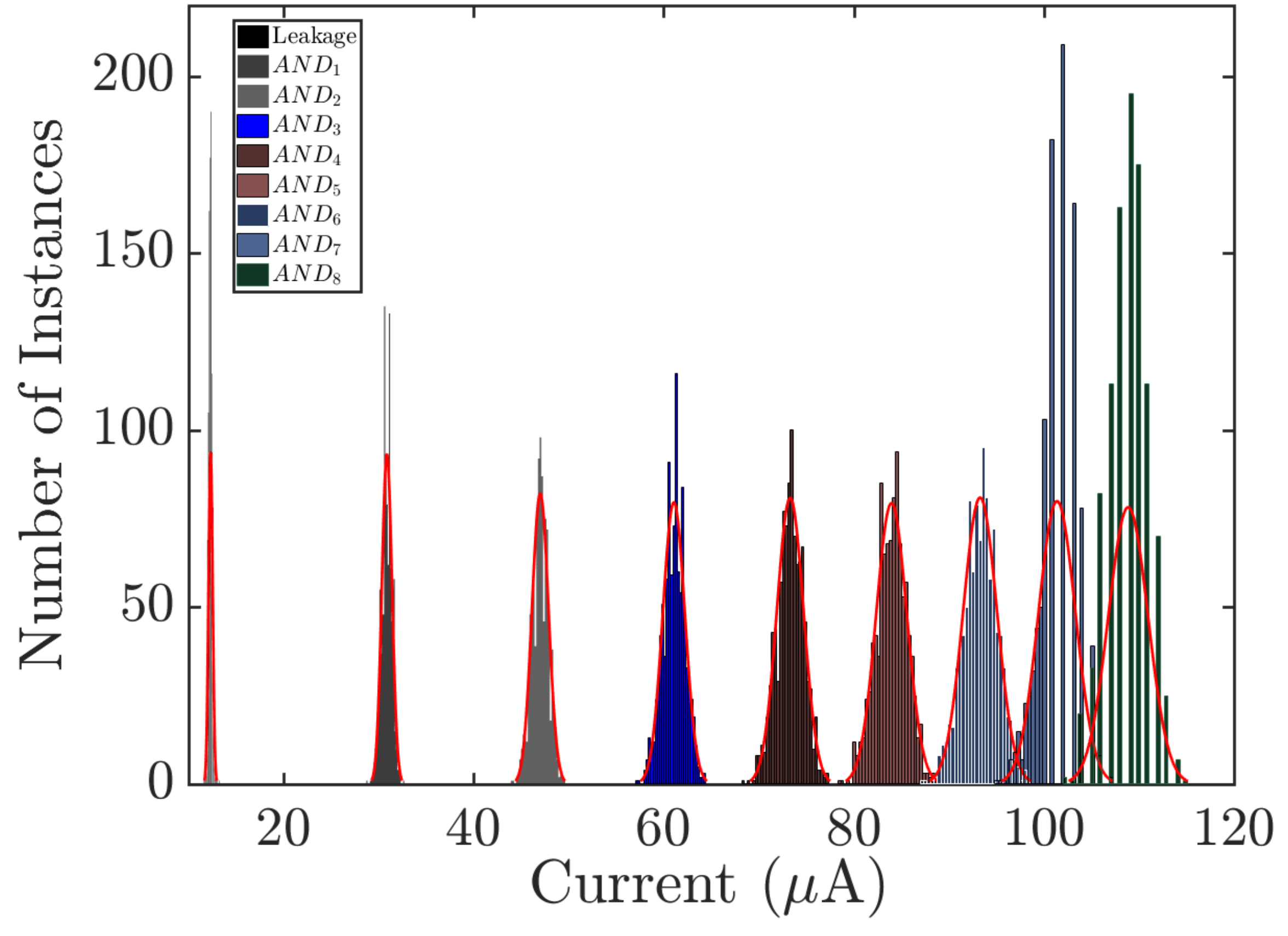}}
                \caption{}
                \label{AND-True-c}
        \end{subfigure}%
        \begin{subfigure}[b]{0.5\linewidth}
                \centering
                \includegraphics[width=0.99\linewidth]{{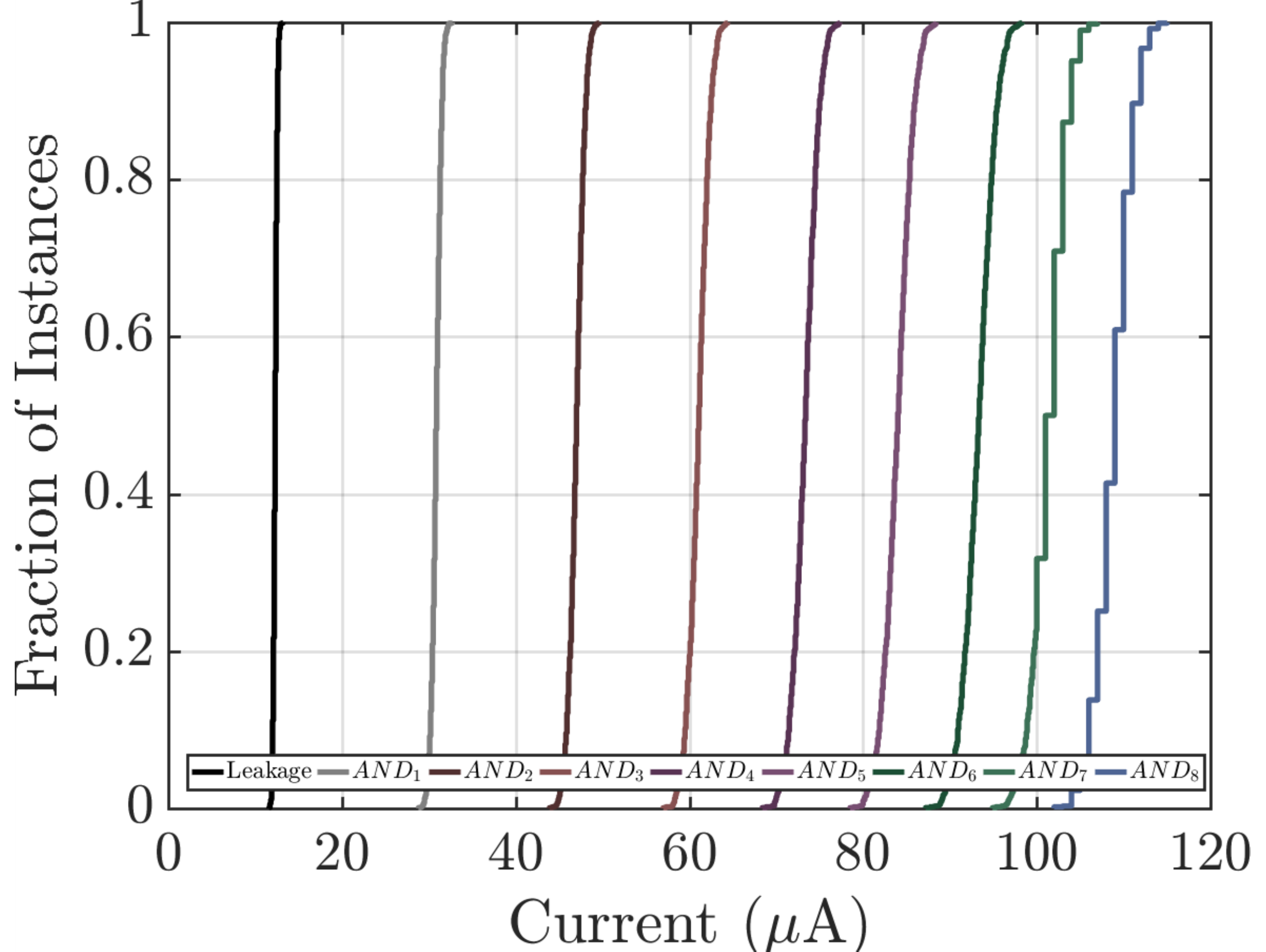}}
                \caption{}
                \label{AND-True-d}
        \end{subfigure}\\
              \begin{subfigure}[b]{0.5\linewidth}
                \centering
                \includegraphics[width=0.99\linewidth]{{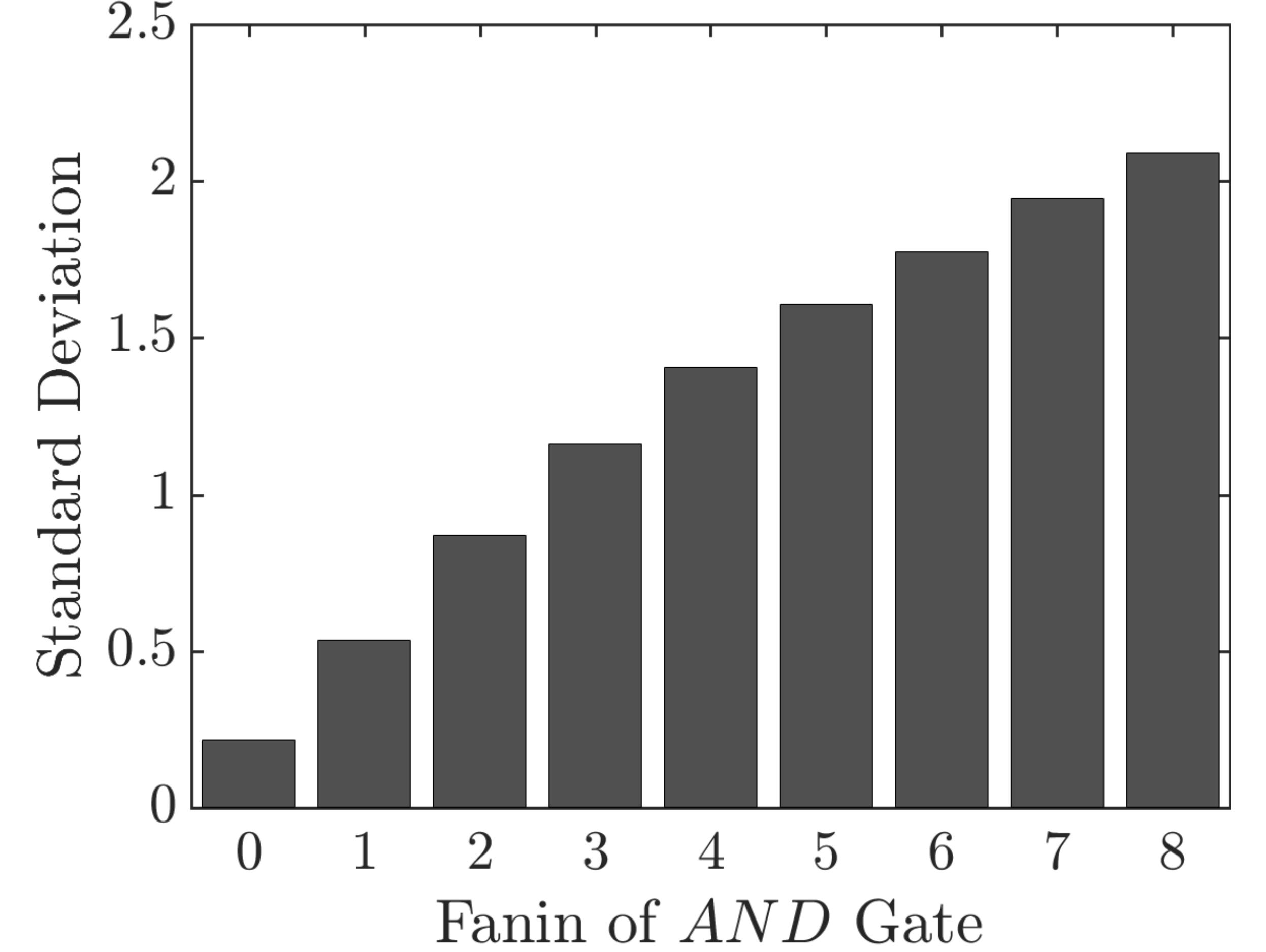}}
                \caption{}
                \label{AND-True-e}
        \end{subfigure}%
        \begin{subfigure}[b]{0.5\linewidth}
                \centering
                \includegraphics[width=0.99\linewidth]{{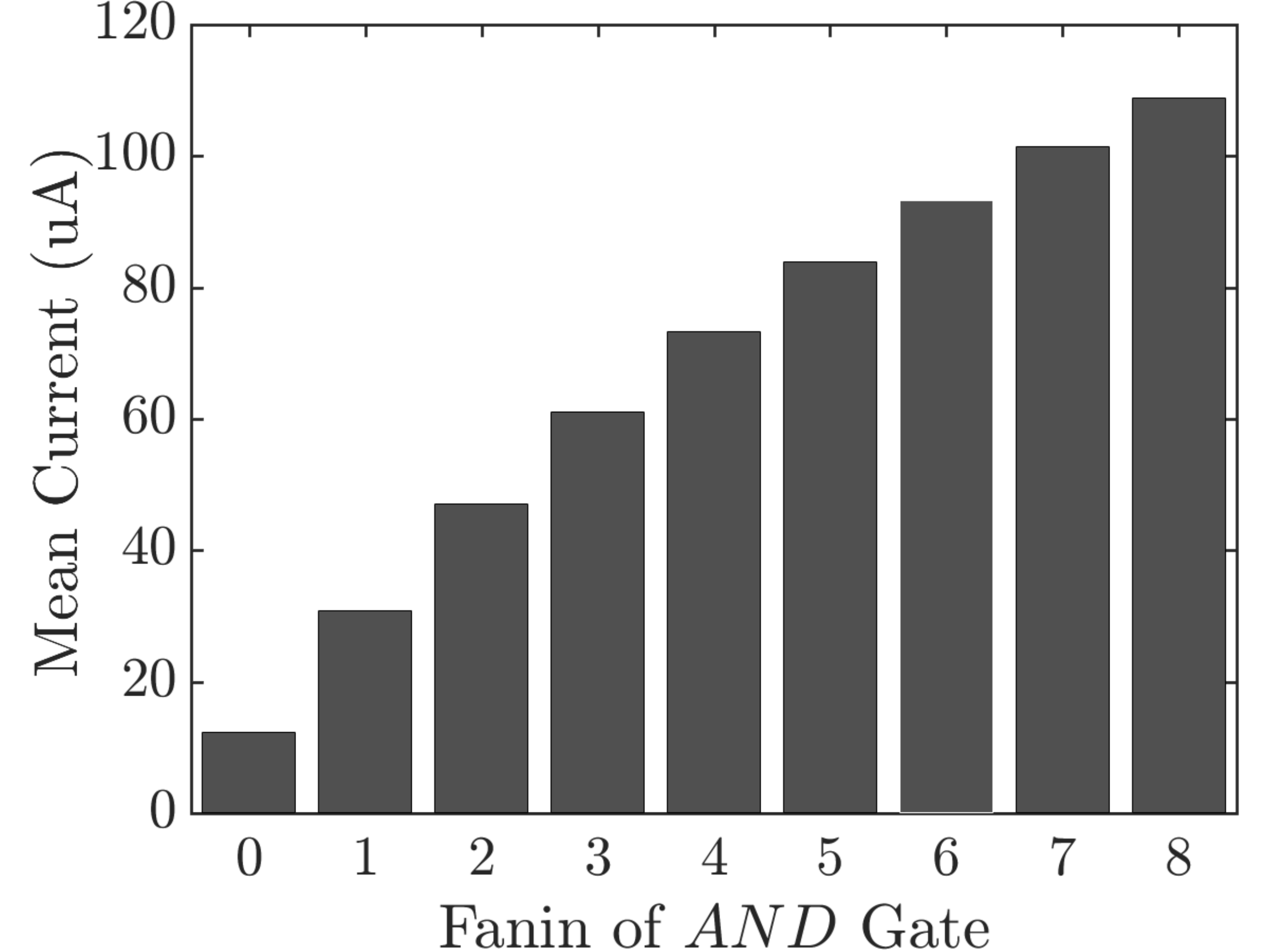}}
                \caption{}
                \label{AND-True-f}
        \end{subfigure} \vspace{-3mm}
        \caption{(a) Analysis of $AND$ gate current profile with various fanins; (b) current difference with fanin of n and n+1 to find the measurement window; (c) PDF, (d) CDF, (e) STD and, (f) mean current of $OR$ gates.}
	\vspace{-6mm}
        \label{DCIM-AND-Current}
\end{figure}

\subsubsection{Simulation Results} \hfill\\ \label{fab-chips}
\textbf{DCIM OR Array:} The current profiles of the $OR$ gate with fanin ranging between 0 to 8 is shown in Fig. \ref{OR-True-a} (time offset is chosen when $EN$ signal is activated). It indicates that $OR$ gates with various fanins charge the BLs at different rates. Since the observed resistance of each BL decreases with the number of LRS inputs (more number of resistors in parallel), current through the array increases with fanin. However, the separation between current profiles of each gate decreases with fanin. With time the current profiles of various fanin gates become indistinguishable. The current profiles merge when the BL is charged up to $V_{DD}-V_{th, Diode}$. Therefore, selection of measurement window is critical. 

Shorter measurement windows increase the resolution for distinguishing fanins of $OR$ gates (the best measurement window is shown in Fig. \ref{OR-True-b}). In the suggested measurement window, the currents of $OR_N$ and $OR_{N+1}$ gates differ by at least 5 $\mu$A. This is evident from the Probability Density Function (PDF) of the current profiles of $OR$ gates (Fig. \ref{OR-True-c}). It is seen that the PDF of $OR_7$ and $OR_8$ have a noticeable overlap. If the recorded current profile (from SCA) falls within this overlap, the adversary might need to consider both possibilities. In such cases, the Cumulative Distribution Function (CDF) may be useful to predict the number of inputs (Fig. \ref{OR-True-d}). Additional properties of the distribution such as the mean and Standard Deviation (STD) may be used to accurately determine the gate fanin. The STD of $OR$ gates with different fanins is presented in Fig. \ref{OR-True-e}. It is noted that the STD does not follow a monotonically increasing or decreasing trend. In contrast, the mean values (Fig. \ref{OR-True-f}) exhibit a monotonically increasing trend with fanin. Thus, the adversary can leverage the CDF, PDF and mean distributions of the $OR$ gate current profiles collected from the test chips/simulation to analyze the current profile recorded by SCA.

\begin{figure}[t]
        \centering 
        \begin{subfigure}[b]{0.5\linewidth}
                \centering
                \includegraphics[width=0.99\linewidth]{{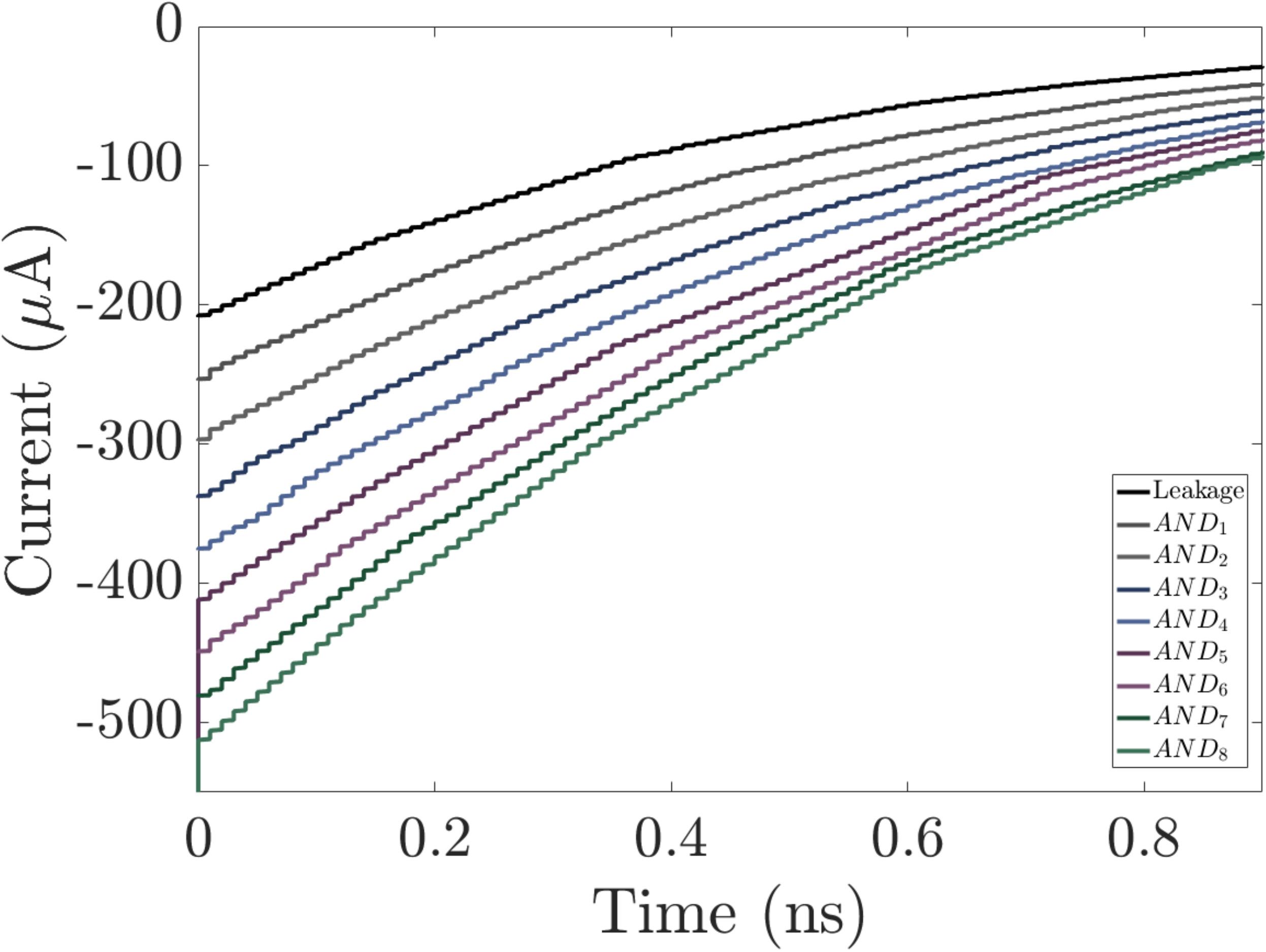}}
                \caption{}
                \label{DCIM-Complementary-a}
        \end{subfigure}%
        \begin{subfigure}[b]{0.5\linewidth}
                \centering
                \includegraphics[width=0.99\linewidth]{{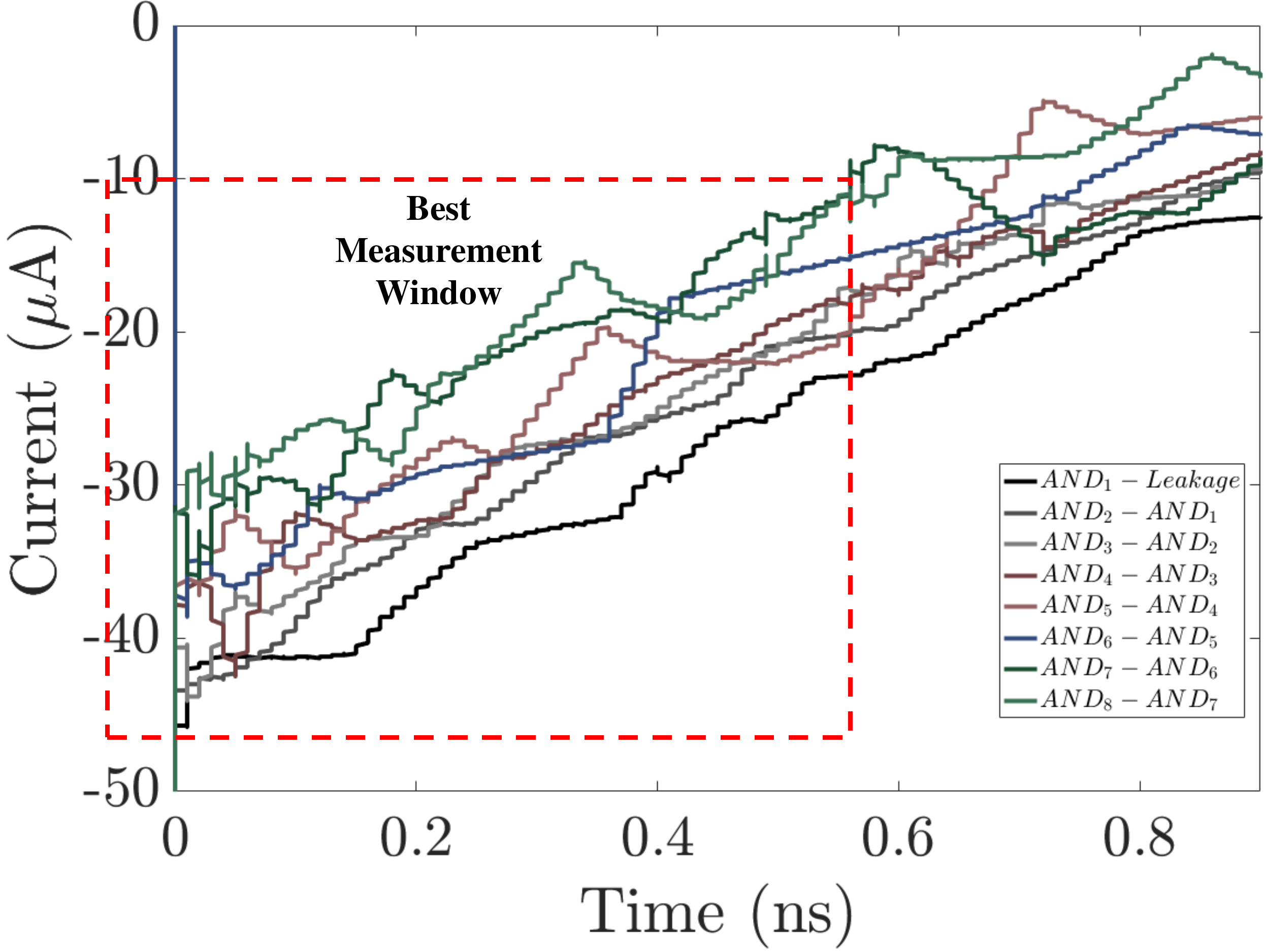}}
                \caption{}
                \label{DCIM-Complementary-b}
        \end{subfigure}\\ 
              \begin{subfigure}[b]{0.5\linewidth}
                \centering
                \includegraphics[width=0.99\linewidth]{{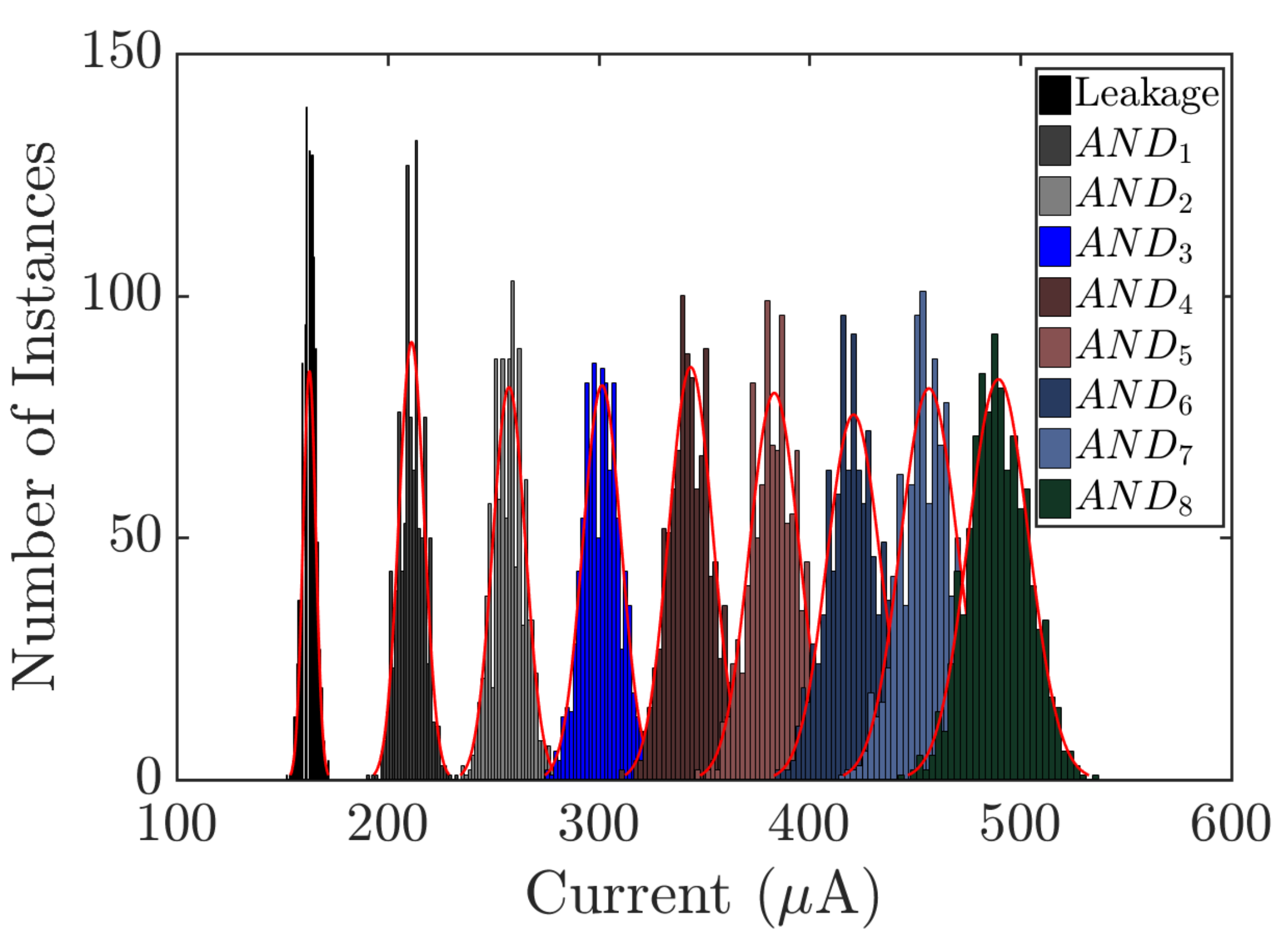}}
                \caption{}
                \label{DCIM-Complementary-c}
        \end{subfigure}%
        \begin{subfigure}[b]{0.5\linewidth}
                \centering
                \includegraphics[width=0.99\linewidth]{{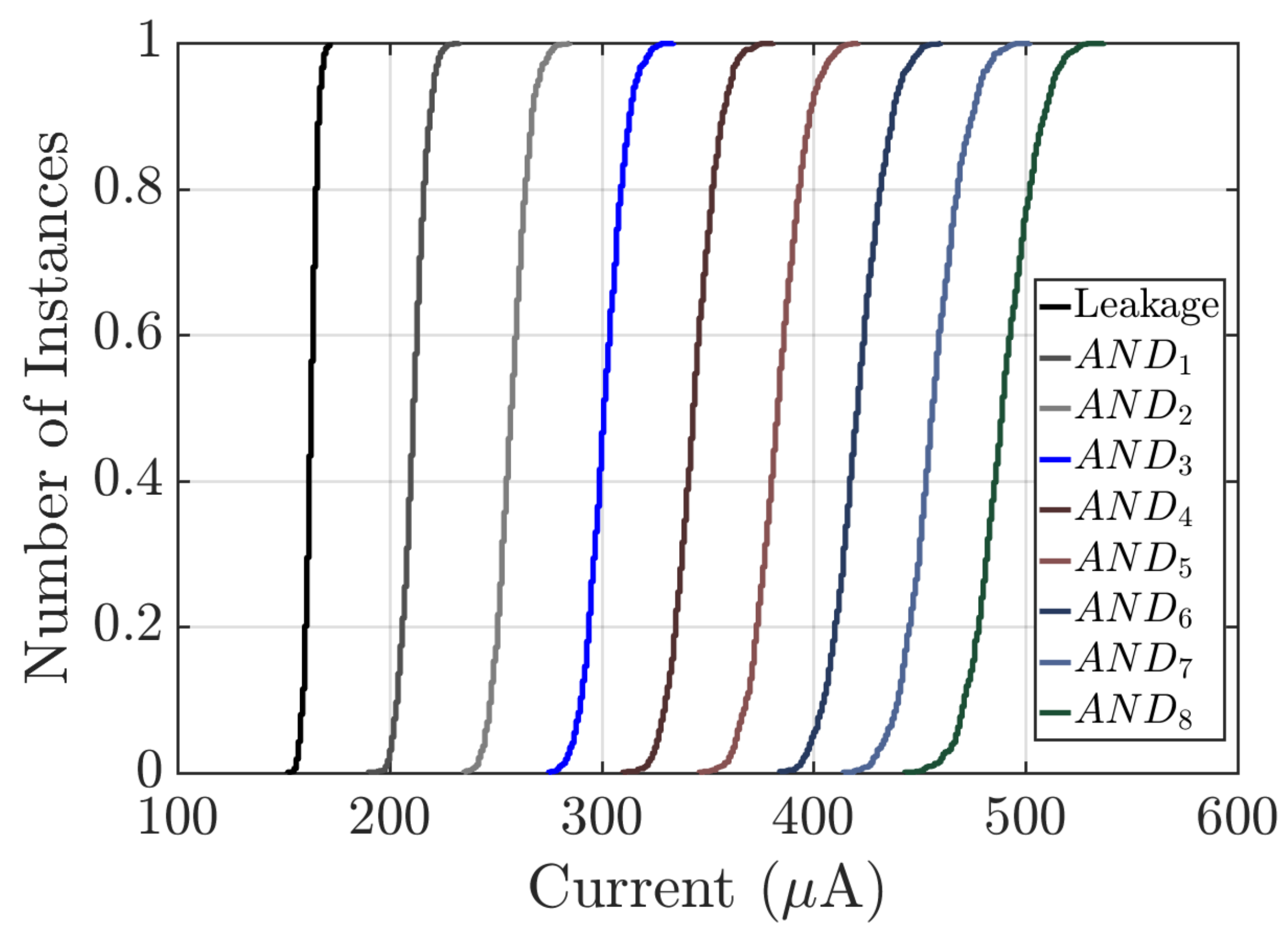}}
                \caption{}
                \label{DCIM-Complementary-d}
        \end{subfigure}\\
              \begin{subfigure}[b]{0.5\linewidth}
                \centering
                \includegraphics[width=0.99\linewidth]{{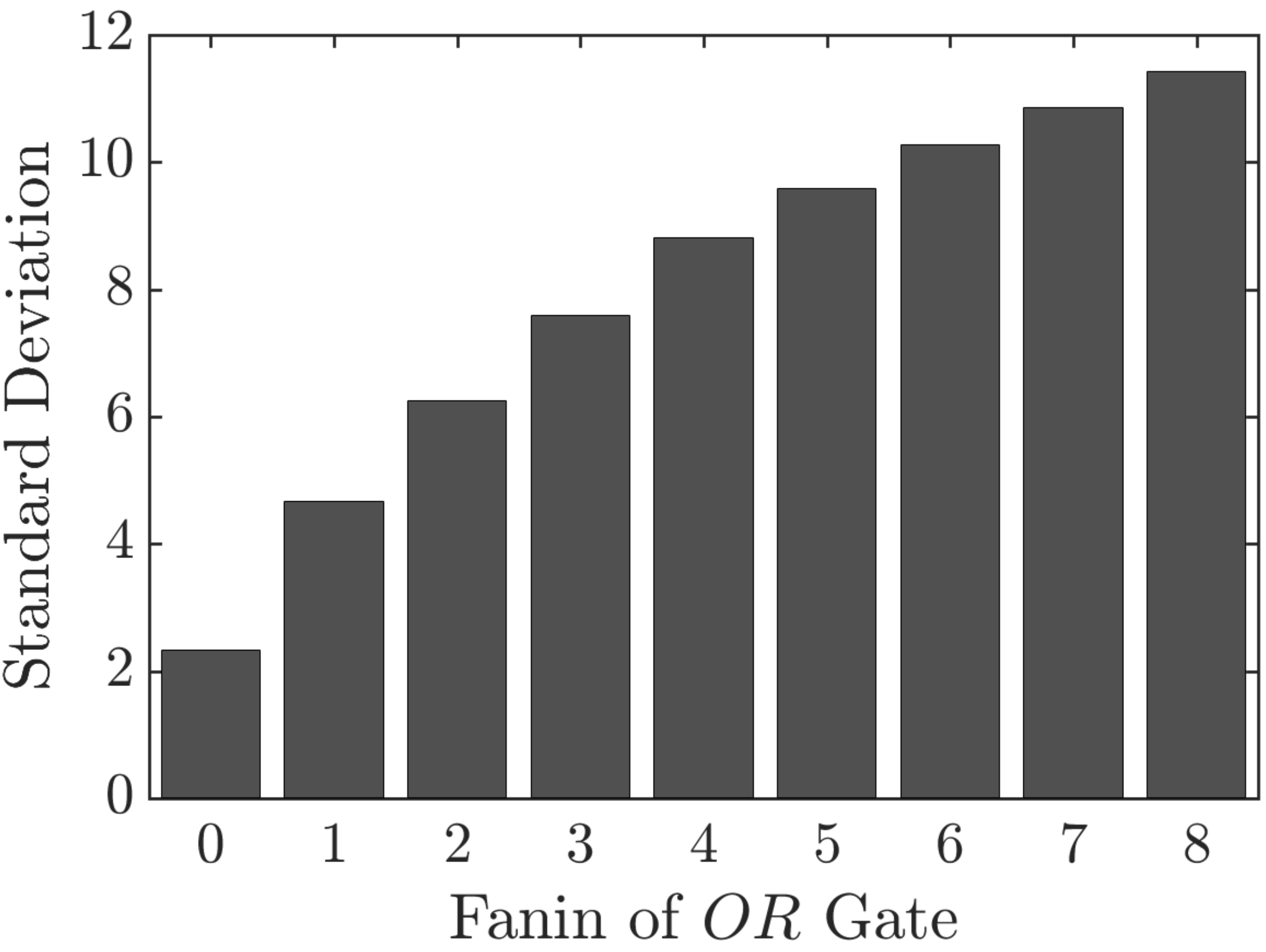}}
                \caption{}
                \label{DCIM-Complementary-e}
        \end{subfigure}%
        \begin{subfigure}[b]{0.5\linewidth}
                \centering
                \includegraphics[width=0.99\linewidth]{{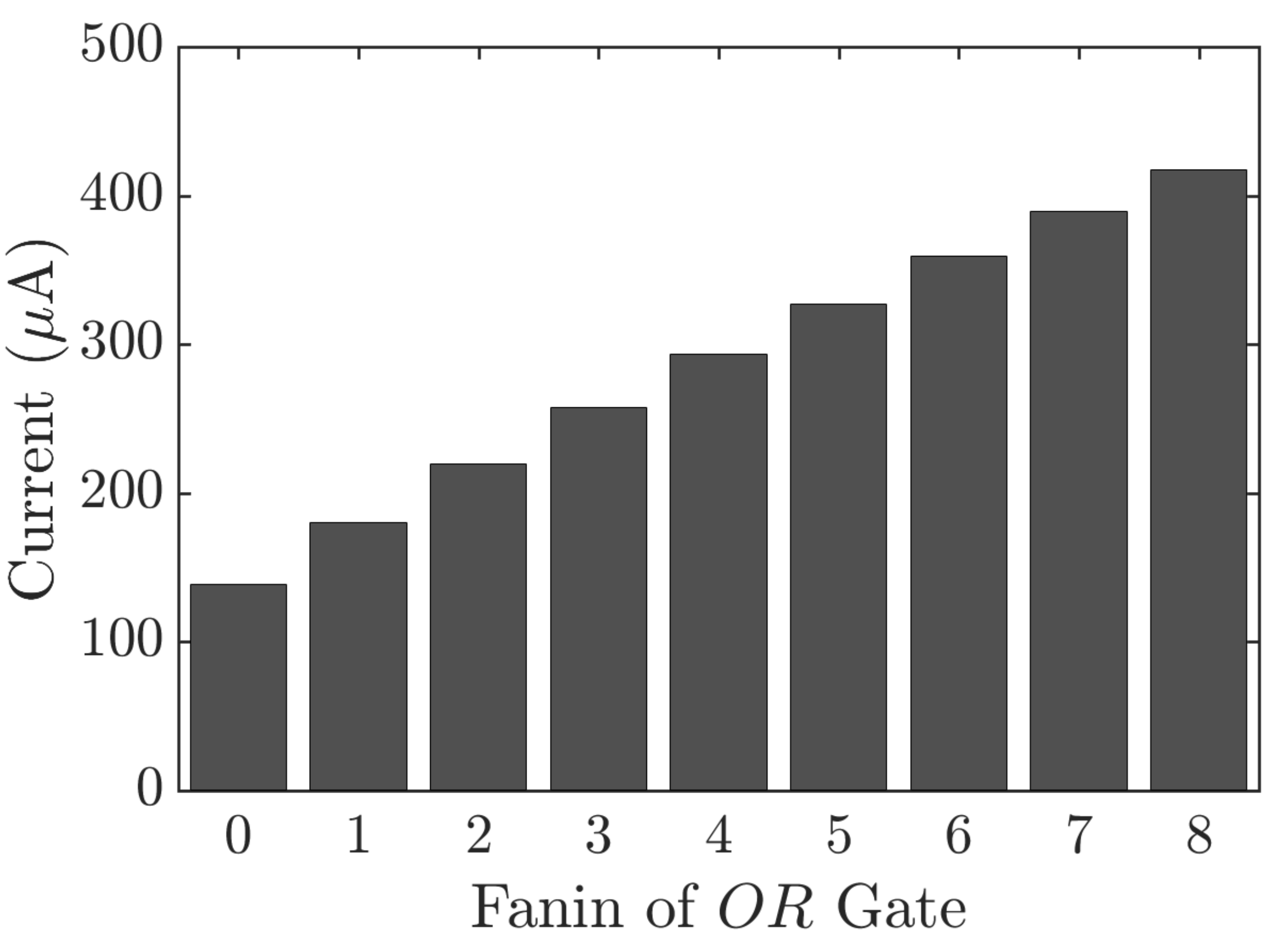}}
                \caption{}
                \label{DCIM-Complementary-f}
        \end{subfigure} \vspace{-3mm}
        \caption{(a) Analysis of $AND$ gate current profile during pre-charge phase; (b) current difference of $AND$ gates with fanin of n and n+1 to find the measurement window; (c) PDF, (d) CDF, (e) STD, and (f) mean current of $AND$ array's BLs currents.}
	\vspace{-6mm}
        \label{DCIM-Complementary}
\end{figure}

\textbf{DCIM AND Array:} Simulation setup for the DCIM $AND$ arrays is similar to the $OR$ array. However, the adversary applies SCA to extract the current profiles by monitoring the ground node. 
Additionally, the $AND$ power is higher than $OR$ power. This can be attributed to the fact that similar sized transistors act as input buffers. The $NMOS$ transistors have higher mobility compared to $PMOS$ transistors and discharge the BL faster.

The current profiles of the $AND$ gate with fanin ranging between 0 to 8 is shown in  Fig. \ref{AND-True-a}. Fig. \ref{AND-True-b} suggests the best measurement window to maximize the difference between power profiles of various fanins. The PDF and CDF of current distribution are shown in Fig. \ref{AND-True-c} and Fig. \ref{AND-True-d}, respectively. It can be noted that the current increases with fanin. Unlike the $OR$ array, both the STD and mean value graphs of the $AND$ array, as shown in Fig. \ref{AND-True-e} and Fig. \ref{AND-True-f}, respectively, increase monotonically with fanin.  Therefore, the PDF, CDF, STD, and mean value distributions collected from the test chips/simulations can be leveraged by the adversary to identify the fanin of the gate from the recorded SCA current profile.

\subsection{Attack Model 2}
\subsubsection{Leveraging the power drawn during pre-charge of AND Array}
Adversary can identify gate fanin by analyzing the current drawn during the pre-charge (pre-discharge) phase of AND (OR) array (Fig. \ref{DCIM-Complementary-a}).
The adversary forces all minterms to `0' by trying multiple patterns of inputs and analyzing OR array current. If OR array current is in the range of leakage current, adversary identifies that as all the minterms are `0'. Every active BL (BLs which participate in the implemented function) is discharged to $V_{DD}-V_{th, Diode}$ if adversary forces all minterms to `0'. This will lead to the maximum current drawn during AND array pre-charge phase. In the next pre-charge phase of AND operation cycle, DCIM has to charge all active BLs up to $V_{DD}$ and based on the capacitor energy ($E=\frac{1}{2}CV^2$) adversary can find $C$ which is the addition of all active BLs capacitance. The adversary will know the capacitance value on each BL by modeling BLs using accurate tools. Subsequently, adversary can find the number of minterms by the analysis proposed in Section {\ref{DCIM-true}}.

To find the pre-charge phase in the extracted power profile, adversary will examine a large peak without a short circuit. All large current peaks range from very large positive values to negative values due to switching of CMOS gates which consume short circuit (when both pull-up and pull-down networks are on). However, pre-charge circuit only includes PMOS transistors that only leads to negative current values for $AND$ array and positive current values for $OR$ array. This attack model works for all the functions (with true and complementary inputs) implemented by DCIM.

\subsubsection{Simulation Results}
The simulation setup for DCIM attack model 2 is similar to DCIM attack model 1. However, the adversary examines the power signature during the pre-charge phase instead of the computation window. The AND array's BLs are charged during the pre-charge phase. The power profiles observed differ based on the number of BLs. The adversary uses the distinct power measured during the pre-charge phase to determine the number BLs, which represents the number of minterms. 
The current profiles during the pre-charge phase of AND gates with different fanins ranging from 0 to 8 are shown in  Fig. \ref{DCIM-Complementary-a}. Fig. \ref{DCIM-Complementary-b} suggests the best measurement window to maximize the difference between these current profiles. Additionally, the PDF and CDF distributions of pre-charge currents are shown in Fig. \ref{DCIM-Complementary-c} and Fig. \ref{DCIM-Complementary-d}, respectively. 

Note that the pre-charge current observed increases with fanin. Both the STD and mean value graphs of the $AND$ array, as shown in Fig. \ref{DCIM-Complementary-e} and (Fig. \ref{DCIM-Complementary-f}, respectively, increase monotonically with fanin.  Therefore, the PDF, CDF, STD, and mean value distributions collected from the test chips/simulations can be leveraged by the adversary to identify the fanin of the gate from the recorded SCA current profile.

\subsection{Analysis of the Impact of Supply Voltage Magnitude on SCARE Performance}
\begin{figure}
        \centering 
        \begin{subfigure}[b]{0.49\linewidth}
                \centering
                \includegraphics[width=0.99\linewidth]{{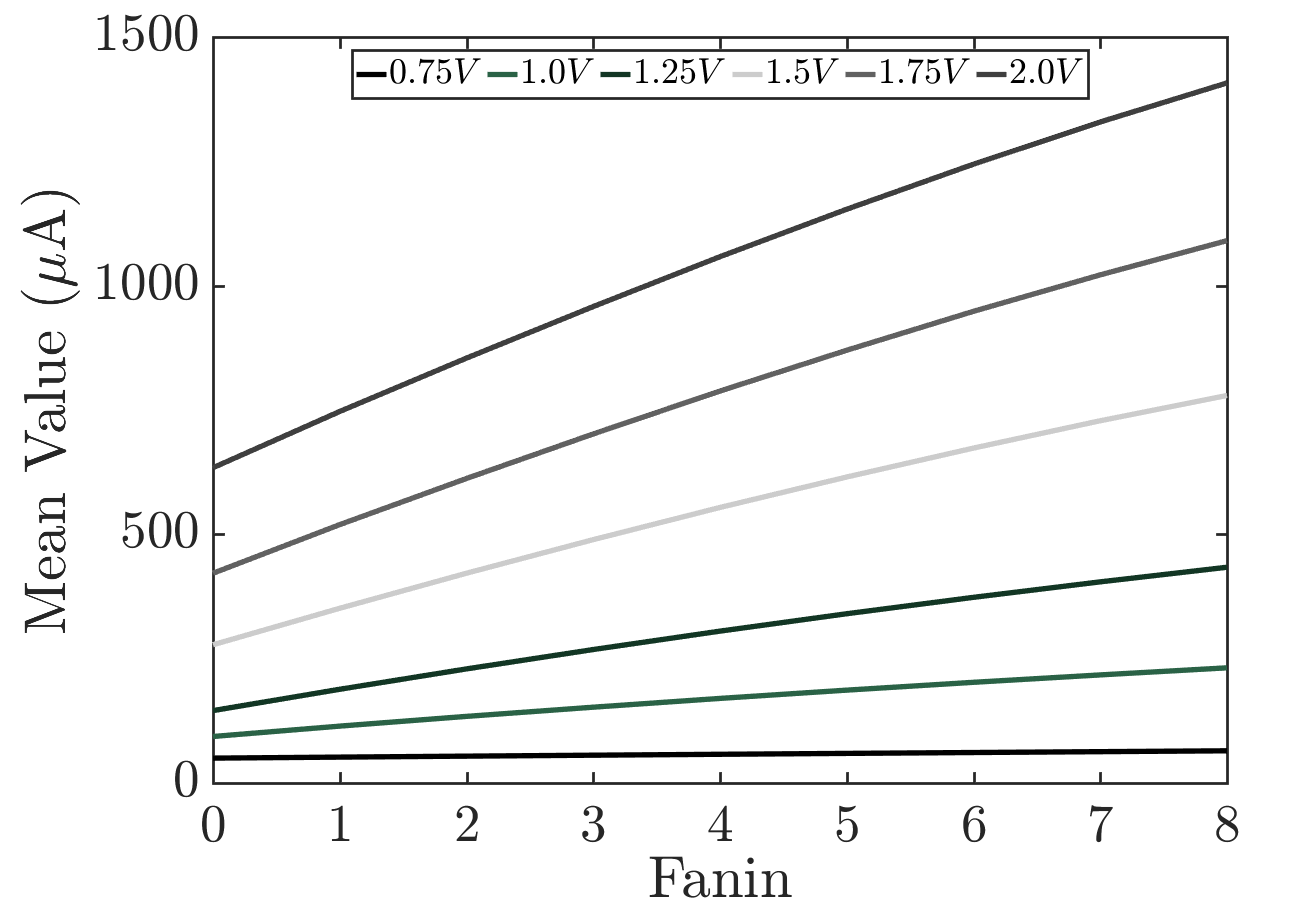}}
                \caption{}
                \label{DCIM-Voltage-a} 
        \end{subfigure} \hspace {-2mm} 
        \begin{subfigure}[b]{0.49\linewidth}
                \centering
                \includegraphics[width=0.99\linewidth]{{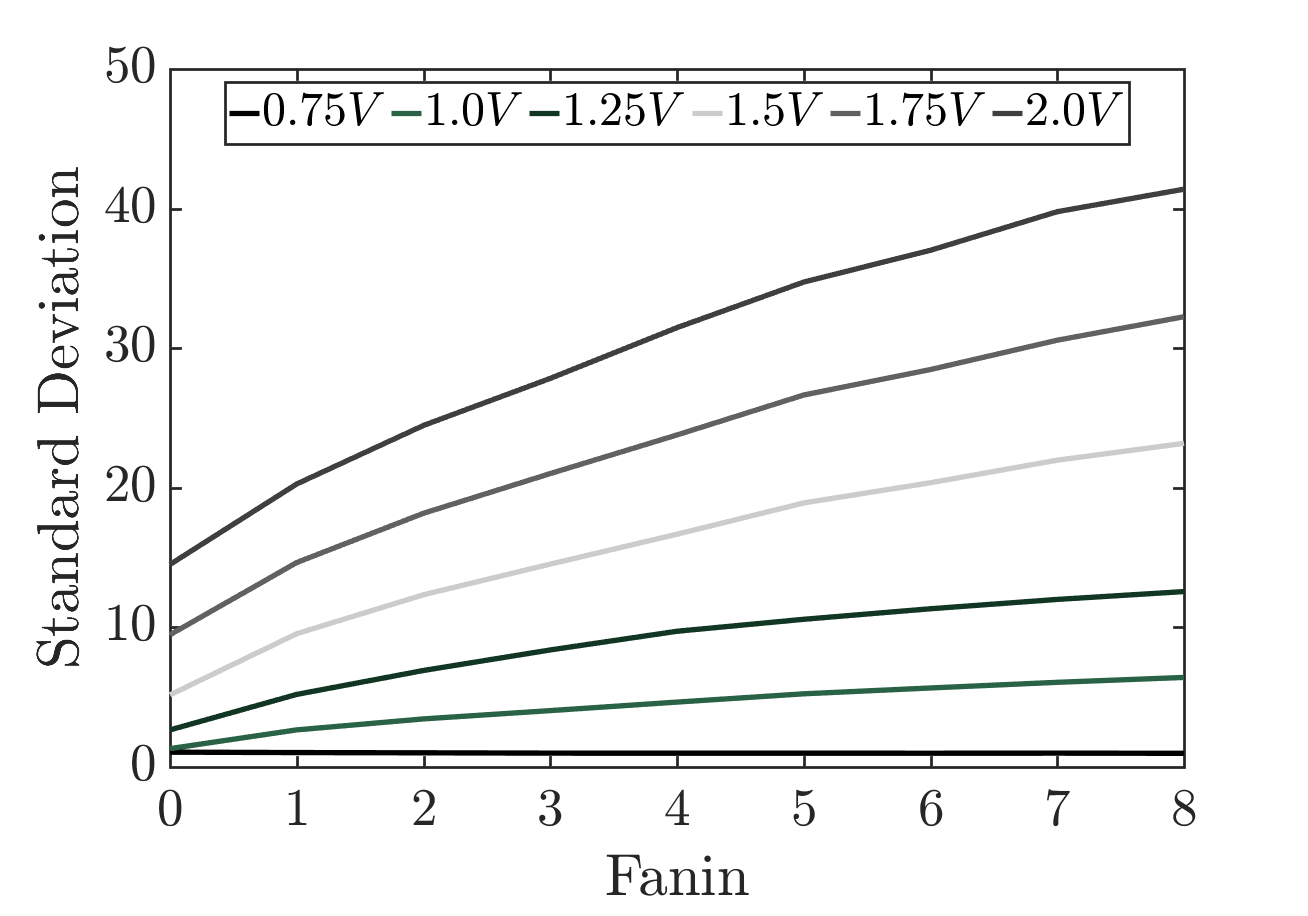}}
                \caption{}
                \label{DCIM-Voltage-b}
        \end{subfigure} \hspace {-2mm} 
        \begin{subfigure}[b]{0.5\linewidth}
                \centering
                \includegraphics[width=0.99\linewidth]{{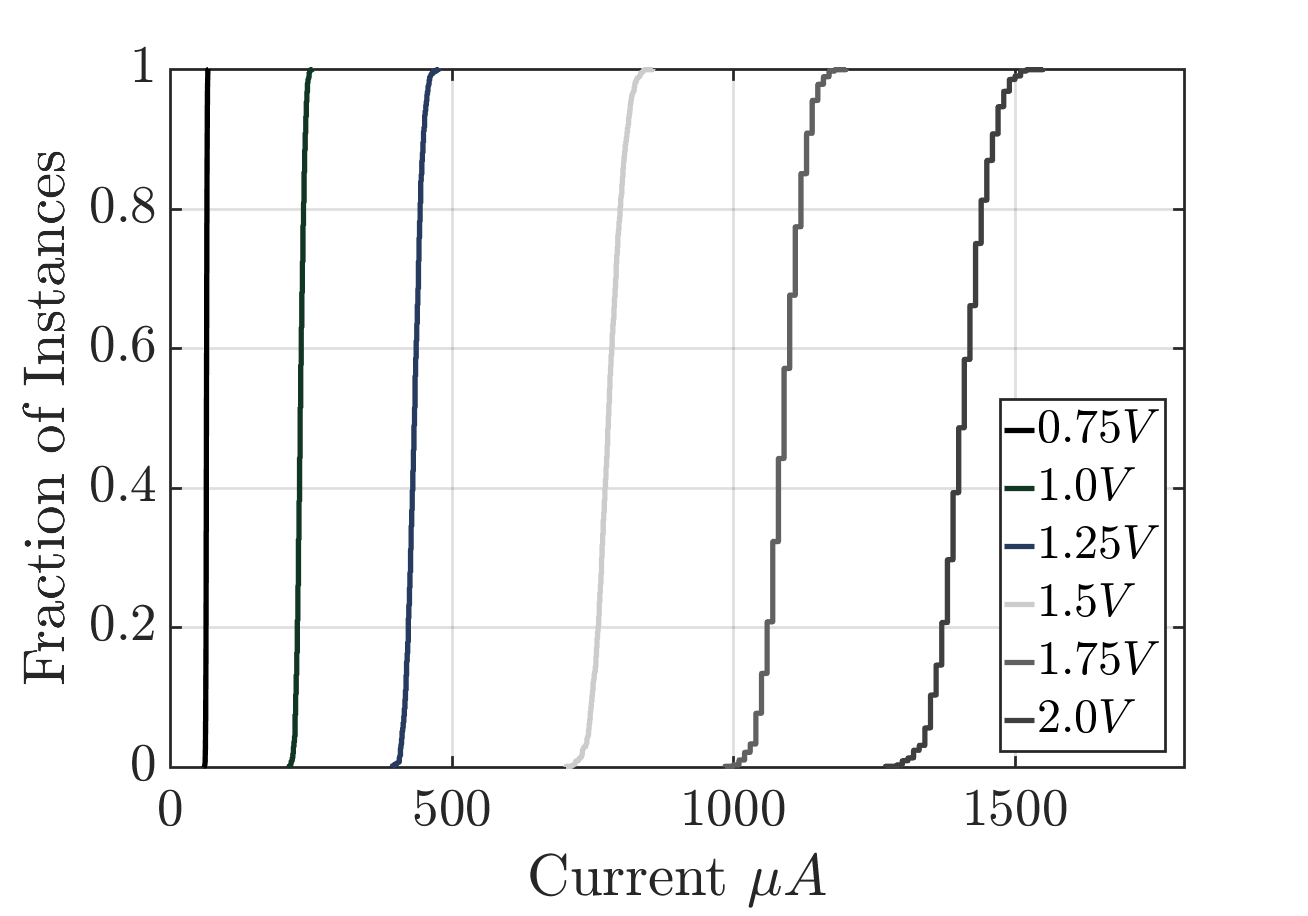}}
                \caption{}
                \label{DCIM-Voltage-c}
        \end{subfigure}\hspace {-2mm}
        \begin{subfigure}[b]{0.5\linewidth}
                \centering
                \includegraphics[width=0.99\linewidth]{{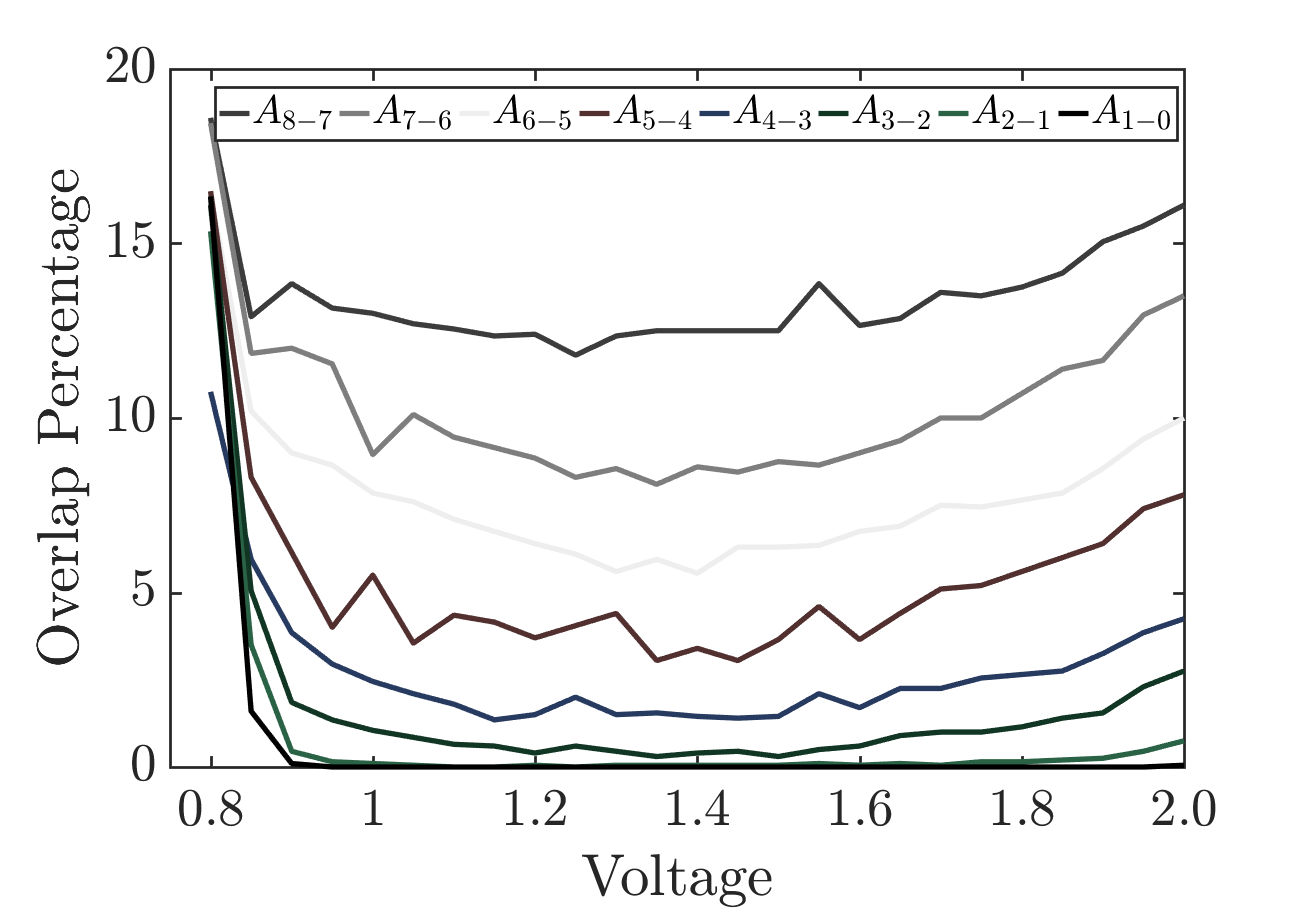}}
                \caption{}
                \label{DCIM-Voltage-d}
        \end{subfigure}\hspace {-2mm}
        \caption{Analysis of current mean value, STD deviations, percentage of distribution overlaps in PDF, and $AND_8$ CDF under different voltage nodes. (a) Mean Values, (b) STD Deviations, (c) Percentage of overlaps, and (d) CDF of $AND_8$.}
	\vspace{-3mm}
        \label{DCIM-Voltage}
\end{figure}

We have also swept the magnitude of supply voltage to analyze its impact on the performance of SCARE. For DCIM, we swept {$V_{DD}$} from {$0.75$}V to {$2$}V in 50mV increments. The result is summarized in Fig. {\ref{DCIM-Voltage-a}}. It is evident that mean value of current of the gates with different fanins increases at higher voltages. The differences in the mean values of currents are increased (higher slope) at higher voltages too, which helps the adversary to distinguish more accurately between different fanins. Sigma analysis of current profiles show that current distribution are wider at higher voltages and sharper at lower voltages, which means sigma increases with voltage (Fig. {\ref{DCIM-Voltage-b}}). As it is shown in Fig. {\ref{DCIM-Voltage-c}}, CDF slope is decreasing at higher voltages, which shows that sigma increases at higher voltages. Note that, under process variation, cases which overlap into adjacent fanins' distribution are important and hard to distinguish. SCARE has calculated overlaps between gates with fanins $n$ and $n+1$ (Fig. {\ref{DCIM-Voltage-d}}), to have an insight on the overlap percentage at various different voltage nodes. Fig. {\ref{DCIM-Voltage-d}} shows that for voltages near the nominal {$V_{DD}$} the overlap is at its lowest and it increases when the supply voltage magnitude increases or decreases. The worst case is for very low supply voltages, when the current magnitude is very small and a small variation in the resistance values of RRAMs can lead to ambiguity of the gates' fanins.

\begin{figure*}[t]
        \centering 
        \begin{subfigure}[b]{0.18\linewidth}
                \centering
                \includegraphics[width=0.99\linewidth]{{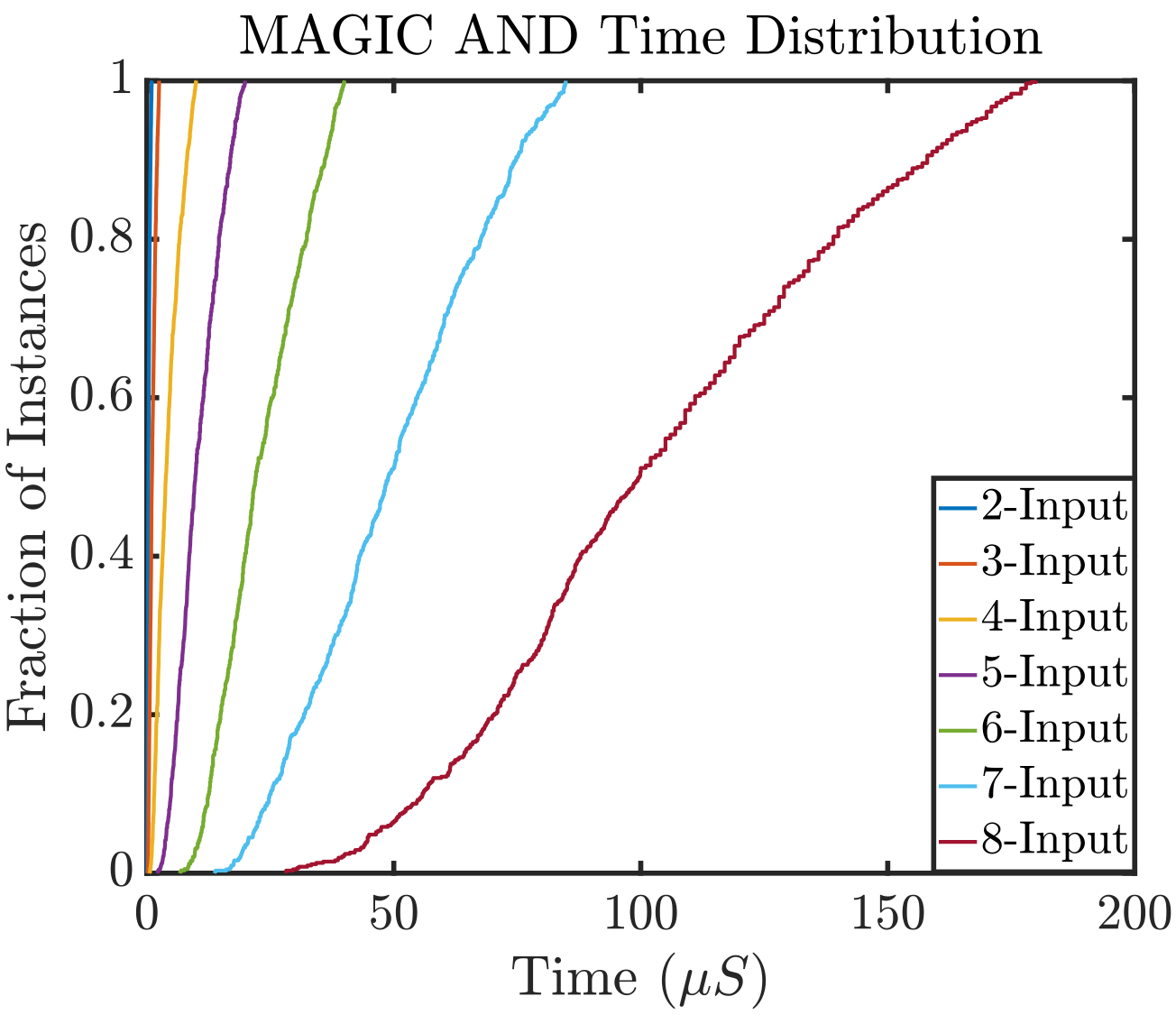}}
                \caption{}
                \label{MAGIC-True-a}
        \end{subfigure} \hspace {-2mm}
        \begin{subfigure}[b]{0.18\linewidth}
                \centering
                \includegraphics[width=0.99\linewidth]{{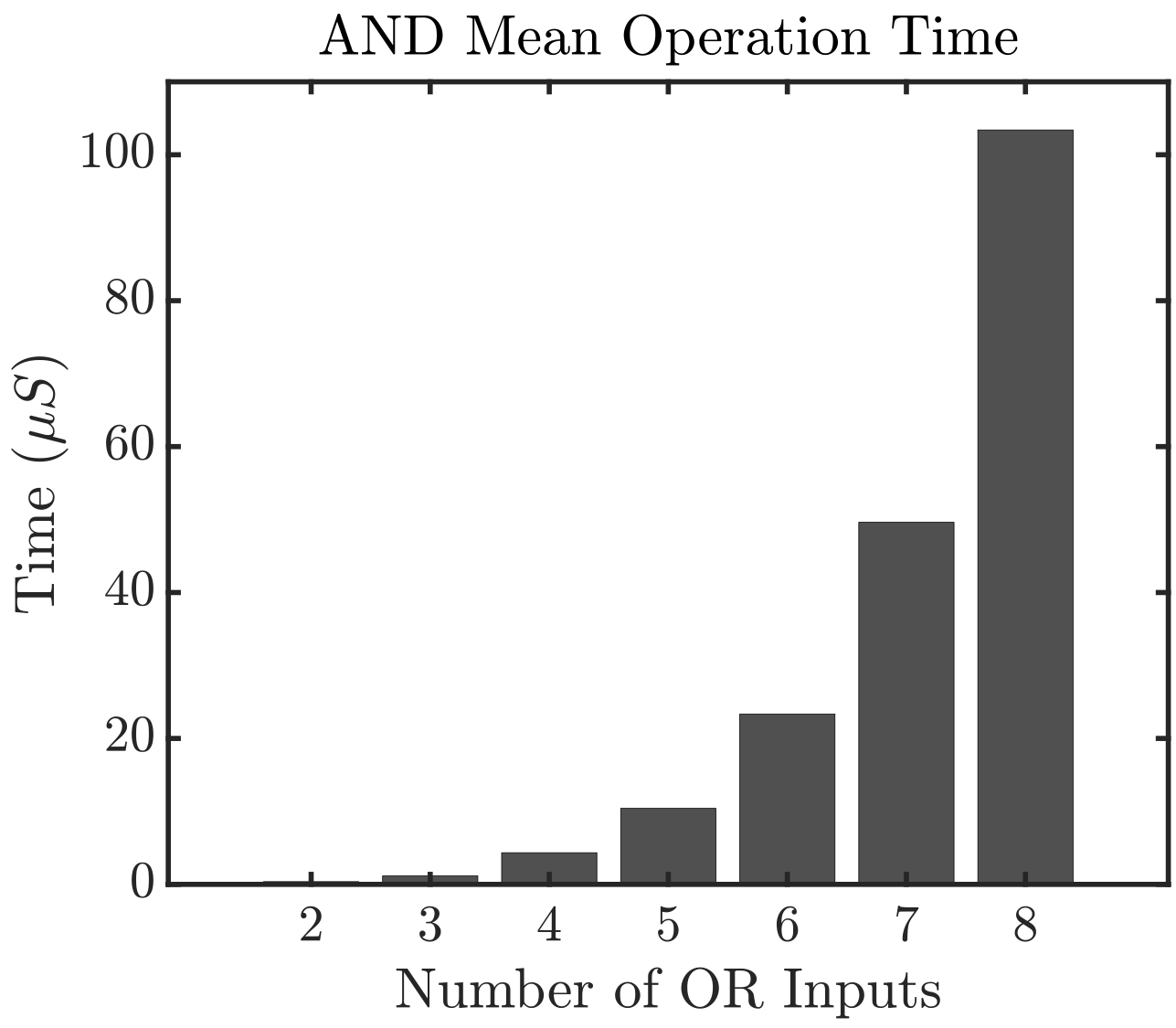}}
                \caption{}
                \label{MAGIC-True-b}
        \end{subfigure} \hspace {-2mm} 
        \begin{subfigure}[b]{0.18\linewidth}
                \centering
                \includegraphics[width=0.99\linewidth]{{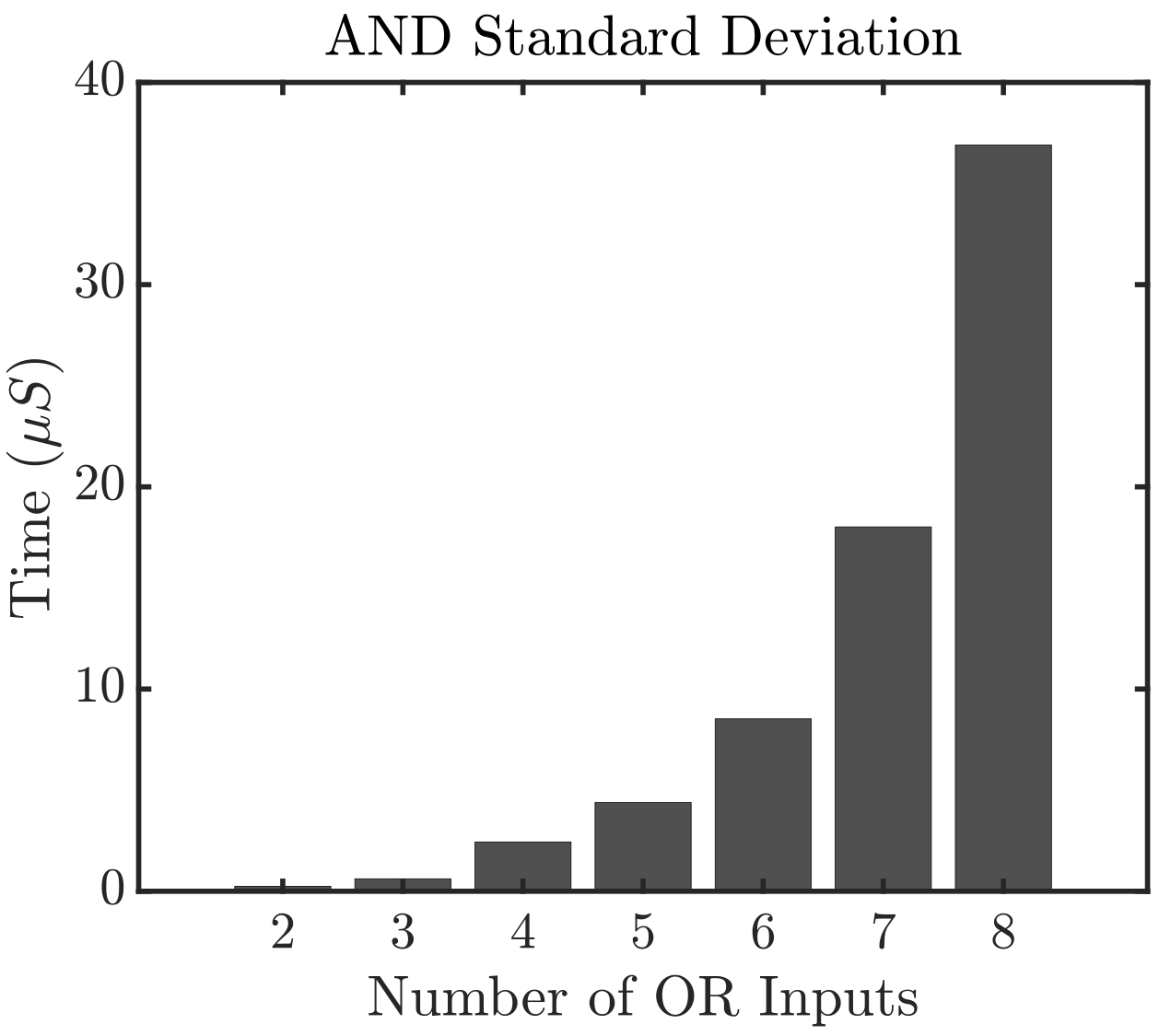}}
                \caption{}
                \label{MAGIC-True-c}
        \end{subfigure}
        \begin{subfigure}[b]{0.19\linewidth}
                \centering
                \includegraphics[width=0.99\linewidth]{{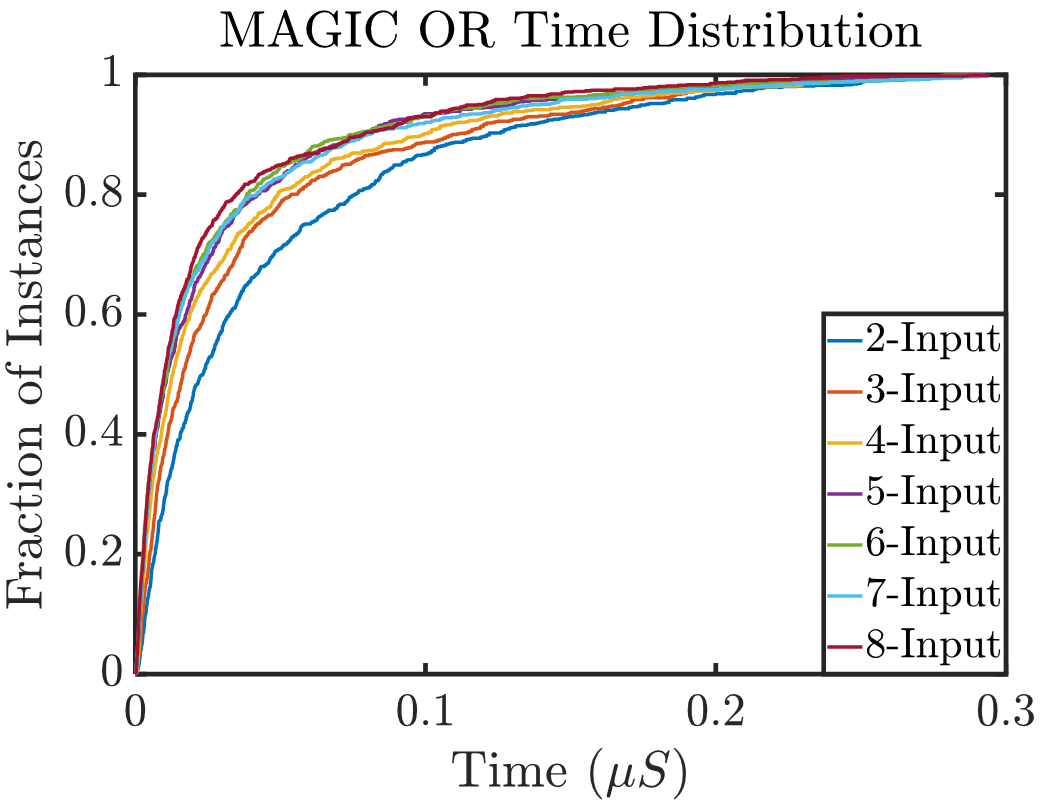}}
                \caption{}
                \label{MAGIC-True-d}
        \end{subfigure} \hspace {-2mm} 
        \begin{subfigure}[b]{0.18\linewidth}
                \centering
                \includegraphics[width=0.99\linewidth]{{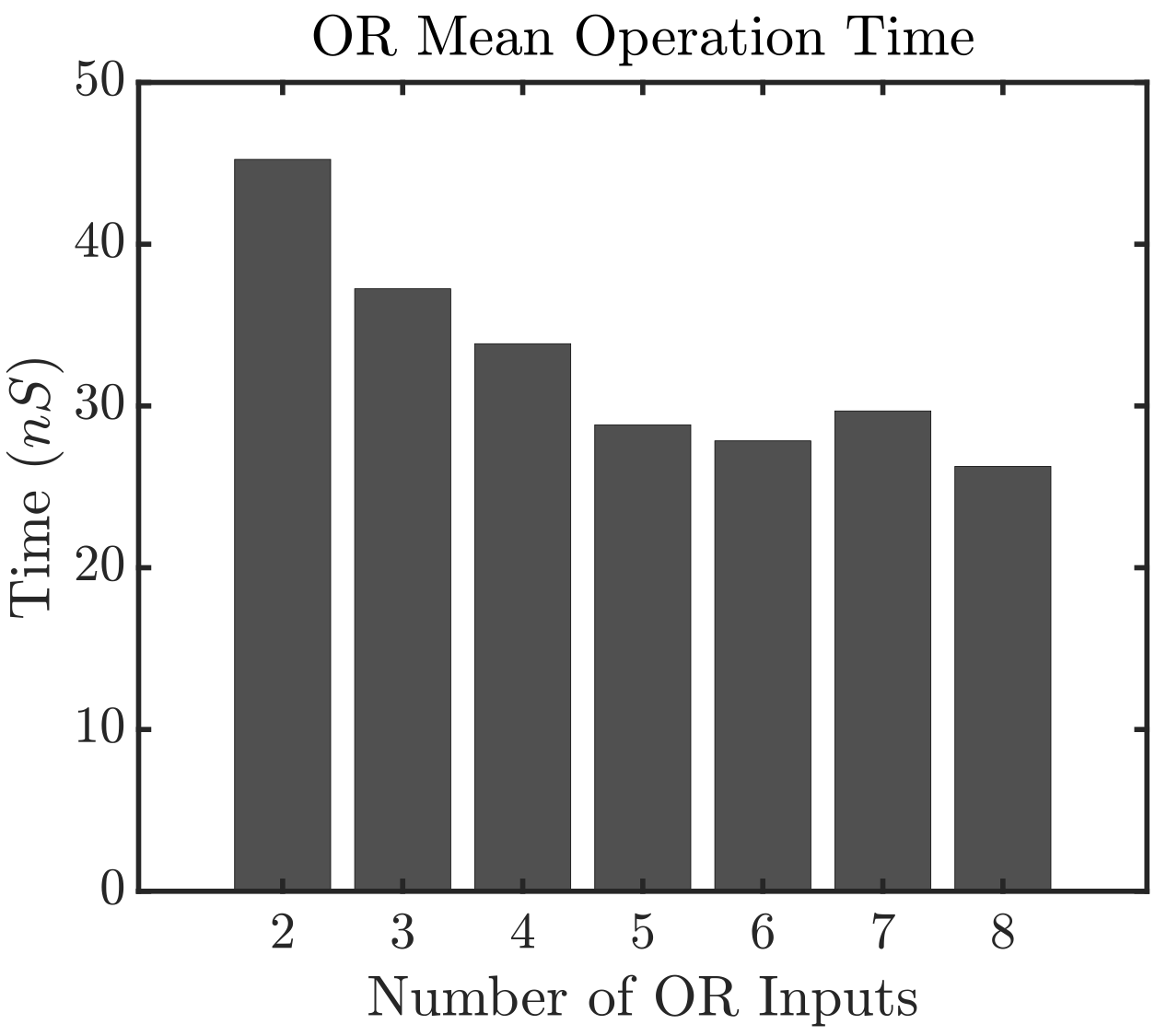}}
                \caption{}
                \label{MAGIC-True-e}
        \end{subfigure} \\
        \begin{subfigure}[b]{0.19\linewidth}
                \centering
                \includegraphics[width=0.99\linewidth]{{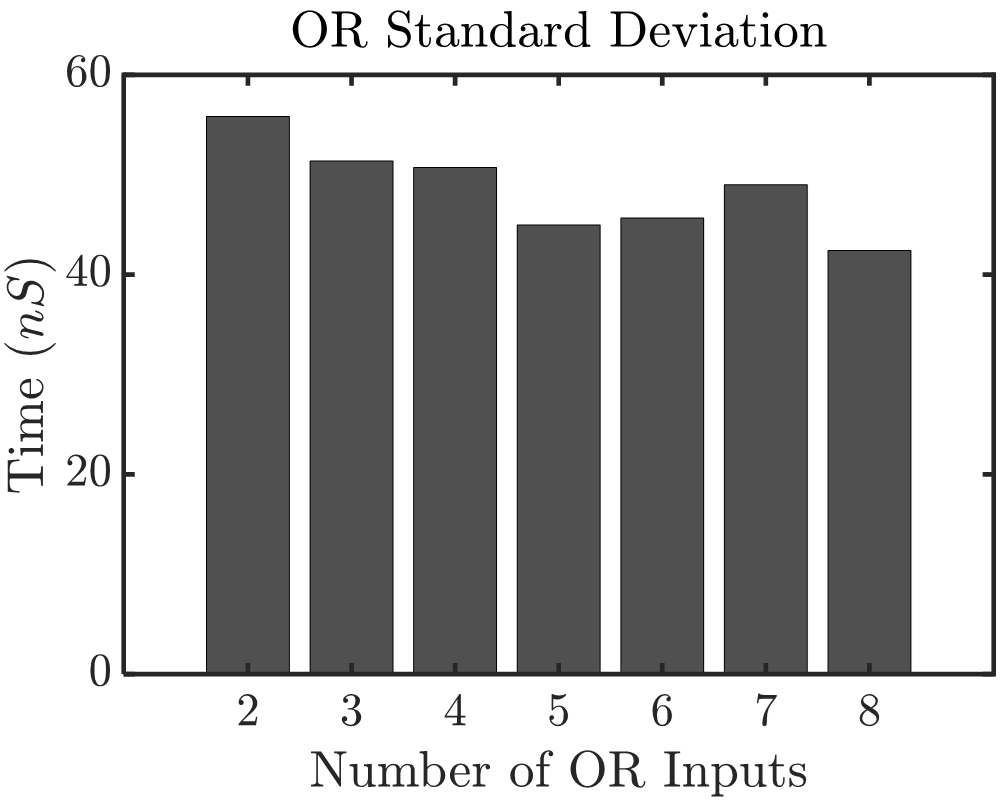}}
                \caption{}
                \label{MAGIC-True-f} 
        \end{subfigure} \hspace {-2mm}
                \begin{subfigure}[b]{0.18\linewidth}
                \centering
                \includegraphics[width=0.99\linewidth]{{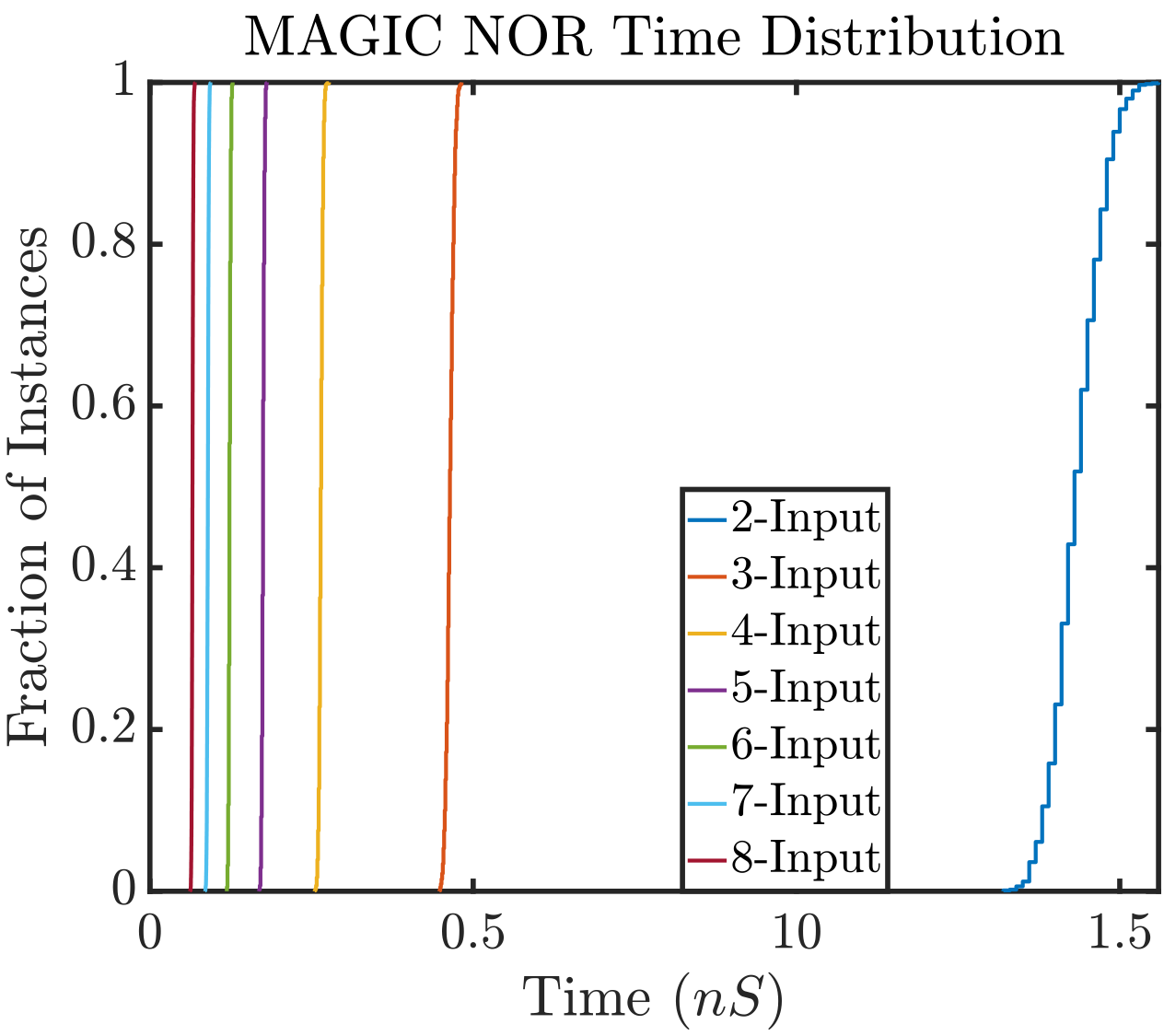}}
                \caption{}
                \label{MAGIC-True-g} 
        \end{subfigure} \hspace {-2mm} 
        \begin{subfigure}[b]{0.18\linewidth}
                \centering
                \includegraphics[width=0.99\linewidth]{{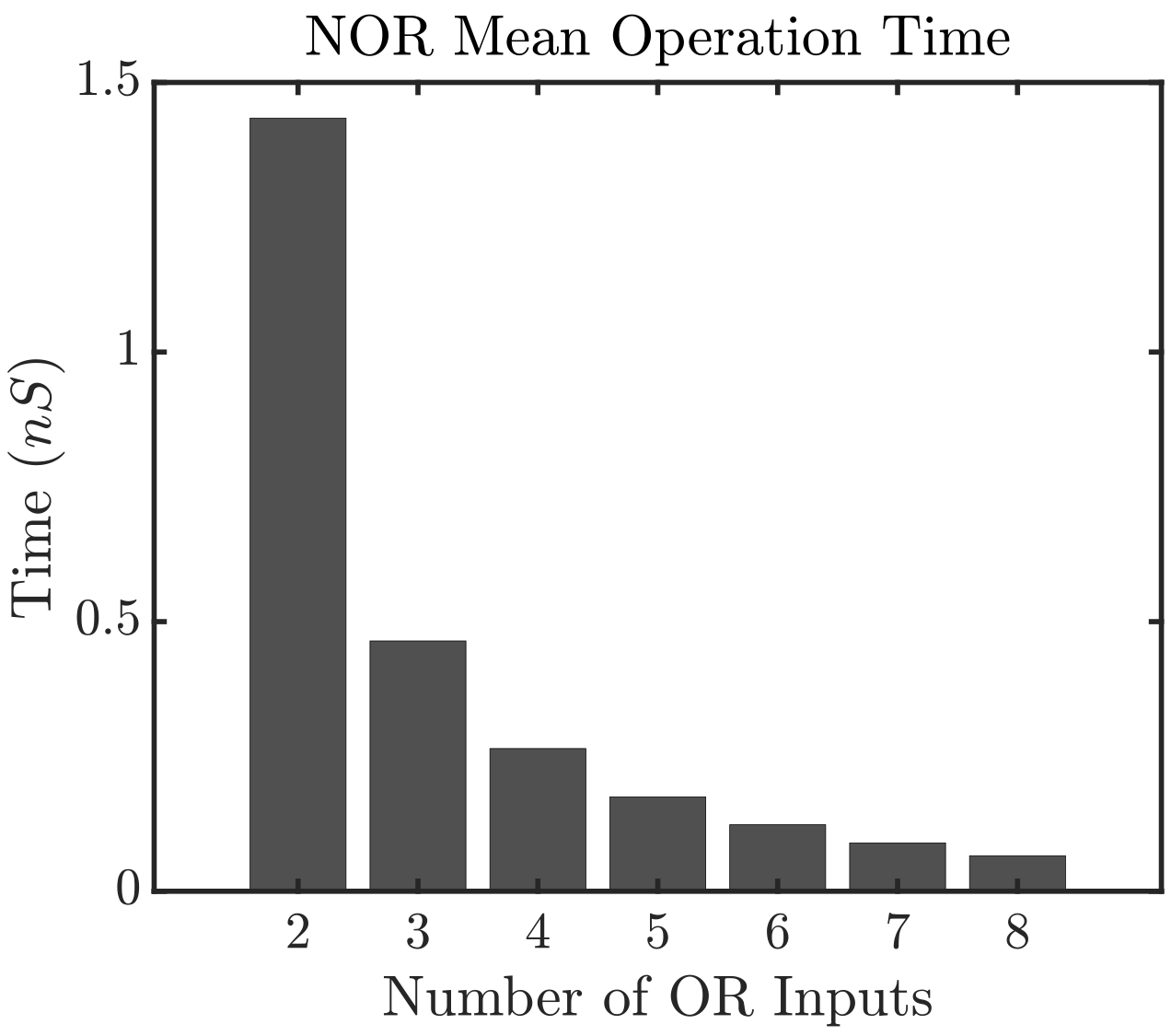}}
                \caption{}
                \label{MAGIC-True-h}
        \end{subfigure} \hspace {-2mm} 
              \begin{subfigure}[b]{0.18\linewidth}
                \centering
                \includegraphics[width=0.99\linewidth]{{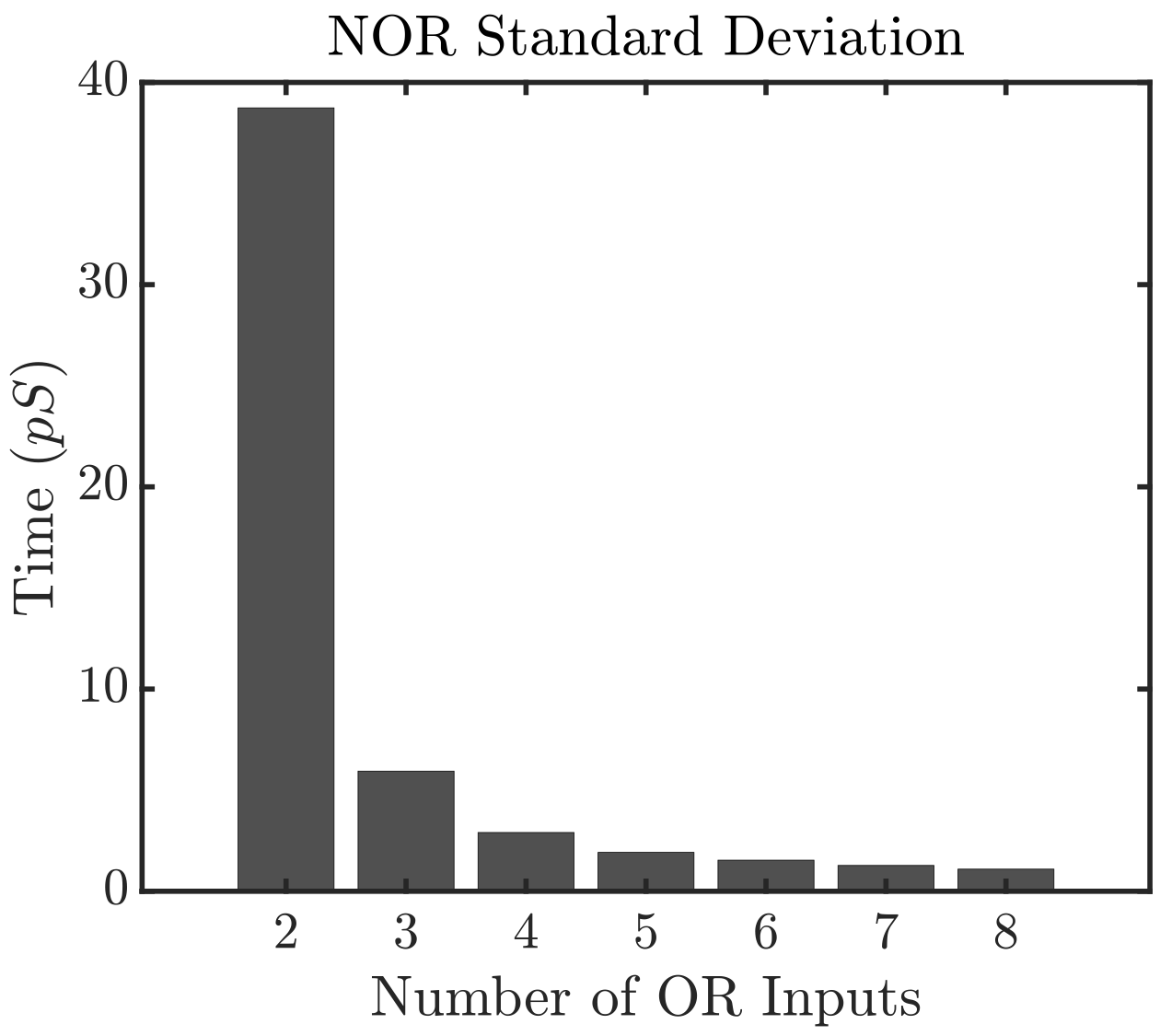}}
                \caption{}
                \label{MAGIC-True-i}
        \end{subfigure} \hspace {-2mm} 
        \caption{Analysis of current profiles with different MAGIC fanins under process variation to obtain CDF distribution of operation times, mean operation times, and STD of operation times for NOR gates ((a),(b), and (c)), OR gates ((d),(e), and (f)), and AND gates ((g), (h), and (i)), respectively.}
	\vspace{-2mm}
        \label{MAGIC_ALL}
\end{figure*}

\section{Attack on MAGIC Architecture}{\label{MAGIC-attack}}
An adversary can distinguish between the power drawn by the OR and AND array of MAGIC by examining the operation time determined using the spike in the power signature that is created during operation. The order of magnitude of difference between OR and AND operation times ranges from 10X to 100X (as seen in Fig. \ref{MAGIC_ALL}).

\subsection{Attack Model 1}{\label{MAGIC-True}}

\subsubsection{Leveraging the power signature/operation time of OR and AND arrays}
Unlike DCIM, MAGIC's computation time depends on the type of gate and the fanin. An adversary can RE the MAGIC functions using the computation time extracted from the power profile.  

MAGIC writes the result of a computation into a designated output RRAM by altering its resistance (HRS $\rightarrow $ `0' and LRS $\rightarrow $ `1'). We observe a significant change in the power profile when the resistance of the output RRAM changes. 
This sharp change in the power profile (during writing the output to RRAM) signifies the end of one MAGIC operation (e.g. 3-input $AND$ operation).  Note that the adversary is capable of finding the computation times for different gates and inputs by implementing known functions in MAGIC test chips and/or simulations and recording their power profiles.

Alternatively, the adversary can observe the constant current passing through output RRAM. In this approach, each of the input literals is set to a logical `$0$' (MAGIC initializes all the input RRAMs to HRS, the output RRAMs of $OR$ and $AND$ gates to HRS, and the output RRAMs of $NOR$ gate to LRS). By measuring the $V_{DD}$ current (minus the leakage current), the adversary can determine the current passing through the output RRAM. This allows the adversary to determine the gate implemented in a particular clock cycle (e.g. differentiate between the $AND$, $NOR$ and $OR$ gates). The observed constant current (I) and the fanin value ($n$) can be attributed to each gate ($AND$, $OR$, and $NOR$) based on the following rules (considering a 8 input system):

NOR: \[\frac{V_{write}}{R_{LRS}+\frac{R_{LRS}}{Max_{in}}} \geq I \geq \frac{V_{write}}{0.5 R_{HRS}}, \;\;\;\; n = \frac{R_{HRS}\times I}{V_{write}}\]

OR: \[\frac{V_{write}}{R_{HRS}} \geq I \geq \frac{V{write}}{1.5 R_{HRS}},\;\;\;\; \frac{n}{n+1} = \frac{R_{HRS} \times I}{V_{write}}\]

AND: \[\frac{V_{write}}{R_{HRS}} \geq I \geq \frac{V_{write}}{(Max_{in}+1)R_{HRS}}, \;\;\;\; n= \frac{V_{write}}{R_{HRS}\times I} -1\]

Note that due to the process variations, $n$ might not be an integer and should be used to approximate to the nearest positive whole number (since fanin should be integer). The proposed attack model works only with true inputs. If the function consists of complementary inputs, the attack will fail since adversary cannot confirm if all the minterms are `1' by forcing all inputs to `1'.

\subsubsection{Simulation Results}
To evaluate the above-mentioned attack model for MAGIC gates, IMC computations are performed for $AND$, $OR$, and $NOR$ gates. Note that MAGIC simulations for $NAND$ are not performed since it requires each of the input RRAMs to be initialized to HRS. Therefore, the output RRAM does not get enough voltage headroom to get written since the input RRAMs (in HRS) consume a high voltage across them. Further study is required to ensure the validity of the MAGIC $NAND$ design. For each of the remaining gates, an increasing ($AND$ gate) or decreasing ($OR$ and $NOR$ gates) trend of computation time is observed when the fanin of the logic array increases from 2 to 8. The computation completion is determined by a sharp change in the current profile during the switching of the output RRAM. 

i) \textbf{MAGIC AND Array}: The distribution of $AND$ operation times for fanin ranging from 2 to 8 is shown in {(Fig. \ref{MAGIC-True-a}}. The computation times for each of the cases is represented by a distinct distribution. Furthermore, the mean and STD of each of the distributions ({Fig. \ref{MAGIC-True-b} and \ref{MAGIC-True-c}}, respectively), show a monotonically increasing trend as the fanin increases. Each of these graphs can be used to accurately determine the fanin of the $AND$ gates. 

ii) \textbf{MAGIC OR Array}: The distribution of $OR$ operation times for fanin ranging from 2 to 8 is shown in Fig. \ref{MAGIC-True-d}. Unlike MAGIC $AND$, we find that the computation time distribution for each of the fanin overlaps. Therefore, the PDF alone cannot be reliably used by an adversary to determine the $OR$ gate fanin.  It is seen that the mean and STD of each of the distributions (Fig. \ref{MAGIC-True-e} and \ref{MAGIC-True-f}, respectively), show a decreasing trend with the fanin. We note that the MAGIC $OR$ implementation is comparatively resilient against SCA compared to $AND$ and $NOR$. But an adversary can still leverage the SCA data to predict the structure of the $OR$ gate and the fanin value with reasonable accuracy.  

iii) \textbf{MAGIC NOR Array}: The distribution of computation times for the $NOR$ operation with fanin ranging from 2 to 8 is shown in Fig. \ref{MAGIC-True-g}. Similar to $AND$, the completion times for each of the cases is represented by distinct CDFs. Furthermore, the mean and STD of each of the distributions (Fig. \ref{MAGIC-True-h} and \ref{MAGIC-True-i}, respectively), show a distinctly decreasing trend with increase in fanin. Each of these graphs can be reliably used to accurately determine the fanin of $NOR$ gate. 

\begin{figure*} [!t] 
        \centering
        \begin{subfigure}[b]{0.23\textwidth}
              \centering
              \includegraphics[width=0.99\linewidth]{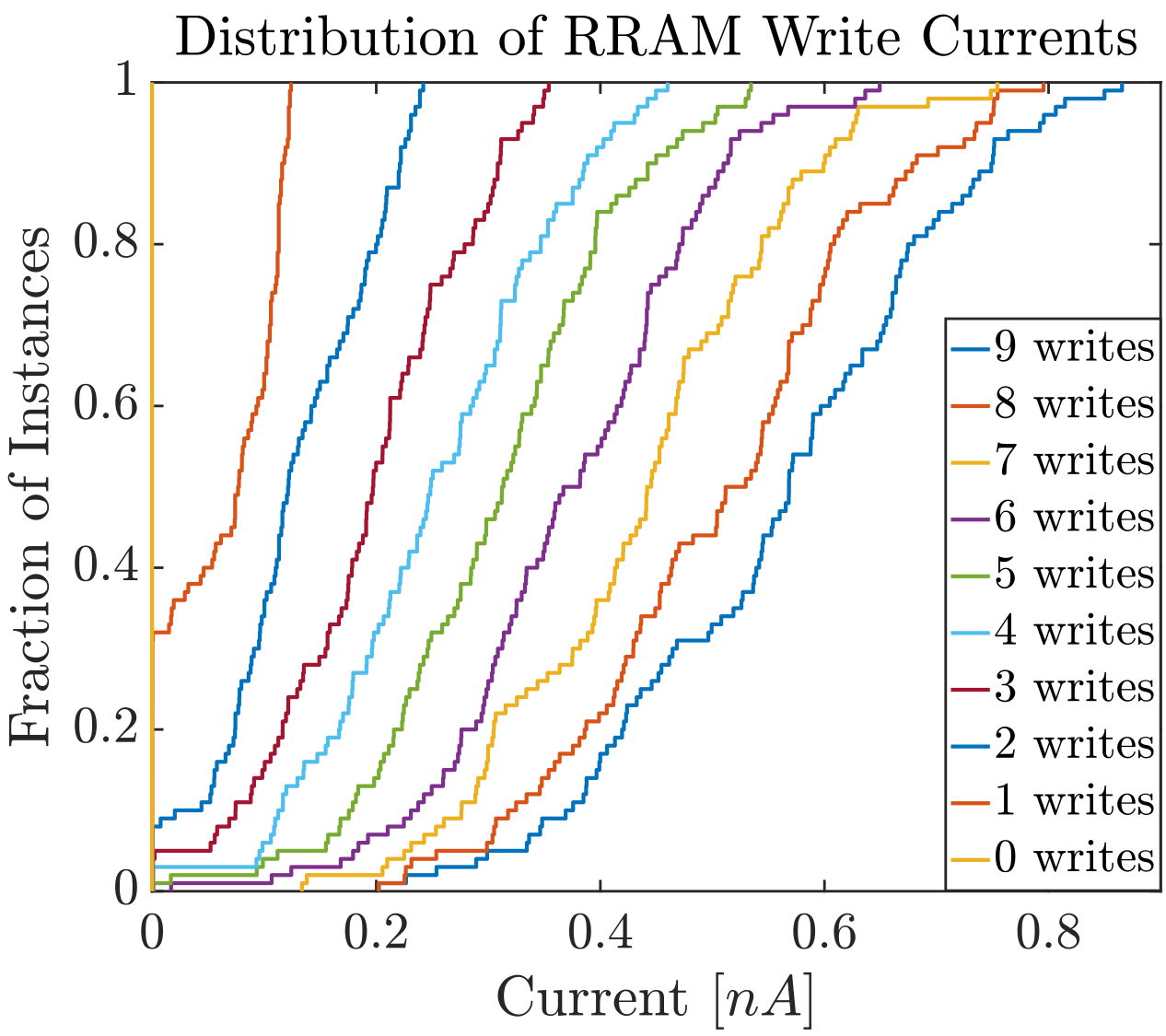}
                \caption{}
                \label{RRAM_write_a}
        \end{subfigure} \hspace {2mm} %
        \begin{subfigure}[b]{0.23\textwidth}
                \centering
              \includegraphics[width=0.99\linewidth]{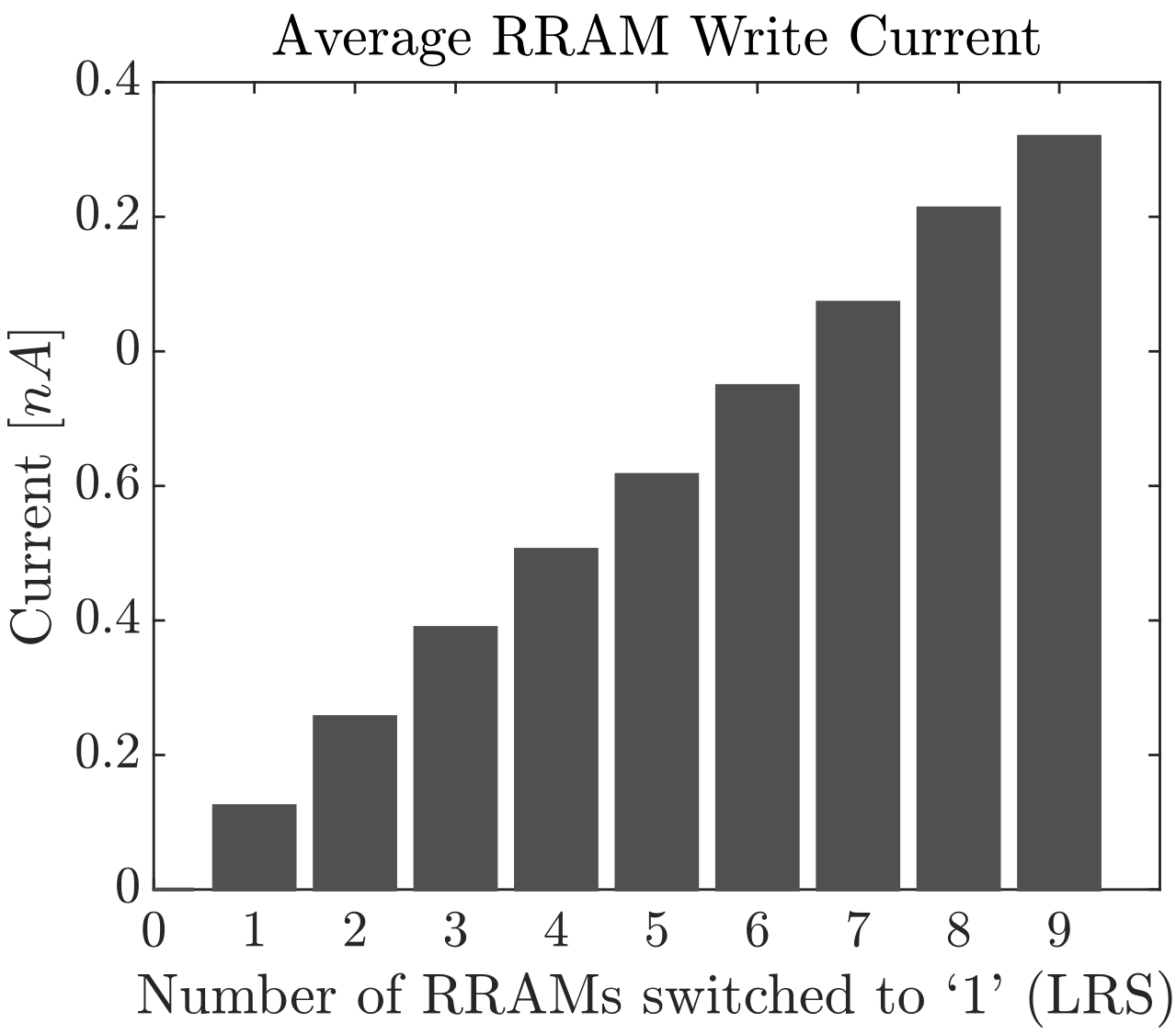}
                \caption{}
                \label{RRAM_write_b}
        \end{subfigure} \hspace {2mm} %
        \begin{subfigure}[b]{0.23\textwidth}
                \centering
                \includegraphics[width=0.99\linewidth]{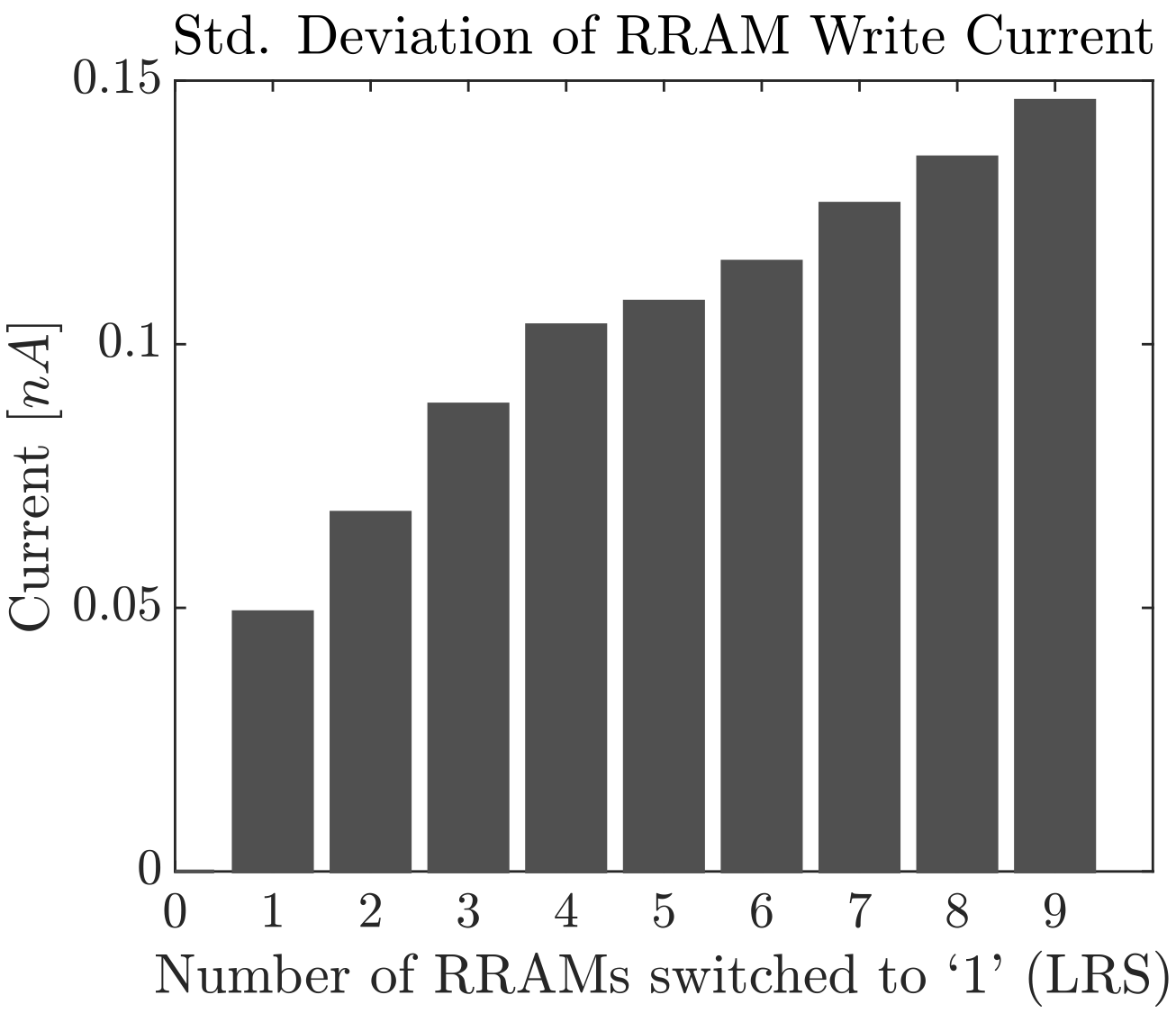}
               \caption{}
                \label{RRAM_write_c}
        \end{subfigure}
        \vspace{-2mm}
        \caption{Simulation results showing (a) write current distribution; (b) average write current; and (c) STD of the distribution for switching 1 to 9 RRAM cells (8 input + 1 output) from `0' to  `1'.}
        \label{RRAM_write}
        \vspace{-2mm}
\end{figure*}

\subsection{Attack Model 2} 
\subsubsection{Leveraging pre-compute RRAM write operation times of AND or OR arrays} 
Since inputs are stored as resistance values in RRAMs, their write operations are asymmetric and the adversary can find the values which are stored in the RRAMs by examining the power profile. The adversary can force each of the inputs to logical `1' (LRS) and examine the RRAM write currents. Based on the write-current observed the adversary can determine the number of RRAM cells switched to `0' (HRS) and the number of cells that remain at `1' (HRS). IMC of any function through MAGIC occurs only after the RRAM cells corresponding to inputs and output are initialized to HRS or LRS values depending on the function and the array operation (i.e. NOR, OR, etc). Assuming a representative example function with 8 input literals, the MAGIC architecture will employ 8 input RRAMs and 1 output RRAM. Furthermore, each of these are preset to `0' (HRS) state. To execute a particular function, some or all of these 9 RRAMs resistances are switched to `1' (LRS state). Note that the power consumed for switching different number of RRAMs (0 to 9 in this case) is distinct and can be extracted through SCA. 

In addition to the case-by-case method explained here, the adversary can determine the number of HRS RRAMs ($n$) by using the following equations (similar to equations mentioned in Section {\ref{MAGIC-True}}):

NOR: \[\frac{V_{write}}{R_{LRS}+\frac{R_{LRS}}{Max_{in}}} \geq I \geq \frac{V_{write}}{0.5 R_{HRS}}, \;\;\;\; n = \frac{R_{HRS}\times I}{V_{write}}\]

OR: \[\frac{V_{write}}{R_{HRS}} \geq I \geq \frac{V{write}}{1.5 R_{HRS}},\;\;\;\; \frac{n}{n+1} = \frac{R_{HRS} \times I}{V_{write}}\]

AND: \[\frac{V_{write}}{R_{HRS}} \geq I \geq \frac{V_{write}}{(Max_{in}+1)R_{HRS}}, \;\;\;\; n= \frac{V_{write}}{R_{HRS}\times I} -1\] 

\subsubsection{Simulation Results}
Simulation setup for the attack model 2 on MAGIC is similar to the one in model 1, but with an added step (looking during RRAM initialization). This added step reveals the complementary inputs. 

A 100-point Monte Carlo analysis is performed on RRAM write current with the setting shown in Table \ref{PV}. The resulting current distribution, average current, and STD of the distribution are shown in Fig. \ref{RRAM_write_a} \ref{RRAM_write_b} and \ref{RRAM_write_c}, respectively. It is evident that the number of RRAMs initialized to `1' and `0' can be found. In order to determine the inputs whose complementary values are used, each input is flipped from its original value (`1') one at a time. If a change in an input value (1$\rightarrow$ 0) leads to an increase in the number of `0's (HRS RRAMs), determined by re-examining the power signature, we can deduce that the original value of the input is used in the function. Alternatively, if the number of `0's decreases, the input's complementary value is used in the function. In this way, adversary can extract the structure of function with true and complementary inputs. 

 \begin{figure} [!t] 
        \centering \hspace{-5mm}
        \begin{subfigure}[b]{0.24\textwidth}
               \centering
              \includegraphics[width=0.99\linewidth]{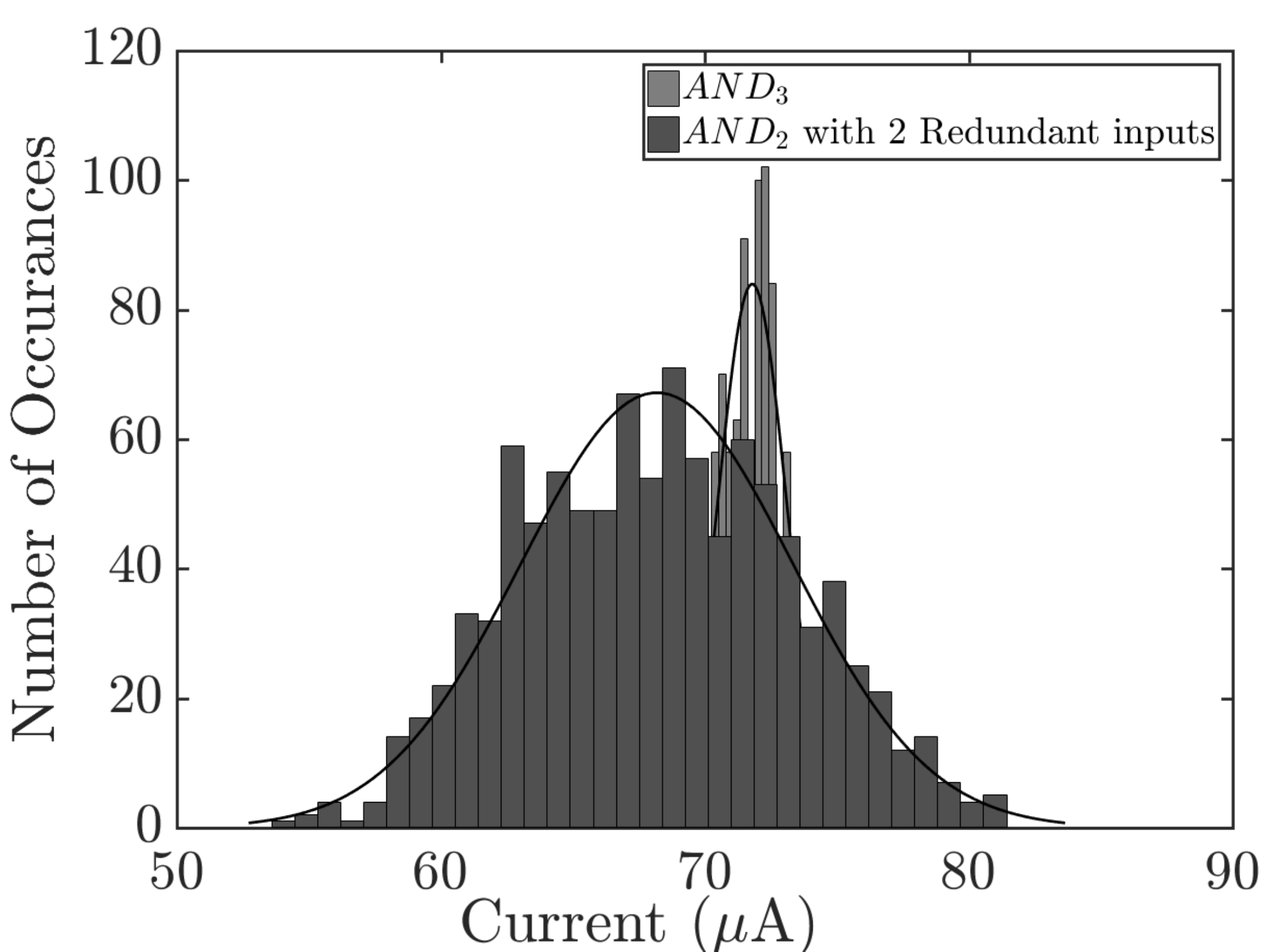}
                \caption{}
                \label{Redundant-Inputs-PDF}
       \end{subfigure}  \hspace{-3mm}
        \begin{subfigure}[b]{0.24\textwidth}
                \centering
             \includegraphics[width=0.99\linewidth]{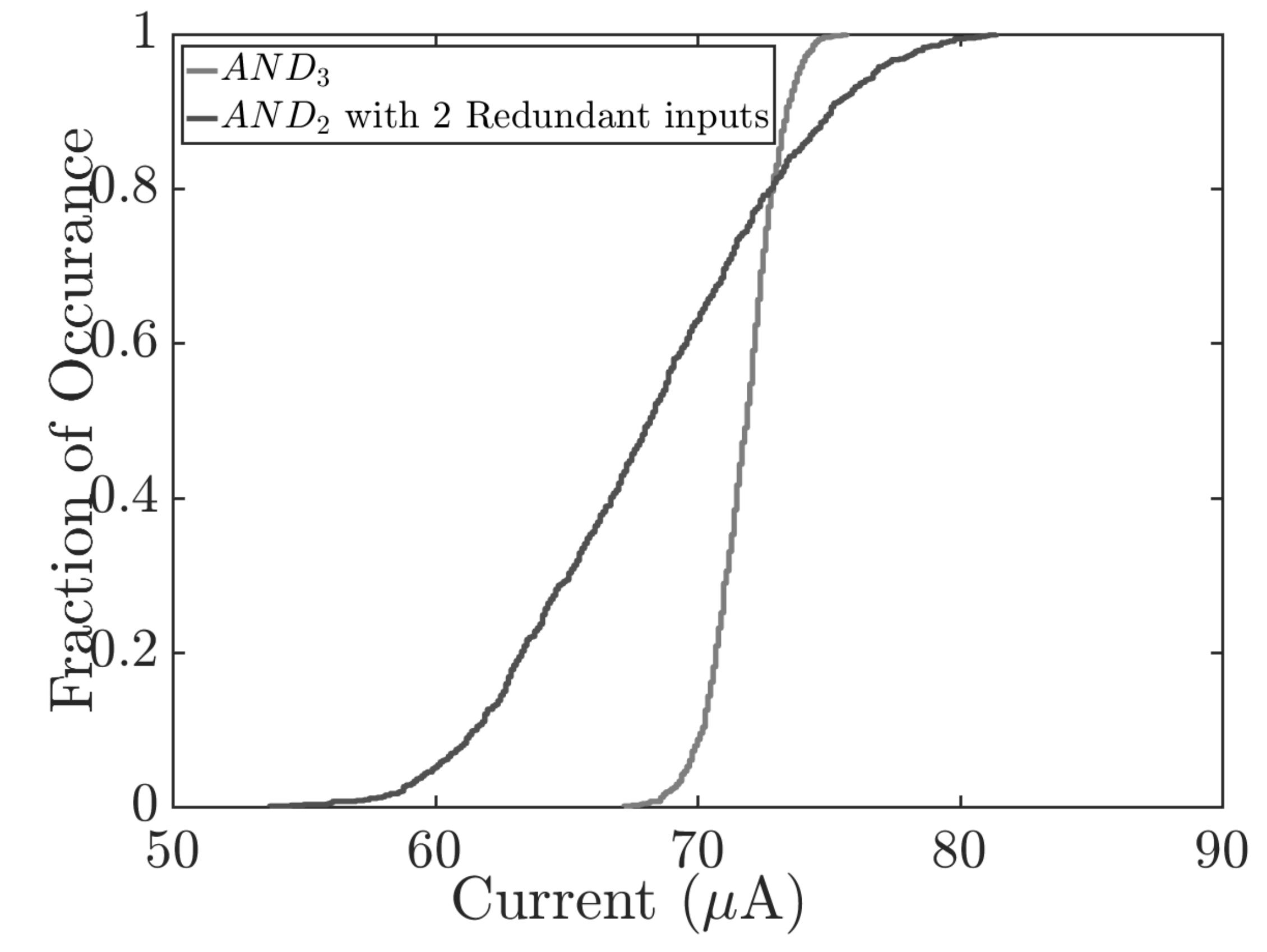}
               \caption{}
               \label{Redundant-Inputs-CDF}
       \end{subfigure}%
        \vspace{-2mm}
        \caption{Analysis of redundant inputs as countermeasure. (a) PDF and (b) CDF power profile of $AND_2$ with two redundant inputs and $AND_3$ without any redundant inputs. There is an overlap between them which obfuscate the power profile.}
        \label{Redundant-Inputs-DCIM}
        \vspace{-3mm}
\end{figure}

\subsection{Analysis of the Impact of Supply Voltage Magnitude on SCARE Performance}

\begin{figure}
        \centering 
        \begin{subfigure}[b]{0.49\linewidth}
                \centering
                \includegraphics[width=0.99\linewidth]{{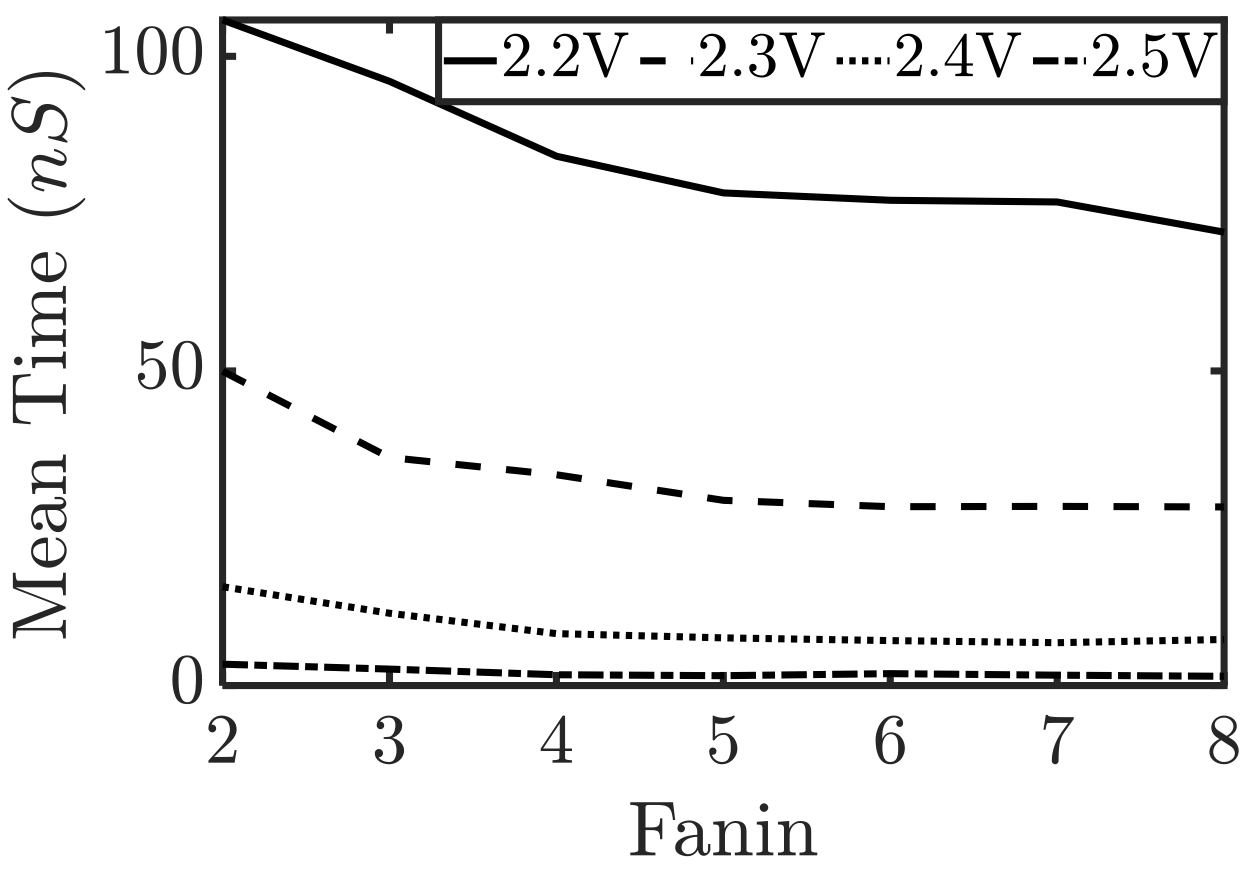}}
                \caption{}
                \label{MAGIC_Voltage_Mean1} 
        \end{subfigure} \hspace {-2mm} 
        \begin{subfigure}[b]{0.49\linewidth}
                \centering
                \includegraphics[width=0.99\linewidth]{{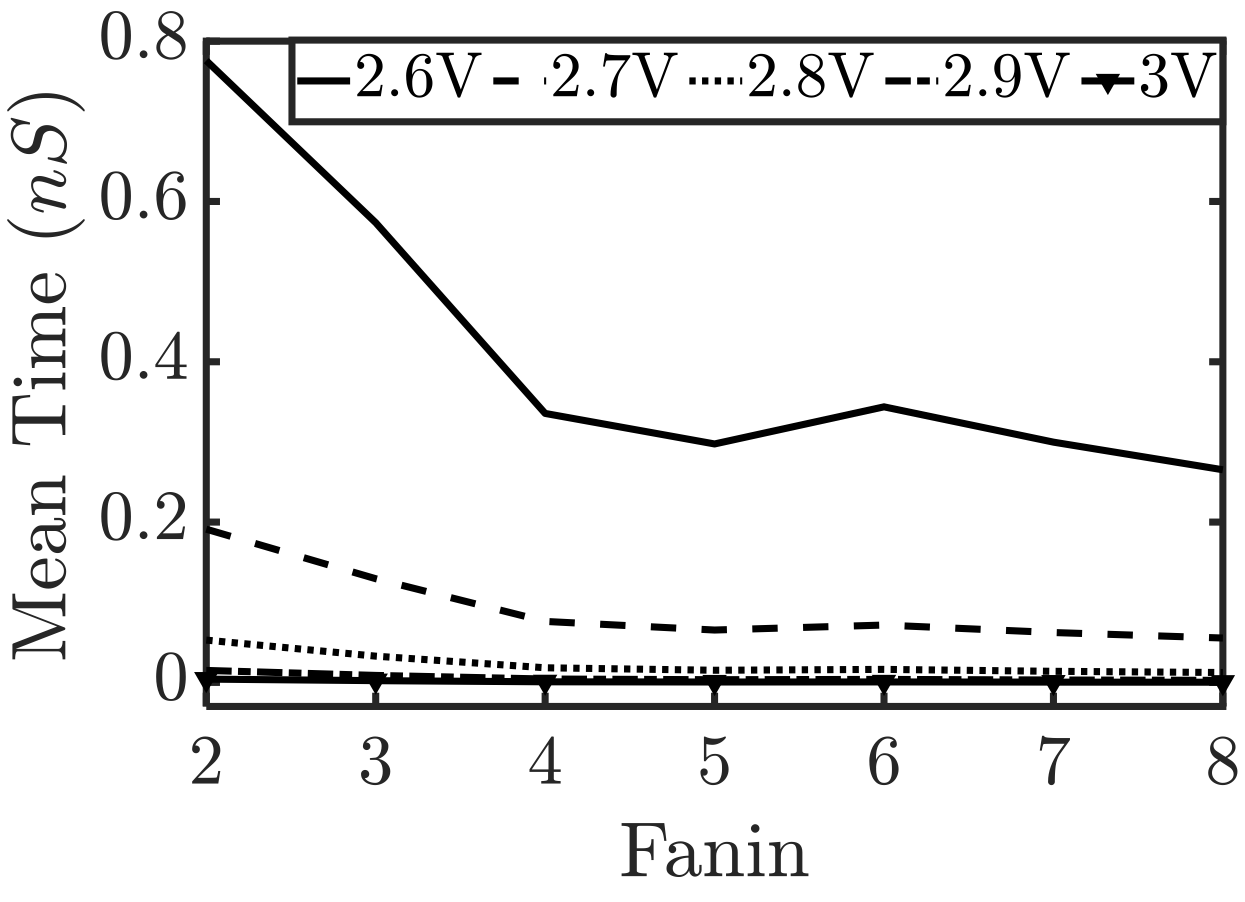}}
                \caption{}
                \label{MAGIC_Voltage_Mean2}
        \end{subfigure} \hspace {-2mm} 
        \begin{subfigure}[b]{0.5\linewidth}
                \centering
                \includegraphics[width=0.99\linewidth]{{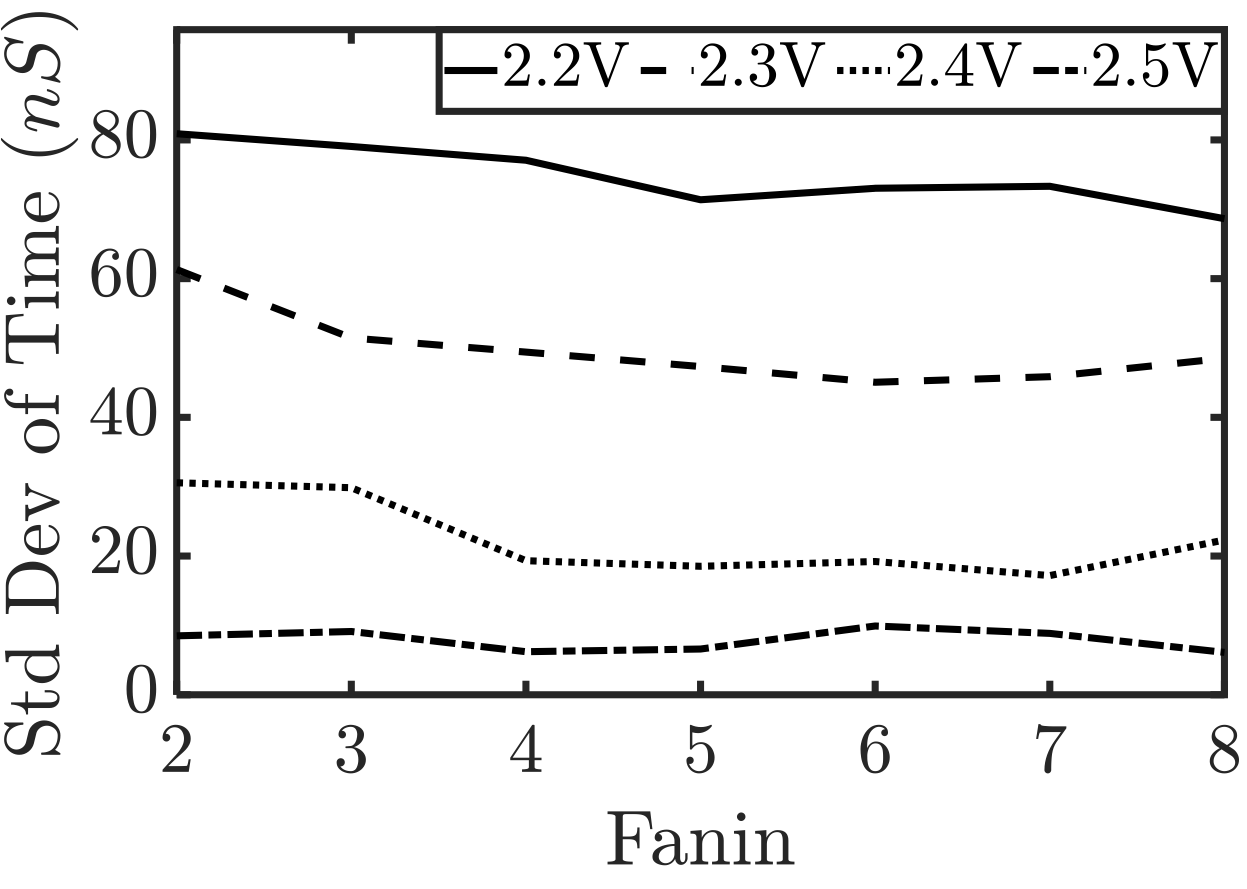}}
                \caption{}
                \label{MAGIC_Voltage_Std1}
        \end{subfigure}\hspace {-2mm}
        \begin{subfigure}[b]{0.5\linewidth}
                \centering
                \includegraphics[width=0.99\linewidth]{{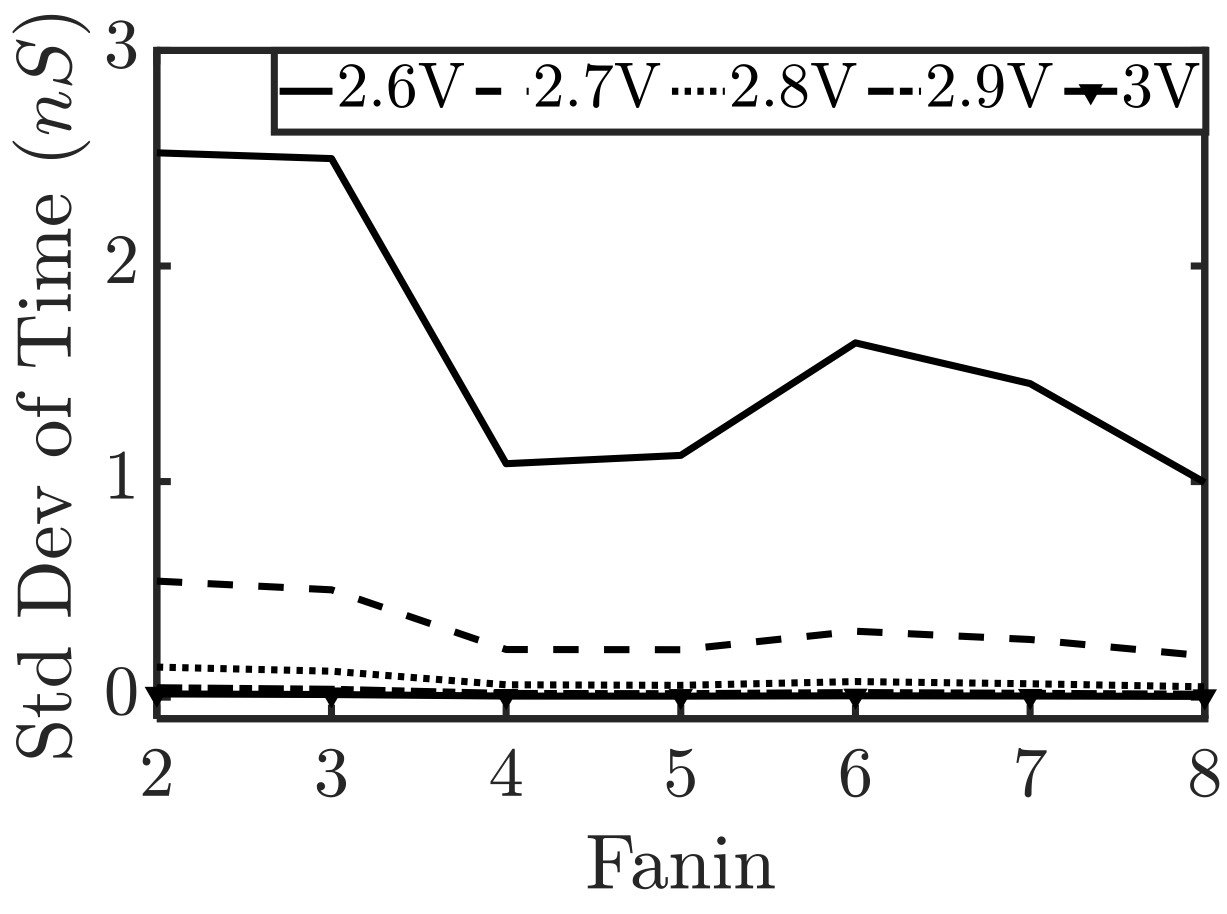}}
                \caption{}
                \label{MAGIC_Voltage_Std2}
        \end{subfigure}\hspace {-2mm}
        \caption{Analysis of operation time mean values, STD deviations, and $OR$ CDF under different voltage nodes. (a),(b) Mean Values, and (c),(d) STD Deviations of $OR_8$.}
	\vspace{-3mm}
        \label{MAGIC-Voltage}
\end{figure}

\begin{figure}
	\centering
    \includegraphics[width=0.7\linewidth]{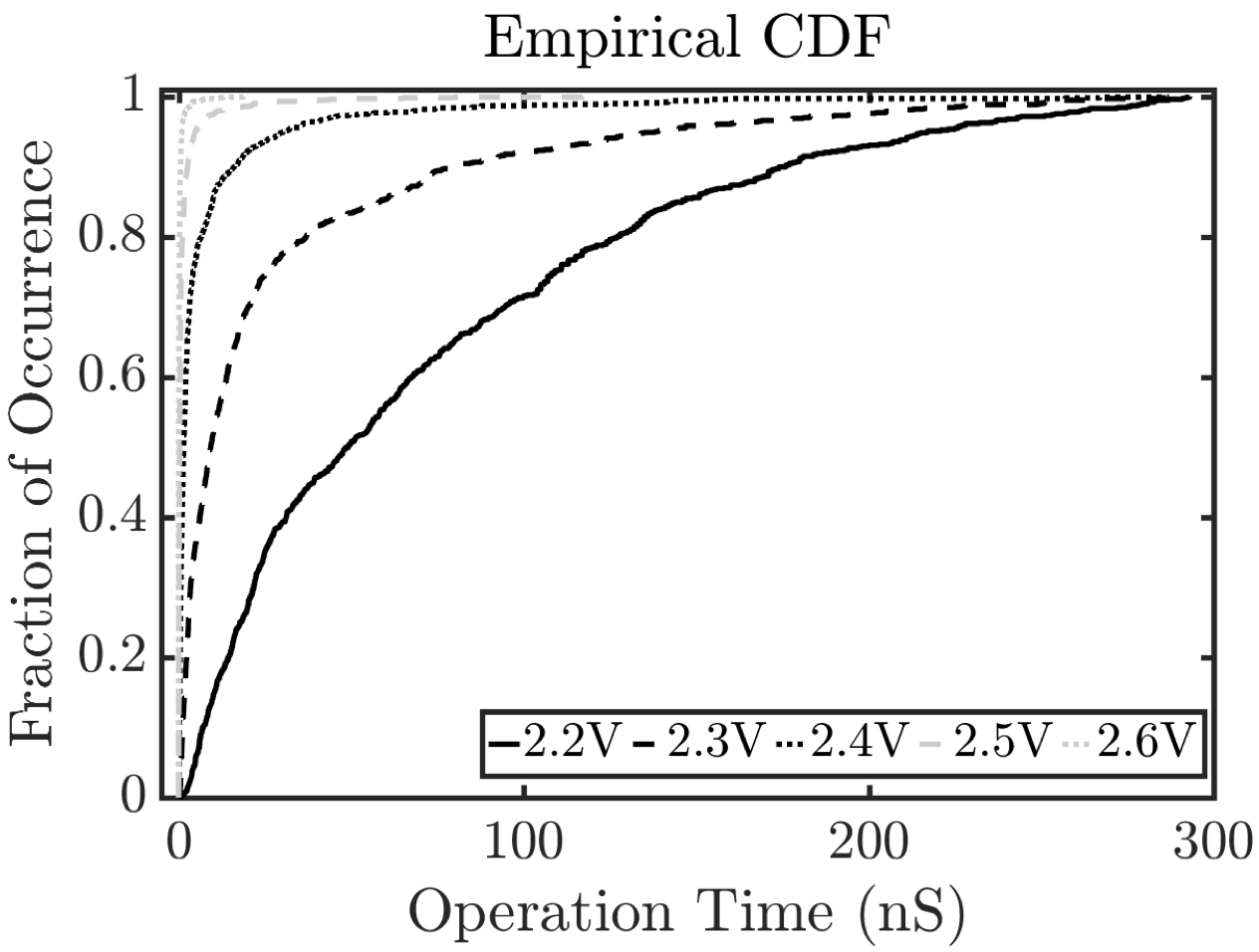}
	\caption{CDF of the operation times of MAGIC $OR_8$ with increasing voltage.}
	\label{MAGIC_Voltage_CDF}
	 \vspace{-6mm}
\end{figure}

In case of MAGIC, the $V_{DD}$ value is swept from 2.2V to 3V in 100mV increments. As shown in Fig. {\ref{MAGIC_Voltage_Mean1}} and {\ref{MAGIC_Voltage_Mean2}}, the mean operation times decreases as the $V_{DD}$ value increases. Furthermore, it is also seen that the standard deviation value decreases under all $V_{DD}$s as the the fanin value increases. Fig. {\ref{MAGIC_Voltage_Std1}} and {\ref{MAGIC_Voltage_Std2}} show that the standard deviation of the operation also decreases with increase in $V_{DD}$ and shows a mostly negative trend with the change in fanin. Fig. {\ref{MAGIC_Voltage_CDF}} shows that the slope increases as the $V_{DD}$ value increases and shows a decrease in sigma value. The slope of $V_{DD}$ $>2.6V$ is extremely high and is therefore not shown in Fig. {\ref{MAGIC_Voltage_CDF}} since they overlap with the CDF at $V_{DD} = 2.6V$. 

\section{Countermeasures}
We propose the following countermeasures in order to protect IMC architectures against SCARE.

\subsection{Redundant Inputs}
\textbf{DCIM:} Few redundant LRS RRAMs on each BL can be implemented which are biased with a fixed voltage. 

For instance, function $a+bc$ can be implemented as: $a.1.1.1.1+b.c.1.1.1+0+0+0+0+0$. In this method area increases and the overhead is based on the number of maximum redundant inputs and number of inputs (e.g. four redundant inputs for a function with eight inputs increases the area by 0.25\%). As long as the number of redundant LRS RRAMs in each BL is less than 8, the SM stays relatively constant. Based on our simulations, eight redundant inputs are enough to mask an array with 64 inputs, which increases the area by 6\%. The number of LRS RRAMs on each BL can be randomly distributed to further obfuscate the structure of the implemented function. Power overhead is completely dependant on the number of redundant inputs in each BL. An example of masking $AND_2$ with two redundant inputs is shown in Fig. \ref{Redundant-Inputs-DCIM}. It can be noted that the $AND_2$ power profile with two redundant inputs completely overlaps with $AND_3$ power profile with no redundant inputs. Therefore, power profile signature gets obfuscated. In this example, the power overhead is 21\%.

Since the selector diode turns off when the voltage across its two terminals is less than $V_{th}$, the AND (OR) array's redundant inputs should not be driven by $V_{DD}$ (`0') instead they should be driven by $\frac{V_{DD}}{3}$ for AND array and $\frac{2V_{DD}}{3}$ for OR array for better obfuscation.

\textbf{MAGIC:} In the case of MAGIC, redundant inputs increases the fanin, and thus increases the number of input RRAM bitcells. As previously shown in Section \ref{MAGIC-attack}, increasing the number of inputs even by `1' literal has a distinguishable change in the operation completion time. This change as depicted in Fig. \ref{MAGIC_ALL}, can be leveraged to mask the true structure of any MAGIC implementation for any of the operations.

\subsection{Minterms with expanded literals}
\textbf{DCIM:} Each minterm in a function can be implemented by the maximum number of inputs. In this scenario, all minterms show the same power profile and SCA alone fails. However, an adversary can still try all the possible input patterns and generate input-output pairs. Next, that can be used to determine the function by using a Karnaugh map to reveal the simplified Boolean expression. 

\begin{figure} [!t] 
        \centering (a)  \hspace{-5mm}
        \begin{subfigure}[b]{0.45\textwidth}
                \centering
              \includegraphics[width=0.99\linewidth]{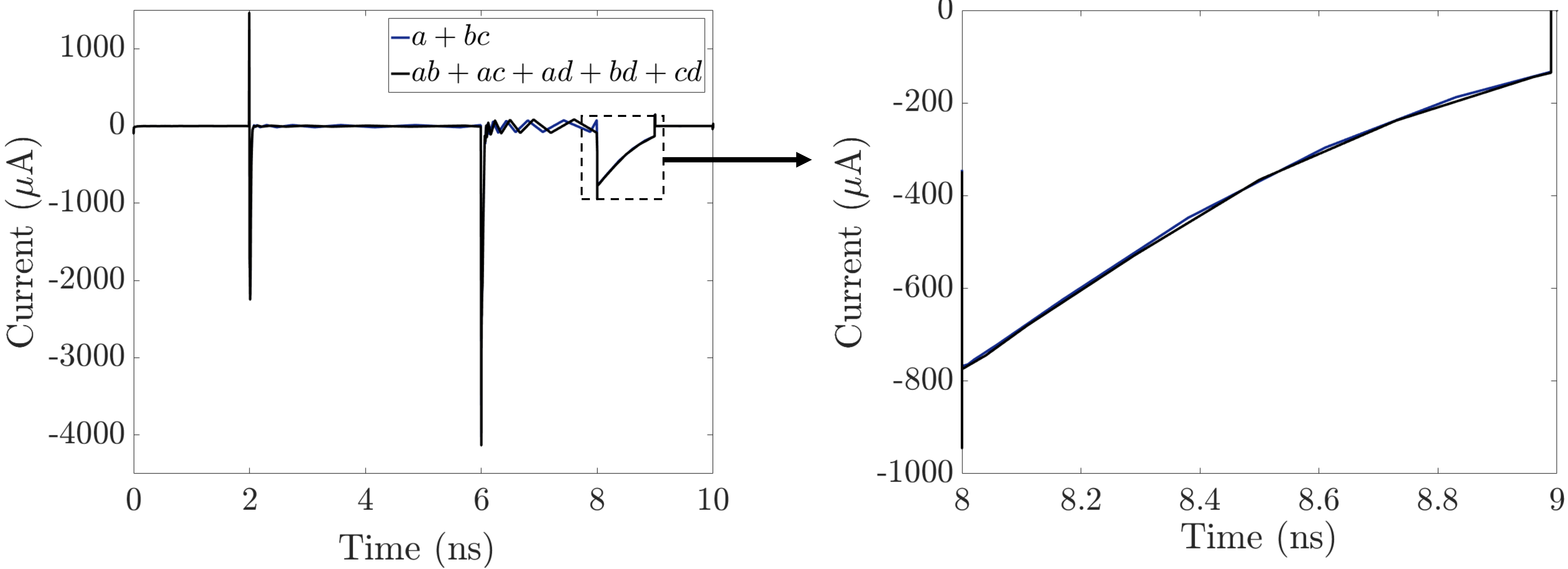}
                \label{Large-Literals-a}
        \end{subfigure} \\ (b) \hspace{-5mm}
        \begin{subfigure}[b]{0.45\textwidth}
                \centering
              \includegraphics[width=0.99\linewidth]{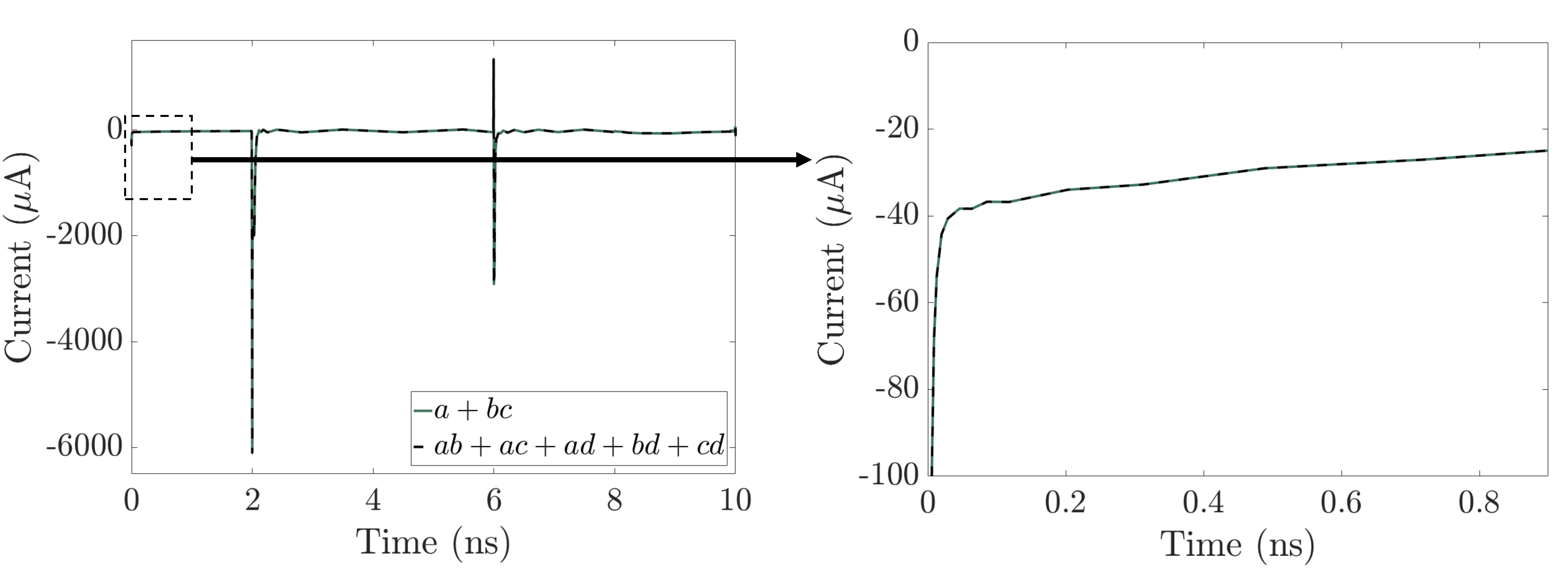}
                \label{Large-Literals-b}
        \end{subfigure}%
        \vspace{-2mm}
        \caption{Analysis of expanding the literals as countermeasure. Power profile of (a) AND during pre-charge and (b) OR during computation for computing $a+bc$ and the expanded version of it. They overlap completely with each other.}
        \label{Large-Literals}
        \vspace{-3mm}
\end{figure}

For example, $a+bc$ can be implemented in a 4 input system as,  $ab\overline{c}\overline{d}+a\overline{b}\overline{c}\overline{d}+ab\overline{c}d+a\overline{b}\overline{c}d+abcd+abc\overline{d}+a\overline{b}cd+a\overline{b}c\overline{d}+\overline{a}bcd+\overline{a}bc\overline{d}$. Furthermore, it will become complicated for the adversary to find the function when $a+bc$ and $ab+ac+ad+bd+cd$ have the same number of minterms in the expanded version and when $a+bc$ has more minterms than $ab\overline{c}+\overline{a}bc+\overline{a}\overline{b}\overline{c}$. This technique can protect the IP at the cost of increased area and power overhead. An example of masking the function by using this technique is shown in Fig. \ref{Large-Literals}, which the two functions consume the same power. Power and area overhead depends on the implemented function and for the example in Fig. \ref{Large-Literals} power consumption increases by 36\% and AND array area stays the same (since crossbar array is already there and it has enough BLs to implement the minterms). However, OR array's area increases by 50\% (number of WLs is changed).

\begin{figure} [!t] 
        \begin{subfigure}[b]{0.23\textwidth}
                \centering
              \includegraphics[width=0.99\linewidth]{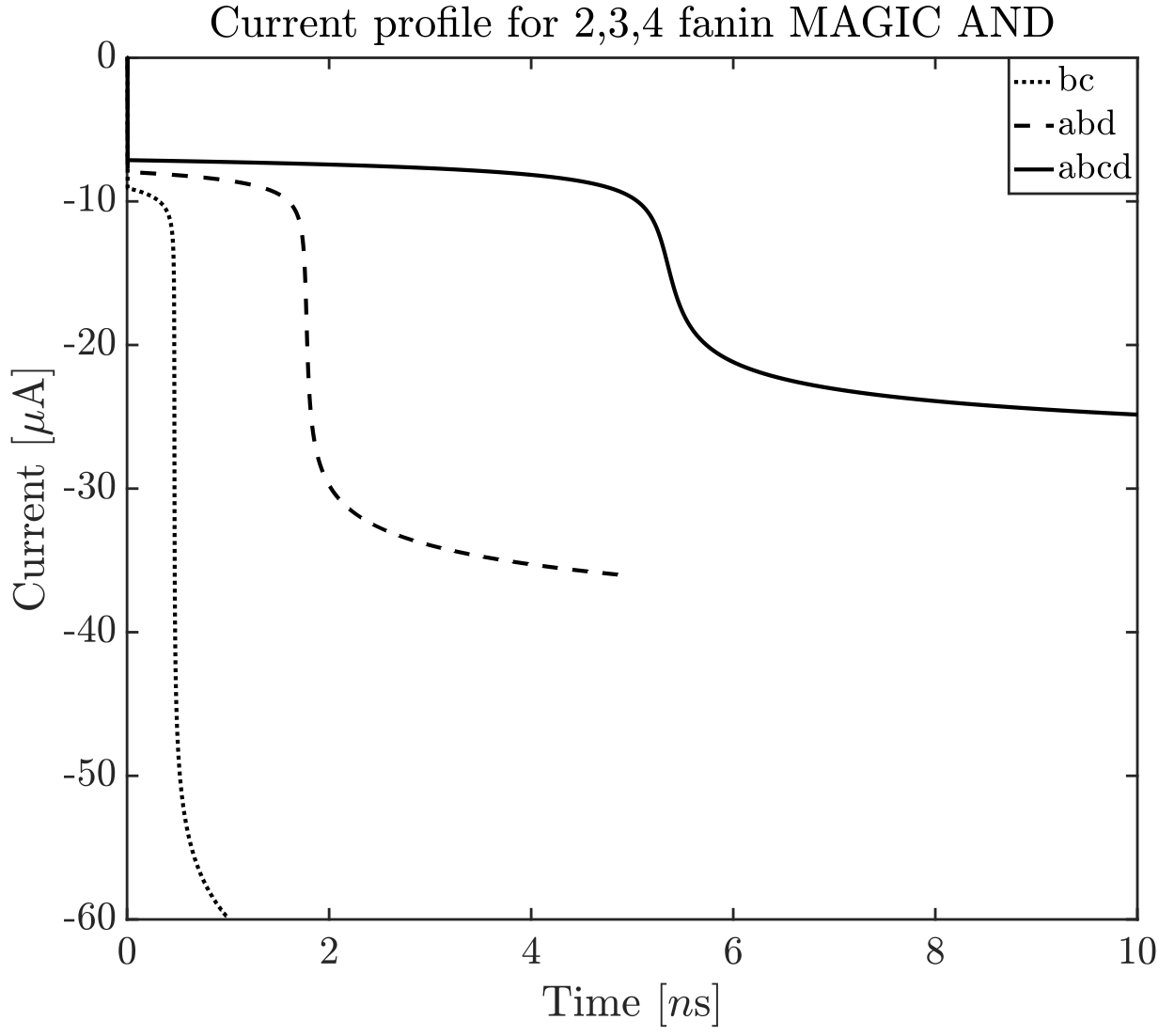}
               \caption{}
                \label{Count_AND}
        \end{subfigure}%
        \begin{subfigure}[b]{0.23\textwidth}
                \centering
              \includegraphics[width=0.99\linewidth]{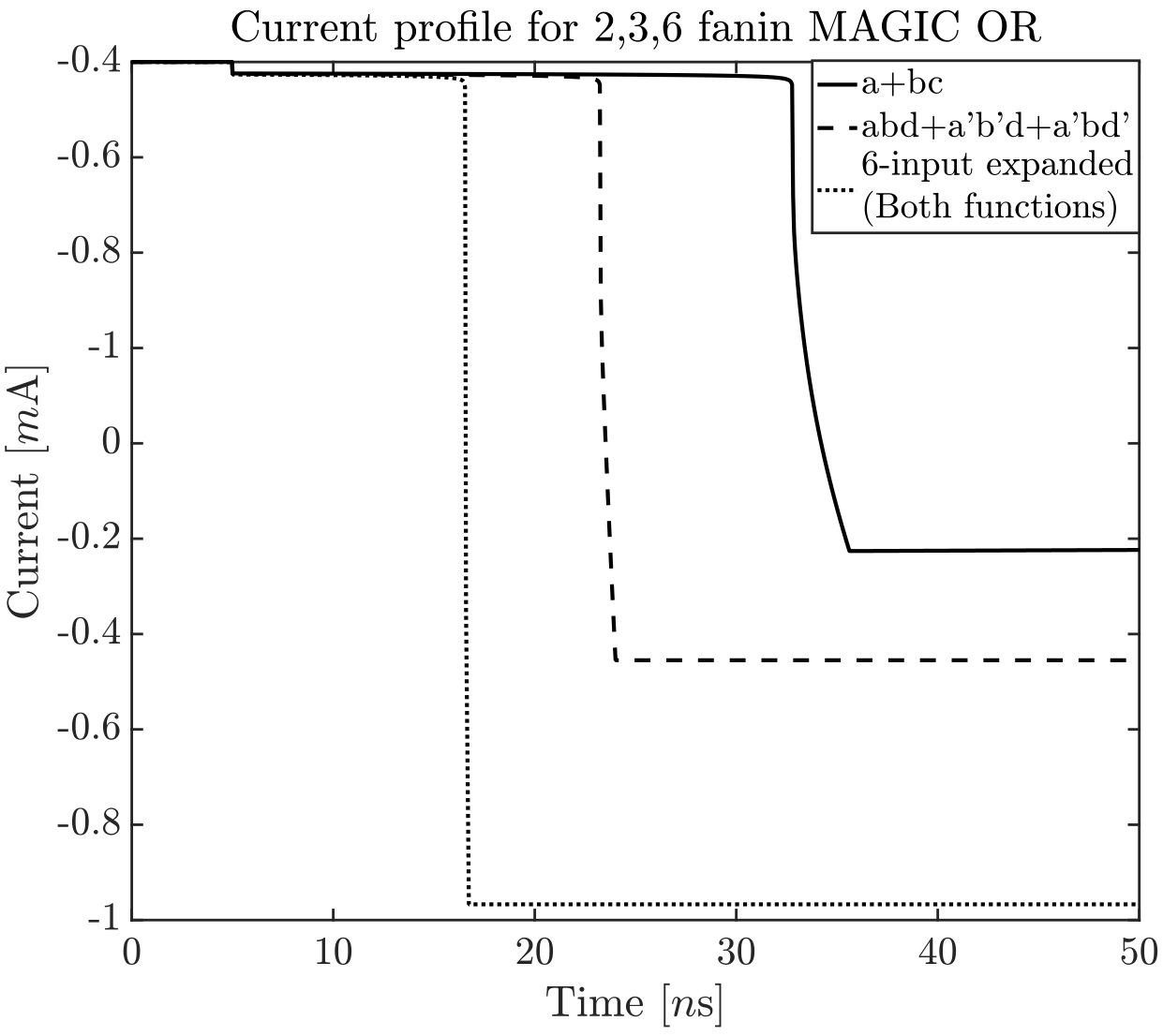}
               \caption{}
                \label{Count_OR}
        \end{subfigure}%
        \vspace{-2mm}
        \caption{Simulation results that show masking of original function structure using maximized SOP form for 2 functions (a) 2, 3, and 4-fanin MAGIC AND current profiles; (b) 2, 3, and 6-fanin MAGIC OR current profiles.}
        \label{MAGIC-masking}
        \vspace{-3mm}
\end{figure}

\textbf{MAGIC:} Similarly for MAGIC, we consider two representative example functions $a+bc$ and $abd+\overline{a}\overline{b}d+\overline{a}b\overline{d}$. The first function requires a 2-fanin AND and a 2-fanin OR operation, while the second function requires three 3-fanin ANDs and one 3-fanin OR operation. Expanding these functions into their maximized SOP form will require six 6-fanin ANDs and one 6-fanin OR operation for both of the functions. Since these operations are identical, SCA delivers the same result and masks the true structure of the operation. Fig. \ref{Count_AND} shows that 2-fanin and 3-fanin ANDs have distinctly different operation completion time as depicted by a sharp change in their current profiles. The 4-fanin AND for both operation in their maximized SOP form is shown to be identical. Similarly, Fig. \ref{Count_OR} shows the distinctly different 2-fanin and 3-fanin current profiles of each function's OR operation and depicts the identical 6-fanin OR current profiles for their maximized form. This attack model will not incur any area overhead since the maximized SOP form will only leverage previously present RRAM cells in the crossbar array for any additional literals. But, it will incur some power overhead due to the increase in the number of SOP minterms to be computed.

The above-mentioned countermeasures increase the RE effort. For example, $ab+cde+fgh$ without any countermeasures would require 84 combinations of inputs to determine the function structure. Implementing the countermeasures increases the required number of combinations to 256 (RE effort increases by 3.04X). This increased effort enhances the resiliency of the IMC architectures against $SCARE$. RE effort increases exponentially with the number of inputs.


\section{Discussions}
\subsection{Extracting the Exact Function}
For MAGIC, in the absence of parallelism, the adversary can find the number of input literals for each minterm and test possible patterns to determine the exact minterm that correlates to a particular function. The adversary can repeat this approach to find all minterms one by one. For DCIM, the adversary can determine the number of input literals per minterm after calculating the number of minterms. Finally, by examining the output, the adversary can try multiple patterns to determine the exact correlating minterms.

\subsection{Extracting Multiple Functions}
For DCIM, the total number of functions implemented can be determined by the number of outputs generated. For each of these functions, the number of minterms per function and the number of input literals per minterm can be extracted by following the method described in this paper. The adversary then proceeds to manipulate the input values to determine the correlation of each minterm with each function output. Since MAGIC does not support parallelism, adversary can easily determine the number of input literals per minterm and try various patterns to correlate each minterm with each function.

\subsection{Number of Test Chips Needed for an Attack}
We performed 1000-point Monte-Carlo simulations to develop SCARE’s power models for DCIM/MAGIC. In absence of models, adversary can fabricate few chips to launch the attack with minor loss in accuracy. The mean value of worst case margin (e.g.,margin between AND7/AND8 for DCIM and OR7/OR8 for MAGIC) is degraded by 3.5\%, 3.2\%, 1.8\% (for DCIM) and 5.2\%, 4.77\%, and 2.8\% (for MAGIC)  for 25, 50 and 100 chips, respectively compared to 1000-point Monte-Carlo. Furthermore, standard deviation increases by 18.5\%, 15.9\%, and 4.9\% for DCIM and 21.3\%, 9\%, and 7.9\% for MAGIC. Adversary can minimize measurement noise by taking multiple samples and averaging them. This approach is possible for IoT/mobile applications where adversary has physical possession.

\subsection{Realistic Attack \& Parallelism}
IMC architectures include MAGIC, DCIM and matrix-vector multipliers (MVM) \cite{MVM}. DCIM and MAGIC can implement arbitrary functions while MVM can only implement dot product. 

Under parallel operations/functions, adversary can find the number of functions by observing the number of outputs in DCIM/MAGIC. In DCIM, number of outputs = number of NOR gates. Power in second-cycle yields NOR gate fanins (e.g., two $OR_2$, one $OR_3$). Power in cycle-1 yields number of AND gates/fanins. After determining the total number of AND/OR gates, adversary can run a limited number of input patterns to relate the input bits to the corresponding observed output bit. Compared to brute-force, SCARE reduces number of test patterns to RE functionality e.g., 62.5\% less patterns to identify $ab+cd$.

In MAGIC, designers need multiple arrays to implement functions in parallel. Power peaks might overlap. The number of functions can be found from the magnitude of power (e.g., for writing 1- vs 2-output RRAMs). While multiple array operations can cause overlapping power spikes due to more than one AND/OR, adversary can determine the individual gates by dividing the total power as a summation of individual powers as modeled before. Timing difference between completion of gate operation can also be exploited.

\subsection{SCARE on conventional memory}
Although not covered in this paper, SCARE is also applicable to SRAM based IMC such as, X-SRAM \cite{X_SRAM} since the power timing profile of the read bitline during computation depends on input patterns (e.g. read bitline discharges faster for function $AB$ if A = 1/B = 1 compared to A = 0/B = 1 or A = 1/ B = 0). These studies are subject of our future research.

\subsection{Hybrid Architectures}
SCARE will experience noise from CMOS-logic for mixture of IMC and CMOS-gates. MAGIC-based IMC (pipelined/non-pipelined) involve high-power and long latency write operations that can be distinguished from CMOS-logic power. Furthermore, the CMOS-logic will compute after IMC for non-pipelined implementations of DCIM/MAGIC and can be separated in time. Pipelined implementations of DCIM will combine CMOS and IMC powers and it could be difficult to distinguish them. This could be a subject of future studies.

\subsection{Prior Knowledge on the Implementation}
Adversary can identify the sequence of gates without any prior knowledge. Note, there are only two efficient methods for function implementation in IMC namely, SOP or Product-of-Sum (POS). For DCIM, adversary can distinguish between SOP/POS by observing the polarity of current drawn during pre-charge phase of each cycle e.g., negative (current drawn from voltage supply) (positive (current drwan by the ground node)) current for AND (OR) array. For MAGIC, adversary can identity function (AND/OR/NOR) due to distinguishable difference in latency of gate operation extracted from the power profile. Therefore, functions do not need to be in AND-OR formats.

If IMC-paradigm (i.e., MAGIC vs DCIM) is unknown, adversary can identify IP by screening the power profiles ($P_{DCIM}<<P_{MAGIC}$). DCIM’s output is sensed by a sense-amplifier (low-power) while MAGIC’s output is written to an output RRAM (high-power). Non-parallel implementation of MAGIC uniquely exhibits multiple peaks in the power profile compared to DCIM.

\subsection{Applicability of Existing SCA Obfuscation Techniques}
SCA obfuscation techniques such as \cite{SCA_obfuscation} propose to inject random code execution to scramble power profile and prevent SCA on cryptographic implementations. Such protection techniques, if extended to IMC architectures, will impose significant throughput overhead since random functions between actual ones will incur extra delay. This is in addition to area and power overheads. In \cite{duplicate_logic}, duplicating logics with complementary operations are proposed to eliminate the asymmetry between power drawn to process 0 and 1. This technique will not protect IMC against SCA since the function and its complement may have different number of minterms. Therefore, they may consume different amount of power.

\subsection{Hspice Modeling \& Fabrication}
Conventional random logic can include various gate flavors (NAND/NOR/AOI/OAI/AND/OR/INV) which makes Hspice-model (or experimental chip)-based attack challenging. However, IMC circuit is systematic and only includes NAND/NOR gates due to SOP/POS implementation for simplicity. Furthermore, IMC using emerging NVMs provide distinct and high amplitude power signatures compared to CMOS gates. Therefore, RE of functionality will be challenging in CMOS even if adversary has accurate power-model of individual gates. In SCARE, we assume (Section {\ref{fab-chips}}) that adversary can fabricate few test-chips (costly) to characterize the power signature of individual gates in IMC if a model is not available. This is achievable by multiplexing power of multiple gates and enabling them one at a time.

Adversary will extract power/timing model from his own fabricated chips with test features to characterize individual gates although it will require high-precision equipment, time (due to multiple measurements). Obtaining PDK from vendor is considered easy (although NDA may be needed). Power/timing analysis of victim chip will be more challenging.


\section{Conclusions}
This paper proposes $SCARE$, a non-invasive RE on IMC using SCA for the first time. 
$SCARE$ is applied to two well-known emerging technology based IMC architectures (DCIM and MAGIC). The adversary extracts power/timing distributions from well-calibrated simulations or IMC test chips with known functions. Next, the functions are extracted by matching the probed power and timing profiles with modeled profiles for various gates and fanins. We also present possible countermeasures to mitigate $SCARE$ attack.

\vspace{+2mm}
\noindent \textbf{Acknowledgement:} This work is supported by SRC (2847.001), and NSF (CNS- 1722557, CCF-1718474, CNS-1814710, DGE-1723687 and DGE-1821766).

\bibliographystyle{IEEEtran}
\bibliography{IEEEabrv,SCA}
\end{document}